\documentclass[pra,aps,showpacs,groupedaddress,12pt,nofootinbib]{revtex4}
\usepackage{amssymb,amsmath,amstext,amsfonts,mathrsfs,bm,accents,mathtools,enumerate}

\newcommand{\cf}{\emph{cf.\;}}

\numberwithin{equation}{section}

\newtheorem{theorem}{Theorem}[section]
\newtheorem{proposition}{Proposition}[section]
\newtheorem{lemma}{Lemma}[section]

\newcommand{\be}{\begin{equation}}
\newcommand{\ee}{\end{equation}}
\newcommand{\ba}{\begin{align}}
\newcommand{\ea}{\end{align}}
\newcommand{\bet}{\begin{theorem}}
\newcommand{\et}{\end{theorem}}
\newcommand{\ed}{\end{document}}

\newcommand{\R}{\mathbb{R}}
\newcommand{\Rn}{\R^n}

\newcommand{\N}{\mathbb{N}}

\newcommand{\p}{\partial}
\newcommand{\zerost}{ (-\infty, 0] \times \R^3 }
\newcommand{\st}{{\R^-\times\R^3}}

\newcommand{\tzero}{{\{0\}\times \R^3}}

\newcommand{\g}{\mathfrak{g}}

\newcommand{\Ad}{\text{Ad} \,}
\newcommand{\ad}{\text{ad} \,}
\newcommand{\Aut}{\text{Aut} \,}
\newcommand{\tr}{\mathrm{tr} \,}

\newcommand\zero[1]{\accentset{(0)}{#1}}

\newcommand{\Ies}{\mathcal{I}_{es}} 
\newcommand{\A}{A} 
\newcommand{\bb}{\mathfrak{B}} 
\newcommand{\stA}{\mathscr{A}} 
\newcommand{\stF}{\mathscr{F}} 
\newcommand{\connLloc}{\mathfrak{A}^{2}_{1;loc}} 

\newcommand{\inst}{(-\infty, 0]\times\R^n}
\newcommand{\threest}{{\R^-\times\R^2}}
\newcommand{\nst}{{\R^-\times\R^n}}
\newcommand{\threetzero}{{\{0\}\times \R^2}}
\newcommand{\ntzero}{{\{0\}\times \R^n}}
 \newcommand{\rinfty}{{\{\Vert  {\bf x}\Vert = R\}\times (-\infty, 0]}}

\newcommand{\ie}{{\mathcal I}_{es}}

\newcommand{\ma}{\mathcal A}
\newcommand{\me}{\mathcal E}
\newcommand{\pp}{\mathcal P}

\newcommand{\fio}{{\Phi}_ {{\varphi}_0}}
\newcommand{\fiL}{{\Phi}_ {\varphi}^L}
\newcommand{\fioL}{{\Phi}_ {{\varphi}_0}^L}

\newcommand{\sn}{\smallskip\noindent}
\newcommand{\mn}{\medskip\noindent}

	\newcommand{\fun}{S_{(0)}}

 \numberwithin{equation}{section}

\begin{document}

\title{A Euclidean Signature Semi-Classical Program}
\author{Antonella Marini}
\affiliation{Department of Mathematics, \\ Yeshiva University, 500 West 185th Street, New York, NY 10033, USA. \\ and \\ Department of Mathematics, \\ University of L'Aquila, Via Vetoio, 67010 L'Aquila, AQ ITALY. \\ E-mail address: marini@yu.edu}
\author{Rachel Maitra}
\affiliation{Department of Applied Mathematics, \\ Wentworth Institute of Technology, 550 Huntington Avenue, Boston, MA 02115-5998, USA. \\ E-mail address: maitrar@wit.edu}
\author{Vincent Moncrief}
\affiliation{Department of Physics and Department of Mathematics, \\ Yale University, P.O. Box 208120, New Haven, CT 06520, USA. \\ E-mail address: vincent.moncrief@yale.edu}

\begin{abstract}
In this article we discuss our ongoing program to extend the scope of certain, well-developed microlocal methods for the asymptotic solution of Schr\"{o}dinger's equation (for suitable `nonlinear oscillatory' quantum mechanical systems) to the treatment of several physically significant, interacting quantum field theories. Our main focus is on applying these `Euclidean-signature semi-classical' methods to self-interacting (real) scalar fields of renormalizable type in 2, 3 and 4 spacetime dimensions and to Yang-Mills fields in 3 and 4 spacetime dimensions. A central argument in favor of our program is that the asymptotic methods for Schr\"{o}dinger operators developed in the microlocal literature are far superior, for the quantum mechanical systems to which they naturally apply, to the conventional WKB methods of the physics literature and that these methods can be modified, by techniques drawn from the calculus of variations and the analysis of elliptic boundary value problems, to apply to certain (bosonic) quantum field theories. Unlike conventional (Rayleigh/ Schr\"{o}dinger) perturbation theory these methods allow one to avoid the artificial decomposition of an interacting system into an approximating `unperturbed' system and its perturbation and instead to keep the nonlinearities (and, if present gauge invariances) of an interacting system intact at every level of the analysis.
\end{abstract}

\maketitle

\section{Introduction}
\label{sec:introduction}

	Perturbation theory, in its various guises, is one of the cornerstone calculational techniques of modern quantum field theory.  Its application to quantum electrodynamics, through its Feynman diagrammatic and path integral incarnations among other approaches, was one of the principal success stories of twentieth century physics and gave rise to numerous, spectacularly confirmed experimental predictions. The quantitative success of this particular application of perturbation methods hinged, in large measure, on the feasibility of decomposing the relevant quantum system (of coupled Dirac and Maxwell fields) into a relativistically invariant and explicitly solvable `unperturbed' system (the corresponding \textit{free} fields) and a complementary, nonlinear interaction that could realistically be treated as a `small' perturbation thereto.

	When one turns however to the quantization of \textit{non-abelian} gauge systems (i.e., to Yang-Mills fields either in isolation or coupled to other quantized objects), some of the advantages of the perturbative approach become seriously compromised. First of all, in order to have an explicitly solvable unperturbed system with which to initiate the analysis, one must dramatically modify the Yang-Mills system itself by discarding the very interaction terms that distinguish it from (several copies of) the Maxwell field. In doing so one necessarily disturbs the non-abelian character of the corresponding gauge group, `truncating' this essential element to the abelian gauge group of the approximating, Maxwell system. Naturally the aim of the associated perturbation theory, at least in part, is to restore these essential features. But it can only realistically hope to accomplish this through formal, asymptotic expansions in the corresponding Yang-Mills `coupling constant'. Elegant techniques for doing this are at the forefront of modern research in quantum field theory of which, for example, the discovery of \textit{asymptotic freedom} represented a crucial advance.

	The Feynman path integral approach to field quantization is not \textit{intrinsically} perturbative in nature (though it is often used in practice to derive elegant, relativistically invariant, perturbative expansions) and can, in its most mathematically coherent, Euclidean-signature formulation be used as a basis for rigorous attacks on quantization problems. Making mathematical sense of the needed (Euclidean-signature) functional integration measures in sufficient generality to handle Yang-Mills theories is however a daunting task, especially in the most physically interesting case of 4 spacetime dimensions wherein even more elementary, interacting quantized fields have continued to defy rigorous construction. Even in ordinary quantum mechanics there are known examples, which gave Feynman himself pause, of explicitly solvable quantum systems that forcefully resisted solution by purely path integral methods alone. Sometimes a direct, `old-fashioned' attack on Schr\"{o}dinger's equation is still the most effective approach.

	Of course the expression of such misgivings about the intrinsic limitations of perturbation theory and even path integral methods must have a hollow ring to it to experts in the field who must continually deal with the inherent difficulties of quantizing interacting systems---all the more so when this expression is coming from outsiders to this field who may have no concrete alternative to put forward. The present authors however have in fact developed an alternative approximation method for solving the Schr\"{o}dinger equations that arise in certain interacting quantum field theories and are confident that their method will shed some non-perturbative light on the nature of the associated quantum states. In particular our method does \textit{not} require that one artificially decompose an interacting quantum system into `free' and `interacting' components and thus, in the case of gauge theories, does \textit{not} require that one, equally artificially, truncate the corresponding gauge group to the abelian gauge group of an associated `free', approximating system. While we do not expect our approach to have the calculational power of conventional perturbation theory we do anticipate that it will allow an attack on some fundamental questions that more conventional  techniques seem inadequate to handle.

	To explain this approach let us backtrack slightly and mention that microlocal analysts have long since developed elegant, semi-classical approximation methods for the analysis of certain `nonlinear oscillatory' purely quantum mechanical systems that are vastly superior to the textbook Wentzel, Brillouin Kramers (or WKB) methods of the physics literature. The latter, which date back to the early days of quantum mechanics (though with precursors that are much older still), posit a formal expansion in Planck's constant of which the leading order term is precisely a solution of the classical Hamilton-Jacobi equation for the system being quantized. But for Schr\"{o}dinger operators of the conventional kinetic+potential type this amounts, among other things, to beginning with a complex oscillatory ansatz for a ground state wave function that is known, a priori, to be effectively real and monotonically decreasing in a suitable sense. That one arrives at the classical Hamilton-Jacobi equation at first approximation was an important step in the early days for convincing one that the then new quantum mechanics incorporated classical mechanics in an appropriate (formal) `correspondence principle' limit. However important this step was conceptually, mathematically it was catastrophic.

	First of all the Hamilton-Jacobi equation, even for the simplest, nontrivial, classical systems virtually never has global smooth solutions, not least of all because of the presence of caustics in the associated families of gradient-flow solution curves on the corresponding configuration manifolds. Even for the more or less tractable one dimensional systems (or those reducible to such through a separation of variables) one is obligated to effect a \textit{matching} of solutions across the boundary between `classically allowed' and `classically forbidden' regions in the associated configuration space. It is hardly surprising that, even for such basic problems as the harmonic oscillator one only gets rather crude approximations to its actual, explicitly known wave functions. The lesser known Einstein, Brillouin Keller (or EBK) approximation method does in fact have the capability of treating higher dimensional, interacting quantum systems but only those that are completely integrable at the classical level.  Maslov and others have developed ingenious ways of handling caustics when they occur but their methods are aimed at treating the finite time propagation of high frequency `wave packets' instead of solving the fundamental eigenvalue problems that form the core of quantum mechanics.

	In contrast to the above, microlocal analysts proposed a much more natural ansatz for the ground state wave function of a (nonlinear oscillatory) quantum system than that of conventional WKB theory and have shown that the leading order approximation in their approach is determined by a certain (nonlinear) `eikonal' equation which can, for many such systems, be proven to have a global smooth solution. Higher order corrections to this leading order term can then be computed through the systematic integration of a sequence of \textit{linear}, `transport' equations along the flow defined by the previously determined solution to the eikonal problem. Excited state wave functions and their corresponding eigenvalues and higher order corrections thereto can in turn be computed from an equally systematic, somewhat similar-in-spirit transport equations analysis. The eikonal equation alluded to above is nothing more than an `imaginary time' (or, in field theoretic language, `Euclidean signature') variant of the classical Hamilton-Jacobi equation for the system actually under study. While one may lose a certain `physical intuition' in dealing with this equation instead of the original, `real time' Hamilton-Jacobi one, the mathematical gain is enormous.

	In the very special case of \textit{harmonic} oscillators, for example, this approach reproduces, exactly, \textit{all} of the correct eigenfunctions and eigenvalues for this fundamental problem and for \textit{anharmonic} oscillatory systems provides, at the very least, provably asymptotic expansions for the solutions thereto. Most importantly, by avoiding the conventional decomposition into an explicitly solvable `unperturbed' system and its perturbation this microlocal approach secures, even at leading order, a much more accurate approximation to the actual wave functions of the interacting system than that provided by standard (Rayleigh/Schr\"{o}dinger) perturbation theory. On the other hand, for reasons that we shall clarify below, its methods for technically carrying out this elegant analysis seem to be essentially limited to finite dimensional, ordinary quantum mechanical systems and, in their current form, not to allow for an extension to quantum field theories.

	The authors, however, not being initially aware of these pre-existing microlocal results, embarked some years ago upon a program to develop semi-classical approximation methods that would in fact be applicable to certain interesting classes of interacting quantum field theories. In our approach the leading order term is a solution to the Hamilton-Jacobi equation for the \textit{Euclidean-signature} variant of the field equations that one is quantizing. This is an enormously more favorable setting than what would be provided by the corresponding Lorentzian-signature Hamilton-Jacobi equation of conventional WKB theory and lends itself to resolution via standard methods drawn from the calculus of variations. For the special cases of massive, polynomially interacting scalar fields of renormalizable type in 2, 3 and 4 spacetime dimensions, for example, we have been able to prove the existence of globally, Fr\'{e}chet smooth `fundamental solutions' to the relevant Euclidean-signature Hamilton-Jacobi equations and to substantially characterize their asymptotic behaviors. These solutions provide the bases for the computation, via linear transport equations, of their higher order `loop' corrections as well as that for the systematic computation of corresponding excited states. These latter steps require a detailed regularization and renormalization procedure that has not yet been carried out in the present context but, being all essentially linear in character, seem analytically less problematic than the ones we have already completed. For the special case of non-interacting (or `free') scalar fields our calculations truncate at finite order reproducing the functional representations of the conventional Fock-space exact solutions for these fundamental problems. The current status of this program is reviewed below in section \ref{subsec:phi}.

	To illustrate in some detail the general features of our \textit{Euclidean-signature semi-classical} program, including its approach to the handling of higher order loop corrections and excited state computations we recall, in section \ref{sec:microlocal}, its `fallback' application to `nonlinear oscillatory' problems in ordinary quantum mechanics and emphasize how our approach thereto differs from that of the microlocal literature---the key distinction that allows us to contemplate its application to quantum field theories. We also discuss therein some specific applications of our version of this microlocal, semi-classical program to certain quantum \textit{anharmonic} oscillators, comparing the results with those derivable from ordinary (Rayleigh/Schr\"{o}dinger) perturbation theory. A particularly interesting case is that of the quartic oscillator for which we found strong evidence that the asymptotic expansion generated for the (logarithm of its) ground state wave function may itself be Borel summable to what is conjecturally the \textit{exact solution} of this paradigmatic model problem. We also briefly discuss the applicability of our methods to the problems of solving the so-called Wheeler-DeWitt equation for Bianchi IX models of (minisuperspace) quantum cosmology and to the Schr\"{o}dinger operators arising in \textit{supersymmetric} quantum mechanics. The latter represents our first excursion into the issue of whether the methods under study can be generalized to include fermionic degrees of freedom.

	The ultimately most ambitious application of our program so far however, is to the case of quantized Yang-Mills fields in (the renormalizable instances of) 3 and 4 spacetime dimensions. In section \ref{subsec:ym} we discuss the current status of this project. In particular we present the essential features of a proof that, in a suitable function space setting, the Euclidean signature Yang-Mills action functional, defined over the Euclidean `half-space' \(\mathbb{R}^- \times \mathbb{R}^3\), admits an absolute minimizing connection, weakly satisfying the Yang-Mills Dirichlet problem, for `arbitrary' boundary data specified on the hypersurface \(\{0\}\times \mathbb{R}^3\). This extremized action, regarded as a functional of the boundary data (essentially a `spatial' connection on \(\mathbb{R}^3\)), provides us with the leading order approximation to (the logarithm of) our proposed ground state wave functional. We also give a preliminary discussion of the differentiability of the action functional so defined and present a sketch of how we hope to extend this argument to a full Fr\'{e}chet smoothness result. A potential application of our anticipated results to the \textit{mass gap} problem for quantum Yang-Mills theory is sketched in section \ref{subsec:b-e}.

We also include, in section \ref{subsec:w-dw}, a brief discussion of how our Euclidean signature semi-classical program might be applied to the (currently \textit{purely formal}) functional Wheeler-DeWitt equation in a non-symmetric, field theoretic setting. A key open question is whether one can develop and analyze a genuine elliptic boundary value problem for this highly nonlinear system and, if so, how the gravitational \textit{positive action theorem} can be successfully applied thereto.

\section{Microlocal Semi-Classical Methods for Quantum Mechanical Systems}
\label{sec:microlocal}
Elegant `microlocal analysis' methods have long since been developed for the study of Schr{\"o}dinger operators of the form
\begin{equation}\label{eq462}
\hat{H} = \frac{-\hbar^2}{2m} \Delta_g + V
\end{equation}
in the special cases for which the configuration manifold \(M \approx \mathbb{R}^n\), the metric \textit{g} is flat and for which the potential energy function \(V:M \rightarrow \mathbb{R}\) is of a suitable `non-linear oscillatory' type \cite{Moncrief:2012,Dimassi:1999,Helfer:1988,Helfer:1984}. These methods\footnote{For reasons to be clarified below we here follow a recent reformulation of the traditional microlocal approach developed by the authors in \cite{Moncrief:2012}.} begin with an ansatz for the ground state wave function of the form
\begin{equation}\label{eq401}
\overset{(0)}{\Psi}_{\!\hbar} (\mathbf{x}) = N_\hbar\; e^{-\mathcal{S}_\hbar (\mathbf{x})/\hbar}
\end{equation}
and proceed to derive asymptotic expansions for the logarithm, \(\mathcal{S}_\hbar:\mathbb{R}^n \rightarrow \mathbb{R}\), expressed formally as a power series in Planck's reduced constant (\(\hbar := h/2\pi\)),
\begin{equation}\label{eq402}
\begin{split}
\mathcal{S}_\hbar (\mathbf{x}) &\simeq \mathcal{S}_{(0)} (\mathbf{x}) + \hbar\mathcal{S}_{(1)} (\mathbf{x}) + \frac{\hbar^2}{2!} \mathcal{S}_{(2)} (\mathbf{x}) \\
 &+ \cdots + \frac{\hbar^n}{n!} \mathcal{S}_{(n)} (\mathbf{x}) + \cdots,
\end{split}
\end{equation}
together with the associated ground state energy eigenvalue \(\overset{(0)}{E}_{\!\hbar}\) expressed as
\begin{equation}\label{eq403}
\overset{(0)}{E}_{\!\hbar} \simeq \hbar (\overset{(0)}{\mathcal{E}}_{\!(0)} + \hbar\overset{(0)}{\mathcal{E}}_{\!(1)} + \frac{\hbar^2}{2!}\overset{(0)}{\mathcal{E}}_{\!(2)} + \cdots
 + \frac{\hbar^n}{n!}\overset{(0)}{\mathcal{E}}_{\!(n)} + \cdots).
\end{equation}
\(N_\hbar\) is a corresponding (for us inessential) normalization constant which one could always evaluate at any (finite) level of the calculation.

When the above ans{\"a}tze are substituted into the time-independent Schr{\"o}dinger equation and the latter is required to hold order-by-order in powers of \(\hbar\) the leading order term in the expansion (\ref{eq402}) is found to satisfy an \textit{inverted-potential-vanishing-energy} `Hamilton-Jacobi' equation given by
\begin{equation}\label{eq404}
\frac{1}{2m} g^{ij} \mathcal{S}_{(0),i} \mathcal{S}_{(0),i} - V = 0.
\end{equation}
For a large class of (non-linear oscillatory) potential energy functions and when \textit{g} is flat (with \(g = \sum_{i=1}^n dx^i \otimes dx^i\)) this equation can be proven to have a globally-defined, smooth, positive `fundamental solution' that is unique up to a (trivial) additive constant. In particular this is true whenever
\begin{enumerate}
\item \textit{V} is smooth, non-negative and has a unique global minimum attained at the origin of \(\mathbb{R}^n\) where \textit{V} vanishes,
\item \textit{V} can be expressed as
\begin{equation}\label{eq405}
V(x^1, \ldots ,x^n) = \frac{1}{2} \sum_{i=1}^n m\; \omega_i^2 (x^i)^2 + A(x^1, \ldots ,x^n)
\end{equation}
where each of the `frequencies' \(\omega_i > 0\) for \(i \in \lbrace 1, \ldots ,n\rbrace\) and wherein the smooth function \(A:\mathbb{R}^n \rightarrow \mathbb{R}\) satisfies
\begin{equation}\label{eq406}
A(0, \ldots ,0) = \frac{\partial A(0, \ldots ,0)}{\partial x^i} = \frac{\partial^2 A(0, \ldots , 0)}{\partial x^i\partial x^j} = 0\quad \forall\; i,j\; \in \lbrace 1, \ldots ,n\rbrace
\end{equation}
and the \textit{coercivity} condition
\begin{equation}\label{eq407}
A(x^1, \ldots ,x^n) \geq -\frac{1}{2} m \sum_{i=1}^n \lambda_i^2 (x^i)^2\quad \forall\; (x^1, \ldots ,x^n)\; \in \mathbb{R}^n
\end{equation}
and for some constants \(\lbrace\lambda_i\rbrace\) such that \(\lambda_i^2 < \omega_i^2\quad \forall\; i\; \in \lbrace 1, \ldots ,n\rbrace\), and
\item \textit{V} satisfies the \textit{convexity} condition
\begin{equation}\label{eq408}
\begin{split}
&\sum_{i,j=1}^n \frac{\partial^2 V(x^1, \ldots ,x^n)}{\partial x^i\partial x^j} \xi^i\xi^j \geq 0\\
&\hphantom{\sum_{i,j=1}^n}\forall\; (x^1, \ldots ,x^n)\; \in \mathbb{R}^n\; \hbox{and all}\\
&\hphantom{\sum_{i,j=1}^n}(\xi^1, \ldots ,\xi^n)\; \in \mathbb{R}^n.
\end{split}
\end{equation}
\end{enumerate}
Since only the sufficiency of these conditions was actually established in \cite{Moncrief:2012} it is quite conceivable that a satisfactory \textit{fundamental solution} to Eq.~(\ref{eq404}) exists under weaker hypotheses on the potential energy.

Our approach to proving the existence of a global, smooth fundamental solution to the (inverted-potential-vanishing-energy) Hamilton-Jacobi equation
\begin{equation}\label{eq409}
\frac{1}{2m} \nabla\mathcal{S}_{(0)} \cdot \nabla\mathcal{S}_{(0)} - V = 0
\end{equation}
is quite different from that developed previously in the microlocal literature but has the advantage of being applicable to certain field theoretic problems whereas it seems the latter does not.\footnote{The reasons for this apparent limitation are clarified in the discussion to follow.}

To establish the existence of \(\mathcal{S}_{(0)}\) we began by proving that the (inverted potential) action functional
\begin{equation}\label{eq410}
\begin{split}
\mathcal{I}_{ip}[\gamma] &:= \int_{-\infty}^0 \left\lbrace\frac{1}{2} m \sum_{i=1}^n \left\lbrack(\dot{x}^i(t))^2 + \omega_i^2 (x^i(t))^2\right\rbrack\right.\\
&{} + A\left.(x^i(t), \ldots, x^n(t))\vphantom{\sum_{i=1}^n}\right\rbrace\; dt,
\end{split}
\end{equation}
defined on an appropriate Sobolev space of curves \(\gamma :(-\infty,0] \rightarrow \mathbb{R}^n\), has a unique minimizer, \(\gamma_{\mathbf{x}}\), for any choice of boundary data
\begin{equation}\label{eq411}
\mathbf{x} = (x^1, \ldots ,x^n) = \lim_{t\nearrow 0} \gamma_{\mathbf{x}}(t)\; \in\; \mathbb{R}^n
\end{equation}
and that this minimizer always obeys
\begin{equation}\label{eq412}
\lim_{t\searrow -\infty} \gamma_{\mathbf{x}}(t) = (0, \ldots ,0).
\end{equation}
We then showed that every such minimizing curve is smooth and satisfies the (\textit{inverted potential}) Euler-Lagrange equation
\begin{equation}\label{eq413}
m \frac{d^2}{dt^2} \gamma_{\mathbf{x}}^i(t) = \frac{\partial V}{\partial x^i} (\gamma_{\mathbf{x}}(t))
\end{equation}
with vanishing (inverted potential) energy
\begin{equation}\label{eq414}
\begin{split}
E_{ip} (\gamma_{\mathbf{x}}(t),\dot{\gamma}_{\mathbf{x}}(t)) &:= \frac{1}{2}m \sum_{i=1}^n (\dot{\gamma}_{\mathbf{x}}^i(t))^2 - V(\gamma_{\mathbf{x}}(t))\\
&= 0\quad \forall\; t\; \in (-\infty,0] := I.
\end{split}
\end{equation}
Setting \(\mathcal{S}_{(0)}(\mathbf{x}) := \mathcal{I}_{ip} [\gamma_{\mathbf{x}}]\) for each \(\mathbf{x}\; \in \mathbb{R}^n\) we proceeded to prove, using the (Banach space) implicit function theorem, that the \(\mathcal{S}_{(0)}:\mathbb{R}^n \rightarrow \mathbb{R}\), so-defined, satisfies the Hamilton-Jacobi equation
\begin{equation}\label{eq415}
\frac{1}{2m} |\nabla\mathcal{S}_{(0)}|^2 - V = 0
\end{equation}
globally on \(\mathbb{R}^n\) and regenerates the minimizers \(\gamma_{\mathbf{x}}\) as the integral curves of its gradient (semi-)flow in the sense that
\begin{equation}\label{eq416}
\begin{split}
\frac{d}{dt} \gamma_{\mathbf{x}}(t) &= \frac{1}{m} \nabla\mathcal{S}_{(0)} (\gamma_{\mathbf{x}}(t))\\
 &\forall\; t\; \in I := (-\infty,0]\; \hbox{ and }\\
 &\forall\; \mathbf{x}\; \in \mathbb{R}^n
\end{split}
\end{equation}
Actually each such integral curve \(\gamma_{\mathbf{x}}:I \rightarrow \mathbb{R}^n\) extends to a larger interval, \((-\infty , t^\ast (\gamma_{\mathbf{x}}))\) with \(0 < t^\ast (\gamma_{\mathbf{x}}) \leq \infty\; \forall\; \mathbf{x}\; \in \mathbb{R}^n\) but since, in general, \(t^\ast (\gamma_{\mathbf{x}}) < \infty\) we only have a semi-flow rather than a complete flow generated by \(\frac{1}{m} \nabla\mathcal{S}_{(0)}\). Purely \textit{harmonic} oscillations on the other hand (for which \(A(x^1, \ldots ,x^n) = 0\)) are an exception, having \(t^\ast (\gamma_{\mathbf{x}}) = \infty\; \forall\; \mathbf{x}\; \in \mathbf{R}^n\).

Among the additional properties established for \(\mathcal{S}_{(0)}\) were the Taylor expansion formulas
\ba
\mathcal{S}_{(0)}(\mathbf{x}) &= \frac{1}{2} m \sum_{i=1}^n \omega_i(x^i)^2 + O(|\mathbf{x}|^3),\label{eq417}\\
\partial_j \mathcal{S}_{(0)}(\mathbf{x})  &= m\omega_jx^j + O(|\mathbf{x}|^2)\label{eq418}\\
\intertext{and}
\partial_j\partial_k\mathcal{S}_{(0)}(\mathbf{x}) &= m\omega_k\delta_j^k + O(|\mathbf{x}|),\label{eq419}
\end{align}
where here (exceptionally) no sum on the repeated index is to be taken, and the global lower bound
\begin{equation}\label{eq420}
\mathcal{S}_{(0)}(\mathbf{x}) \geq \mathcal{S}_{(0)}^\ast := \frac{1}{2} m \sum_{i=1}^n \nu_i (x^i)^2
\end{equation}
where \(\nu_i := \sqrt{\omega_i^2 - \lambda_i^2} > 0\; \forall\; i\; \in \lbrace 1, \ldots ,n\rbrace\). Note especially that this last inequality guarantees that, in particular, \(e^{-\mathcal{S}_{(0)}/\hbar}\) will always be normalizable on \(\lbrace\mathbb{R}^n,g = \sum_{i=1}^n dx^i \otimes dx^i\rbrace\).

The higher order `quantum corrections' to \(\mathcal{S}^{(0)}\) (i.e., the functions \(\mathcal{S}_{(k)}\) for \(k = 1, 2, \ldots\)) can now be computed through the systematic integration of a sequence of (first order, linear) `transport equations', derived from Schr{\"o}dinger's equation, along the integral curves of the gradient (semi-)flow generated by \(\mathcal{S}_{(0)}\). The natural demand for global smoothness of these quantum `loop corrections' forces the (heretofore undetermined) energy coefficients \(\lbrace\overset{(0)}{\mathcal{E}}_{\!(0)}, \overset{(0)}{\mathcal{E}}_{\!(1)}, \overset{(0)}{\mathcal{E}}_{\!(2)}, \ldots\rbrace\) all to take on specific, computable values.

Excited states can now be analyzed by substituting the ansatz
\begin{equation}\label{eq421}
\overset{(\ast)}{\Psi}_{\!\hbar} (\mathbf{x}) = \overset{(\ast)}{\phi}_{\!\hbar} (\mathbf{x}) e^{-\mathcal{S}_\hbar(\mathbf{x})/\hbar}
\end{equation}
into the time independent Schr{\"o}dinger equation and formally expanding the unknown wave functions \(\overset{(\ast)}{\phi}_{\!\hbar}\) and energy eigenvalues \(\overset{(\ast)}{E}_{\!\hbar}\) in powers of \(\hbar\) via
\ba
\overset{(\ast)}{\phi}_{\!\hbar} &\simeq \overset{(\ast)}{\phi}_{\!(0)} + \hbar\overset{(\ast)}{\phi}_{\!(1)} + \frac{\hbar^2}{2!}\overset{(\ast)}{\phi}_{(2)} + \cdots\label{eq422}\\
\overset{(\ast)}{E}_{\!\hbar} &\simeq \hbar\overset{(\ast)}{\mathcal{E}}_{\!\hbar} \simeq \hbar \left(\overset{(\ast)}{\mathcal{E}}_{\!(0)} + \hbar\overset{(\ast)}{\mathcal{E}}_{\!(1)} + \frac{\hbar^2}{2!}\overset{(\ast)}{\mathcal{E}}_{\!(2)} + \cdots\right) \label{eq423}
\end{align}
while retaining the `universal' factor \(e^{-\mathcal{S}_\hbar (\mathbf{x})/\hbar}\) determined by the ground state calculations.

From the leading order analysis one finds that these excited state expansions naturally allow themselves to be labelled by an \textit{n}-tuple \(\mathbf{m} = (m_1, m_2, \ldots , m_n)\) of non-negative integer `quantum numbers', \(m_i\), so that the foregoing notation can be refined to
\ba
\overset{(\mathbf{m})}{\Psi}_{\!\hbar} (\mathbf{x}) &= \overset{(\mathbf{m})}{\phi}_{\!\hbar} (\mathbf{x}) e^{-\mathcal{S}_\hbar (\mathbf{x})/\hbar}\label{eq424}\\
\intertext{and}
\overset{(\mathbf{m})}{E}_{\!\hbar} &= \hbar\overset{(\mathbf{m})}{\mathcal{E}}_{\!\hbar}\label{eq425}
\end{align}
with \(\overset{(\mathbf{m})}{\varphi}_{\!\hbar}\) and \(\overset{(\mathbf{m})}{\mathcal{E}}_{\!\hbar}\) expanded as before. Using methods that are already well-known from the microlocal literature \cite{Dimassi:1999} but slightly modified to accord with our setup \cite{Moncrief:2012} one can now compute all the coefficients \(\lbrace\overset{(\mathbf{m})}{\phi}_{\!(k)}, \overset{(\mathbf{m})}{\mathcal{E}}_{\!(k)}, k = 0, 1, 2 \ldots\rbrace\) through the solution of a sequence of linear, first order transport equations integrated along the semi-flow generated by \(\mathcal{S}_{(0)}\).

A key feature of this program, when applied to an n-dimensional \textit{harmonic} oscillator, is that it regenerates all the well-known, \textit{exact} results for both ground and excited states, correctly capturing not only the eigenvalues but the \textit{exact eigenfunctions} as well \cite{Moncrief:2012,Dimassi:1999,Helfer:1988}. One finds for example that the fundamental solution to the relevant (inverted-potential-vanishing-energy) Hamilton-Jacobi equation, for an n-dimensional oscillator (with mass \textit{m} and (strictly positive) oscillation frequencies \(\lbrace\omega_i\rbrace\)) is given by
\begin{equation}\label{eq426}
\mathcal{S}_{(0)} (\mathbf{x}) = \frac{1}{2} m \sum_{i=1}^n \omega_i (x^i)^2
\end{equation}
and that all higher order corrections to the logarithm of the ground state wave function vanish identically leaving the familiar gaussian
\begin{equation}\label{eq427}
\overset{(0)}{\Psi}_{\!\hbar} (\mathbf{x}) = \overset{(0)}{N}_{\!\hbar}\; e^{-\frac{m}{2\hbar} \sum_{i=1}^n \omega_i (x^i)^2}
\end{equation}
where \(\mathbf{x} = (x^1, \ldots ,x^n)\) and \(\overset{(0)}{N}_{\!\hbar}\) is a normalization constant.

The construction of excited states begins with the observation that the only globally regular solutions to the corresponding, leading order `transport equation' are composed of the monomials
\begin{equation}\label{eq428}
\overset{(\mathbf{m})}{\phi}_{\!(0)} (\mathbf{x}) = (x^1)^{m_1} (x^2)^{m_2} \cdots (x^n)^{m_n},
\end{equation}
where \(\mathbf{m} = (m_1, m_2, \ldots , m_n)\) is an n-tuple of non-negative integers with \(|m| := \sum_{i=1}^n m_i > 0\), and proceeds after a finite number of unequivocal steps, to assemble the exact excited eigenstate prefactor
\begin{equation}\label{eq429}
\begin{split}
\overset{(\mathbf{m})}{\phi}_{\!\hbar} (\mathbf{x}) &= \overset{(\mathbf{m})}{N}_{\!\hbar} H_{m_1} \left(\sqrt{\frac{m\omega_1}{\hbar}} x^1\right) H_{m_2} \left(\sqrt{\frac{m\omega_2}{\hbar}} x^2\right)\\
 &\cdots H_{m_n} \left(\sqrt{\frac{m\omega_n}{\hbar}} x^n\right)
\end{split}
\end{equation}
where \(H_k\) is the Hermite polynomial of order \textit{k} (and \(\overset{(\mathbf{m})}{N}_{\!k}\) is the corresponding normalization constant) \cite{Moncrief:2012,Dimassi:1999,Helfer:1988}.

While there is nothing especially astonishing about being able to rederive such well-known, exact results in a different way, we invite the reader to compare them with those obtainable via the textbook WKB methods of the physics literature \cite{Brack:2008,Ozorio:1988}. Even for purely \textit{harmonic} oscillators conventional WKB methods yield only rather rough approximations to the wave functions and are, in any case, practically limited to one-dimensional problems and to those reducible to such through a separation of variables. The lesser known Einstein Brillouin Keller (or EBK) extension of the traditional semi-classical methods does apply to higher (finite-)dimensional systems but only to those that are completely integrable at the classical level \cite{Stone:2005}. In sharp contrast to these well-established approximation methods the (Euclidean signature\footnote{The significance of this qualifying expression will become clear when we turn to field theoretic problems.}) semi-classical program that we are advocating here requires neither classical integrability nor (as we shall see) finite dimensionality for its implementation.

As was discussed in the concluding section of Ref.~\cite{Moncrief:2012} our fundamental solution, \(\mathcal{S}_{(0)} (\mathbf{x})\), to the (inverted-potential-vanishing-energy) Hamilton-Jacobi equation for a coupled system of nonlinear oscillators has a natural geometric interpretation. The graphs, in the associated phase space \(T^\ast\mathbb{R}^n\), of its positive and negative gradients correspond precisely to the stable \((W^{\mathrm{s}}(p) \subset T^\ast\mathbb{R}^n)\) and unstable \((W^{\mathrm{u}}(p) \subset T^\ast\mathbb{R}^n)\) Lagrangian submanifolds of the assumed, isolated equilibrium point \(p \in T^\ast\mathbb{R}^n\):
\ba
W^{\mathrm{u}}(p) &= \left\lbrace (\mathbf{x},\mathbf{p}):\mathbf{x} \in \mathbb{R}^n, \mathbf{p} = \nabla\mathcal{S}_{(0)} (\mathbf{x})\right\rbrace\label{eq430}\\
W^{\mathrm{s}}(p) &= \left\lbrace (\mathbf{x},\mathbf{p}):\mathbf{x} \in \mathbb{R}^n, \mathbf{p} = -\nabla\mathcal{S}_{(0)} (\mathbf{x})\right\rbrace\label{eq431}
\end{align}

Another result established for the aforementioned nonlinear oscillators of Ref.~\cite{Moncrief:2012} is that the first quantum `loop correction', \(\mathcal{S}_{(1)} (x^1, \ldots , x^n)\), to the (`tree level') fundamental solution, \(\mathcal{S}_{(0)} (x^1, \ldots , x^n)\), also has a natural geometric interpretation in terms of `Sternberg coordinates' for the gradient (semi-)flow generated by this fundamental solution. Sternberg coordinates, by construction, linearize the Hamilton-Jacobi flow equation
\ba
m \frac{dx^i (t)}{dt} &= \frac{\partial\mathcal{S}_{(0)}}{\partial x^i} (x^1(t), \ldots , x^n(t))\label{eq432}\\
\intertext{to the form}
\frac{dy^i(t)}{dt} &= \omega_i y^i(t)\quad \hbox{(no sum on \textit{i})}\label{eq433}
\end{align}
through, as was proven in Ref.~\cite{Moncrief:2012}, the application of a global diffeomorphism
\ba
&\mu:\mathbb{R}^n \rightarrow \mu(\mathbb{R}^n) \subset \mathbb{R}^n = \left\lbrace (y^1, \ldots , y^n)\right\rbrace,\label{eq434}\\
&\mathbf{x} \mapsto \mu(\mathbf{x}) = \left\lbrace y^1(\mathbf{x}), \ldots , y^n(\mathbf{x})\right\rbrace\label{eq435}
\end{align}
that maps \(\mathbb{R}^n\) to a star-shaped domain \(K = \mu(\mathbb{R}^n) \subset \mathbb{R}^n\) with \(\mu^{-1}(K) \approx \mathbb{R}^n = \left\lbrace (x^1, \ldots , x^n)\right\rbrace\).

Though not strictly needed for the constructions of Ref.~\cite{Moncrief:2012}, Sternberg coordinates have the natural feature of generating a Jacobian determinant for the Hilbert-space integration measure that \textit{exactly cancels} the contribution of the first quantum `loop correction', \(\mathcal{S}_{(1)}(\mathbf{x})\), to inner product calculations, taking, for example,
\ba
\begin{split}
\left\langle\overset{(\mathbf{m})}{\Psi},\overset{(\mathbf{m})}{\Psi}\right\rangle &:= \int_{\mathbb{R}^n} \left|\overset{(\mathbf{m})}{\Psi}(\mathbf{x})\right|^2\; d^n x\\
&= \int_{\mu(\mathbb{R}^n)} \left|\overset{(\mathbf{m})}{\Psi} \circ \mu^{-1}(\mathbf{y})\right|^2\; \sqrt{\det{g_{\ast\ast}}(\mathbf{y})}\; d^n y\label{eq436}
\end{split}
\intertext{to the form}
\begin{split}
\left\langle\overset{(\mathbf{m})}{\Psi},\overset{(\mathbf{m})}{\Psi}\right\rangle &= \int_{\mu(\mathbb{R}^n)} \left|\left\lbrack\overset{(\mathbf{m})}{\varphi} e^{\frac{-\mathcal{S}_{(0)}}{\hbar}-\frac{\hbar}{2!}\mathcal{S}_{(2)}+\cdots}\right\rbrack \circ \mu^{-1}(\mathbf{y})\right|^2\\
&\hphantom{=} \sqrt{\det{g_{\ast\ast}}(\mathbf{0})}\; d^n y\label{eq437}
\end{split}
\end{align}
where, in the last integral, the contribution of \(\mathcal{S}_{(1)} \circ \mu^{-1}(\mathbf{y})\) to the wave function
\begin{equation}\label{eq438}
\overset{(\mathbf{m})}{\Psi} \circ \mu^{-1} (\mathbf{y}) = \overset{(\mathbf{m})}{\varphi} e^{\frac{-\mathcal{S}_{(0)}}{\hbar}-\mathcal{S}_{(1)} - \frac{\hbar}{2!}\mathcal{S}_{(2)}\cdots}\; \circ \mu^{-1} (\mathbf{y})
\end{equation}
has precisely cancelled the non-Cartesian measure factor \(\sqrt{\det{g_{\ast\ast}} (\mathbf{y})}\), leaving the constant (Euclidean) factor \(\sqrt{\det{g_{\ast\ast}}(\mathbf{0})}\) in its place. Roughly speaking therefore, this role of \(\mathcal{S}_{(1)}\) is to `flatten out' the Sternberg coordinate volume element, reducing it to ordinary Lebesgue measure (albeit only over the star-shaped domain \(\mu(\mathbb{R}^n)\)), by exactly cancelling the Jacobian determinant that arises from the coordinate transformation.

For the nonlinear oscillators discussed in Ref.~\cite{Moncrief:2012}, Sternberg coordinates also have the remarkable property of allowing the leading order transport equation for \textit{excited states} to be solved in closed form. Indeed, the regular solutions to this equation are comprised of the monomials
\begin{equation}\label{eq439}
\overset{(\mathbf{m})}{\varphi}_{\!(0)} (\mathbf{y}) = (y^1)^{m_1} (y^2)^{m_2} \cdots (y^n)^{m_n}
\end{equation}
wherein, precisely as for the harmonic case, the \(m_i\) are non-negative integers with \(|\mathbf{m}| := \sum_{i=1}^n m_i > 0\). On the other hand the higher order corrections, \(\left\lbrace\overset{(\mathbf{m})}{\varphi}_{\!(k)}(\mathbf{y}); k = 1, 2, \ldots\right\rbrace\), to these excited state prefactors will not in general terminate at a finite order as they do for strictly \textit{harmonic} oscillators but they are nevertheless systematically computable through the sequential integration of a set of well-understood linear transport equations \cite{Moncrief:2012,Dimassi:1999}. Formal expansions (in powers of \(\hbar\)) for the corresponding (ground and excited state) energy \textit{eigenvalues} are uniquely determined by the demand for global regularity of the associated eigenfunction expansions. More precisely one finds, upon integrating the relevant transport equation at a given order, that the only potential breakdown of smoothness for the solution would necessarily occur at the `origin' \(\mathbf{x} = 0\) (chosen here to coincide with the global minimum of the potential energy) but that this loss of regularity can always be uniquely avoided by an appropriate choice of eigenvalue coefficient at the corresponding order.

A number of explicit calculations of the eigenfunctions and eigenvalues for a family of 1-dimensional \textit{anharmonic} oscillators of quartic, sectic, octic, and dectic types were carried out in Ref.~\cite{Moncrief:2012} and compared with the corresponding results from conventional Rayleigh/Schr{\"o}dinger perturbation theory. To the orders considered (and, conjecturally, to all orders) our eigenvalue expansions agreed with those of Rayleigh/Schr{\"o}dinger theory whereas our wave functions, even at leading order, more accurately captured the more-rapid-than-gaussian decay known rigorously to hold for the exact solutions to these problems. For the quartic oscillator in particular our results strongly suggested that both the ground state energy eigenvalue expansion and its associated wave function expansion are Borel summable to yield natural candidates for the actual exact ground state solution and its energy.

Remarkably all of the integrals involved in computing the quantum corrections \(\left\lbrace\mathcal{S}_{(1)}, \mathcal{S}_{(2)}, \mathcal{S}_{(3)}, \cdots\right\rbrace\) to \(\mathcal{S}_{(0)}\) (up to the highest order computed in \cite{Moncrief:2012}, namely \(\mathcal{S}_{(25)}\)) were expressible explicitly in terms of elementary functions for the \textit{quartic} and \textit{sectic} oscillators whereas for the octic and dectic cases some (but not all) of the quantum corrections required, in addition, hypergeometric functions for their evaluation. It seems plausible to conjecture that these patterns persist to all orders in \(\hbar\) and thus, for the quartic and sectic\footnote{These results were subsequently extended to significantly higher orders by P.~Tang~\cite{misc-orbit02}.} cases in particular, lead to formal expansions for \(\mathcal{S}_\hbar\) in terms of elementary functions. The evidence supporting the conjectured Borel summability of this formal expansion in the quartic case is discussed in detail in Section~V.A. of \cite{Moncrief:2012}.

For the Lagrangians normally considered in classical mechanics it would not be feasible to define their corresponding action functionals over (semi-)infinite domains, as we have done, since the integrals involved, when evaluated on solutions to the Euler-Lagrange equations, would almost never converge. It is only because of the special nature of our problem, with its inverted potential energy function and associated boundary conditions, that we could define a convergent action integral for the class of curves of interest and use this functional to determine corresponding minimizers.

A remarkable feature of our construction, given the hypotheses of convexity and coercivity imposed upon the potential energy \(V(\mathbf{x})\), is that it led to a \textit{globally smooth} solution to the corresponding Hamilton-Jacobi equation. Normally the solutions to a Hamilton-Jacobi equation in mechanics fail to exist globally, even for rather elementary problems, because of the occurrence of caustics in the associated families of solution curves. For our problem however caustics were non-existent for the (semi-)flow generated by the gradient of \(\mathcal{S}_{(0)}(\mathbf{x})\). The basic reason for this was the inverted potential character of the forces considered which led to the development of diverging (in the future time direction) solution curves having, in effect, uniformly positive Lyapunov exponents that served to prevent the occurrence of caustics altogether.

By contrast, the more conventional approach (in the physics literature) to semi-classical methods leads instead to a standard (non-inverted-potential-non-vanishing-energy) Hamilton-Jacobi equation for which, especially in higher dimensions, caustics are virtually unavoidable and for which, even in their absence, a nontrivial matching of solutions across the boundary separating classically allowed and classically forbidden regions must be performed. While Maslov and others have developed elegant methods for dealing with these complications \cite{Maslov:1981} their techniques are more appropriate in the short wavelength limit wherein wave packets of highly excited states are evolved for finite time intervals. On the other hand our approach is aimed at the ground and lower excited states though, in principle, it is not limited thereto.

As we have already mentioned though, our approach is a natural variation of one that has been extensively developed in the microlocal analysis literature but it also differs from this innovative work in fundamental ways that are crucial for our ultimate, intended application to field theoretic problems. In the microlocal approach \cite{Dimassi:1999,Helfer:1988,Helfer:1984} one begins by analyzing the (classical, inverted potential) dynamics locally, near an equilibrium, by appealing to the stable manifold theorem of mechanics \cite{Abraham:1978}. One then shows, by a separate argument, that, for an equilibrium \textit{p} (lying in some neighborhood \(U \subset \mathbb{R}^n\)) the corresponding stable (\(W^s(p) \subset T^\ast U\)) and unstable (\(W^u(p) \subset T^\ast U\)) submanifolds of the associated phase space \(T^\ast U\) are in fact Lagrangian submanifolds that can be characterized as graphs of the (positive and negative) gradients of a smooth function \(\phi:U \rightarrow \mathbb{R}\):
\ba
W^s(p) &= \left\lbrace (\mathbf{x},\mathbf{p}) | \mathbf{x} \in U, \mathbf{p} = \nabla\phi (\mathbf{x})\right\rbrace\label{eq440}\\
W^u(p) &= \left\lbrace (\mathbf{x},\mathbf{p}) | \mathbf{x} \in U, \mathbf{p} = -\nabla\phi (\mathbf{x})\right\rbrace .\label{eq441}
\end{align}
This function is shown to satisfy a certain `eikonal' equation (equivalent to our inverted-potential-vanishing-energy Hamilton-Jacobi equation restricted to \textit{U}) and \(\phi (\mathbf{x})\) itself is, of course, nothing but the (locally defined) analogue of our action function \(\mathcal{S}_{(0)}(\mathbf{x})\). A further argument is then needed to extend \(\phi(x)\) to a solution globally defined on \(\mathbb{R}^n\).

The potential energies, \(V(\mathbf{x})\), dealt with in the microlocal literature often entail multiple local minima, or ``wells'', for which our global convexity and coercivity hypotheses are not appropriate. Much of the detailed analysis therein involves a careful matching of locally defined approximate solutions (constructed on suitable neighborhoods of each well) to yield global asymptotic approximations to the eigenvalues and eigenfunctions for such problems. Since, however, we are focussed primarily on potential energies having single wells (corresponding to unique classical ``vacuum states''), many of the technical features of this elegant analysis are not directly relevant to the issues of interest herein.

For the case of a single well, however, we have essentially unified and globalized several of the, aforementioned, local arguments, replacing them with the integrated study of the properties of the (inverted potential) action functional (\ref{eq410}). When one turns from finite dimensional problems to field theoretic ones \cite{Marini:2016,Maitra:inprep} this change of analytical strategy will be seen to play an absolutely crucial role. For the typical (relativistic, bosonic) field theories of interest to us in this context, the Euler Lagrange equations for the corresponding, inverted potential action functionals that now arise are the \textit{Euclidean signature}, elliptic analogues of the Lorentzian signature, hyperbolic field equations that one is endeavoring to quantize. While generalizations of the aforementioned stable manifold theorem do exist for certain types of infinite dimensional dynamical systems, the elliptic field equations of interest to us do not correspond to well-defined dynamical systems \textit{at all}. In particular their associated Cauchy initial value problems are never well-posed. This is the main reason, in our opinion, why the traditional microlocal methods have not heretofore been applicable to quantum field theories.

On the other hand the direct method of the calculus of variations is applicable to the Euclidean signature action functionals of interest to us here and allows one to generalize the principle arguments discussed above to a natural infinite dimensional setting.

Before turning to such field theoretic generalizations however, we wish to draw attention to two particular, purely quantum mechanical extensions of the methods sketched above. The first of these entails the application of (a suitably extended version of) the methods in question to so-called SUSY QM or `supersymmetric quantum mechanics'. In SUSY QM a set of commuting, `bosonic' quantum degrees of freedom \(\lbrace x^1, \ldots , x^n\rbrace\) is introduced and matched by an equal number of anticommuting, `fermionic' degrees of freedom \(\lbrace\Psi^1, \ldots , \Psi^n\rbrace\). In terms of these a pair (in the simplest case of so-called \(N = 2\) supersymmetry) of self-adjoint supersymmetry `charge' operators \(\lbrace Q_1, Q_2\rbrace\) is defined through the introduction of a `superpotential' function \(W (x^1, \ldots , x^n)\) and the Hamiltonian \textit{H} for the system expressed algebraically as a certain anticommutator of these charges. Depending upon the choice of the superpotential such systems may exhibit either unbroken or broken supersymmetry and thus provide simplified, finite dimensional models for corresponding field theoretic systems.

The Schr\"{o}dinger equation for such a SUSY QM system can be concretely represented in terms of an ordinary matrix partial differential operator acting upon a multicomponent, `spinorial' wave function. In Ref. \cite{Moncrief:inprep2} we have begun to investigate the applicability of the `microlocal' methods sketched above to such equations. Since fermion number is exactly conserved for such systems the corresponding Schr\"{o}dinger operator assumes a block-diagonal form wherein each block corresponds to a fixed number of fermions which, thanks to the Pauli exclusion principle, varies from 0 to \textit{n}. The associated Hilbert space thus includes a completely empty and a fully filled femionic sector upon each of which the corresponding Schr\"{o}dinger operator reduces to an elementary, single component form to which the methods discussed above are, with only slight modifications, immediately applicable. Indeed, one can readily apply them to analyze, in depth, both the ground and excited states for the cases of both unbroken and broken supersymmetry.

The (anti-) commutation relations satisfied by the charge operators \(\lbrace Q_1, Q_2\rbrace\) and the Hamiltonian allow one to relate, to some extent, the eigenstates and eigenvalues corresponding to different fermionic subspaces of the full Hilbert space by, roughly speaking, using \(Q^\dagger = 1/2 (Q_1 - iQ_2)\)  and  \(Q = 1/2 (Q_1 + i Q_2)\) as (fermion number) raising and lowering operators. However, the nilpotency of these SUSY charges prevents one, when \(n > 2\), from generating all of the eigenvalues and eigenstates through the actions of \(Q^\dagger\) and \textit{Q} on the fermion vacuum and fermion-filled eigenspaces. Thus for such higher dimensional problems one must address directly the solution of nontrivial matrix Schr\"{o}dinger eigenvalue problems. Since our analysis of these multicomponent problems is not yet complete we shall only remark here that, in the corresponding, `microlocal' approach to their solution, the fundamental solution \(S_{(0)}(x)\) to the \textit{inverted-potential-vanish-energy} `Hamilton-Jacobi' equation (\ref{eq404}) continues to play a key role but must now be supplemented with the solution of a sequence of multi-component, linear transport equations integrated along the (semi-) flow generated by the fundamental solution \(S_{(0)}\).

Though these SUSY QM problems are only rough models for their field theoretic analogues they represent our first excursion into the realm of fermionic degrees of freedom wherein, for us at least, the main open question is whether our (Euclidean-signature-semi-classical) methods can indeed be applied to genuine fermionic field theories.

The second quantum mechanical extension of the microlocal semi-classical methods that we wish to mention is that to a problem in `quantum cosmology' --- namely the problem of solving the relevant Wheeler-DeWitt equation for spatially homogeneous, Bianchi type IX (or `Mixmaster') universes. Though the (partial differential) Wheeler-DeWitt equation for this model problem was first formulated nearly a half century ago, techniques for its solution that bring to light the discrete, quantized character naturally to be expected for its solution, have only recently been developed. In particular the microlocal analytical methods that we have already sketched for the study of conventional Schr\"{o}dinger operators can be modified in such a way as to apply to this equation \cite{Moncrief:2015,Bae:2015}.

That some essential modification of the microlocal methods is needed for the analysis of this equation is evident from the fact that the Wheeler-DeWitt equation does not define an eigenvalue problem, in the conventional sense, \textit{at all}. For spatially closed universe models, such as those of Mixmaster type, all of the would-be eigenvalues of the Wheeler-DeWitt operator, whether for `ground' or `excited' quantum states are required to \textit{vanish identically}. But a crucial feature of standard microlocal methods, when applied to conventional Schr\"{o}dinger eigenvalue problems, exploits the flexibility to adjust the eigenvalues being generated, order-by-order in an expansion in Planck's constant, to ensure the global smoothness of the eigenfunctions, being constructed in parallel, at the corresponding order. But if, as in the Wheeler-DeWitt problem, there are no eigenvalues to adjust, wherein lies the flexibility needed to ensure the required smoothness of the hypothetical eigenfunctions? And, by the same token, where are the `quantum numbers' that one would normally expect to have at hand to label the distinct quantum states? Remarkably however, as was shown in detail in Refs. \cite{Moncrief:2015,Bae:2015}, the scope of microlocal methods can indeed, in spite of this apparent impasse, be broadened to provide creditable, aesthetically appealing answers to the questions raised above.

While it is far from clear whether such methods could ever be further generalized to apply to the field theoretic Wheeler-DeWitt operator of (formal) canonical quantum gravity, a sketch of how that might be carried out (in a Euclidean-signature semi-classical expansion) by appealing to the `positive action theorem' is given in the concluding section of Ref. \cite{Moncrief:2015} and reviewed in Section~\ref{subsec:w-dw} below. In particular, we draw attention there to several remarkably attractive features of such an approach and show how it avoids some of the serious complications that obstructed progress on the, somewhat similar-in-spirit, \textit{Euclidean-path-integral} approach to quantum gravity. 

\section{Modified Semi-Classical Methods for Bosonic fields}
\label{sec:bosonic_fields}

\subsection{An application to polynomial scalar field theories}
\label{subsec:phi}

For technical reasons, the elegant `microlocal' methods developed in the past for the analysis of conventional Schr\"{o}dinger eigenvalue problems, have not heretofore been applicable to quantum field theories. In this section we describe a `Euclidean signature semi-classical' program to extend the scope of these analytical techniques to encompass the study of self-interacting (massive) scalar fields with polynomial renormalizable self-interaction in \(1+1\), \(2 + 1\) and \(3 + 1\) dimensions. The basic microlocal approach entails the solution of a single, nonlinear equation of Hamilton-Jacobi type followed by the integration (for both ground and excited states) of a sequence of \textit{linear} `transport' equations along the `flow' generated by the `fundamental solution'. As for the finite-dimensional problems described in the previous section, the authors' approach naturally splits into a single nonlinear, but essentially `classical'  problem (the solution of the basic inverted-potential-vanishing-energy Hamilton-Jacobi equation) and a sequence of linear calculations of quantum corrections.

In this section, we apply our method to suitable polynomial field theories and establish existence, uniqueness and global regularity for a globally defined `fundamental solution' of the Hamilton-Jacobi equation mentioned above, via an application of the direct method of the calculus of variations, elliptic theory and a Banach space version of the implicit function theorem.

In particular, such polynomial field theories include the cases of $\Phi^4$ scalar fields in $4$-dimensional Minkowski spacetime,  $\Phi^p$ fields with exponents $p = 4,$ and $6$ in $3$-dimensional Minkowski space, and with $p$ an arbitrary, positive even integer greater than $2$ in $2$-dimensional Minkowski space.

For each case, global existence and smoothness (in a suitable function space setting) of the relevant Hamilton-Jacobi functional, $S_{(0)} [\varphi]$ is proven and it is discussed how our tree approximation for these ground state wave functionals captures the more-rapid-than-gaussian decay that should hold for the exact solutions. Since, in our setup, the squared modulus of the ground state wave functional provides the natural integration measure for the associated Hilbert space of quantum states, it seems encouraging that our approach exhibits this non-Fock-like behavior already at leading order. By contrast note that, to any finite order, conventional Rayleigh/Schr\"odinger theory would generate instead an approximate wave functional that decays more \emph {slowly} than the corresponding gaussian. Because our techniques, even when applied to quantum mechanics problems, do not simply reproduce the standard results of  Rayleigh/Schr\"odinger perturbation theory, we expect that one should be able to generate much more accurate approximations to the Hilbert space of quantum states for certain quantized fields.


Although in this section we only deal with the aforementioned fundamental solutions, the fields considered are rigorously proven to be non-trivial  in lower dimensions; \cite{GlimmJaffeI, GlimmJaffeII, GlimmJaffeIII, GlimmJaffeIV, Feldman:1976}. On the contrary,
$\Phi^4$  fields in $4$ dimensions are often believed (though still not rigorously proven \cite{Podolsky:2010, Suslov:2008}) to renormalize to (trivial) free fields. If such a conclusion were proven to be correct, it would only emerge at the level of the higher order quantum, `loop' corrections, because the necessity to regularize otherwise ill-defined functional Laplacians only arises in our approach at the level of the transport equations for these higher order corrections. The study of the higher order, `loop' corrections, for both ground and excited states, is work currently in progress.

The formal Schr\"odinger operator we consider for a scalar field in $n + 1$ dimensions, with $n=1, 2$, or $3$,  is
\be
\label{23}
\hat H = \int_{\R^n} \left\{-\frac{\hbar^2}{2}\,\frac{\delta^2}{\delta \varphi^2( {\bf x^\prime})} + \frac{1}{2}\nabla^\prime\varphi({\bf x^\prime})\cdot\nabla^\prime\varphi({\bf x^\prime}) + \mathcal P(\varphi({\bf x^\prime})) \right\} \,d{\bf x^\prime}\,,
\ee
in which ${\bf x^\prime}\in\R^n$, $\nabla^\prime$ is the `spatial'  gradient on $\R^n$, and the self-interaction polynomial $\mathcal P(\cdot)\equiv\sum_{j=2}^k a_j(\cdot)^j$ is assumed to be convex, to possess no linear term,  to include a `mass' term,  $a_2>0$, and to be of even degree $k$,
with
\be\label{expcrit}   k\leq\frac{2d}{d-2}\equiv \frac{2n+2}{n-1} \;\mbox { if } \,n=2, 3\,,
\ee
in which $d\equiv n+1$ is the dimension of the domain. The number ${2d}/{(d-2)}$ above is  the \emph{critical exponent} for the elliptic theory applied to the
 \emph{Euclidean signature} action functional
\be
\label{29}
\mathcal I_{es} [\Phi] \equiv \int_{\R^n} \int_{-\infty}^0\left\{ 1/2\, (\partial_t\Phi)^2  + 1/2\, |\nabla^\prime\Phi|^2+ \mathcal P(\Phi)\right\} \,dt\,d{\bf x^\prime}\equiv \int_\nst
\left(\frac{\vert\nabla\Phi\vert^2}{2} + \mathcal P(\Phi)
\right) d{\bf x}\,,
\ee
which is guaranteed to be coercive by the hypotheses imposed upon $\mathcal P$. (These assumptions are satisfied by the  `massive'  $\Phi^4$ theory, for which $\mathcal P(\Phi)=1/2 \,m^2 \,\Phi^2 +\lambda \,\Phi^4$ with $\lambda, m^2>0$.)

The functional Laplacian,  formally defined as `trace of the Hessian',  will require some `regularization' in order to be well-defined on the (wave) functionals of interest; in fact, although the Hessians of the latter functionals will be smooth, they need  not be of trace class.
The needed regularization however would not affect the determination of the `fundamental solution', $S_{(0)} [\varphi(\cdot)]$, to the Euclidean signature-vanishing-energy variant of the Hamilton-Jacobi equation,
\be
\label{24}
\int_{\R^n} \left(\frac{1}{2}\frac{\delta S_{(0)}}{\delta\varphi( {\bf x^\prime}) }\,\frac{\delta S_{(0)}}{\delta\varphi( {\bf x^\prime}) } -\frac{1}{2}\nabla^\prime\varphi({\bf x^\prime})\cdot\nabla^\prime\varphi({\bf x^\prime}) - \mathcal P(\varphi({\bf x^\prime}) ) \right) \,d{\bf x^\prime}=0\,,
\ee
that arises at lowest order from substituting our ansatz
\be
\label{25}
\zero\Psi_\hbar [\varphi(\cdot)] = N_\hbar \,e^{-{S}_\hbar[\varphi(\cdot)]/\hbar}
\ee
for the ground state wave functional into the Schr\"odinger equation
\be
\label{26}
\hat H\,\zero\Psi_\hbar = \zero E_\hbar \zero\Psi_\hbar
\ee
and demanding that the latter hold order-by-order in powers of $\hbar$ relative to the formal expansions
\be
S_{\hbar}[\varphi(\cdot)] \simeq S_{(0)}[\varphi(\cdot)] + \hbar S_{(1)}[\varphi(\cdot)] + \frac{\hbar^{2}}{2!} S_{(2)}[\varphi(\cdot)] + \dots + \frac{\hbar^{k}}{k!} S_{(k)}[\varphi(\cdot)] + \dots
\ee
and
\be
\label{E}
	\zero{E}_{\hbar} \simeq \hbar \left(\zero{\mathcal{E}}_{(0)} + \hbar \,\zero{\mathcal{E}}_{(1)} + \frac{\hbar^{2}}{2!} \,\zero{\mathcal{E}}_{(2)} + \dots + \frac{\hbar^{k}}{k!} \,\zero{\mathcal{E}}_{(k)} + \dots\right)\,.
\ee
In the formulas above $\varphi(\cdot)$ denotes a real-valued distribution on  $\R^n$,  boundary data for a real spacetime scalar field $\Phi$ defined on   $(-\infty, 0] \times\R^n$.

As for the finite-dimensional problem described in Section \ref{sec:microlocal},   a `fundamental solution' to Eq. \eqref{24}  is constructed by proving the existence of unique minimizers $\Phi_\varphi$ for the \emph{Euclidean signature} action functional \eqref{29}
for arbitrary boundary data $\varphi$ specified at $t = 0$, and setting
\be
\label
{H-J funct}
S[\varphi] \equiv S_{(0)}[\varphi] = \Ies[\Phi_\varphi]= \inf_{\Phi\in\ma(\varphi)} \Ies[\Phi]\,.
\ee
The minimization procedure is carried out for $\Phi$ in the space of distributions
\be
\label{space}
\ma(\varphi)\equiv\{\Phi\in H_1 (\nst)\,:\, \Phi= \varphi \mbox{ at }
\ntzero\,\},
\ee
for fixed arbitrary $\varphi\in\bb\equiv \{\varphi = \tr \, \tilde\varphi\,, \mbox { with } \tilde \varphi\in  H_{{3}/{2}}(\nst)\}\simeq H_{1}(\ntzero)$. The latter isomorphism of spaces holds because the trace maps,
 $\tr: W_{k;p}(\nst)\to W_{k-1/p;p}(\ntzero)$, are surjective if $k>1/p$; \cf Theorem 7.58 in \cite{Adams}. In particular,
one may choose an extension $\tilde \varphi$ which is smooth in the interior; \cf Lemma 5.1 in \cite{Marini:2016}.
   \footnote{The above choice of the space of boundary data $\bb$, although not the \emph{natural} one for a minimization procedure taking place in $H_1(\nst)$, is required for the Hamilton-Jacobi equation to hold strongly, besides being instrumental to implementing the first step in the regularity theory, and  constitutes a non-substantial restriction. Weakening the requirement on the boundary space $\bb$, for example by imposing the boundary data to be in $H_{1/2} (\ntzero)$,  would not in substance yield a  stronger theorem. In that case in fact, the absolute minimizer for $\Ies$  would then be in $H_{1}(\nst)$ rather than in $H_{3/2}(\nst)$, and the Hamilton-Jacobi equations would not hold strongly. An analogous remark can be made for the case of the Dirichlet problem for Yang-Mills connections analyzed in the next section.}

 Coerciveness, weak lower sequential semicontinuity, and strict convexity of $\ie$ on $\ma(\varphi)$ (\cf Theorem 1.1.3 of Ref.~\cite{Blanchard1992}, and Chapter 3 of \cite{Marini:2016}) guarantee existence and uniqueness for the
absolute minimizer $\Phi_\varphi$; \cite{Marini:2016}.  Uniqueness of the absolute minimizer, although not necessary to guarantee that $\fun [\varphi]$ be well-defined, is required for `everywhere' differentiability of $\fun$ with respect to the boundary data to hold.\footnote{In the
next section we will argue for the case of Yang-Mills connections that non-uniqueness is an obstruction to everywhere differentiability for the suitable analogue of $\fun$; with regard to this, see  also the discussion under the title ``Smoothness of the Hamilton-Jacobi functional" further below in the present section.}

A candidate for the \emph{tree approximation} to the ground state wave functional for the $\mathcal P (\Phi)$ theory is the functional
\be\label{Omega}
\Omega_0(\varphi) \equiv \mathcal N e^{- {\fun}[\varphi]/ \hbar}\,,
\ee
in which $\mathcal N$ is a normalization constant.

\noindent{\bf\emph{The weak Dirichlet problem satisfied by the absolute minimizer.}}
With regard to existence and regularity for the elliptic theory associated to the functional $\Ies$ via the calculus of variations, in \cite{Marini:2016} the authors follow the essential lines of the classical references \cite{Agmon, Gilbarg-Trudinger},  with some perhaps minor modifications of those. Here, we highlight the general outline and the main differences in comparison to the classical literature.
We start by providing
 fundamental definitions based on the standard theory of the calculus of variations. Once these well-founded definitions are established and acquired, one can develop the resulting theory, with no need for repetitive reproduction of those in each specific  context. 
What perhaps may well be simply a slight shift in the philosophical approach to boundary value problems, affects the coherence and elegance of the theory and, in some instances, its content.

With that in mind, we define a weakly differentiable function $\Phi$ to be a \emph{weak solution} to the \emph{Dirichlet Problem} \be
\label{wdp}\quad {(\mathcal{D^{\,\prime}  })}\qquad\left\{\begin{array}{ll}
\Delta \Phi = f \quad&\mbox{on }\nst\\
\Phi\in  \ma(\varphi)\,,
\end{array}\right.
\ee
with $f$ prescribed in $H_{-1}(\nst)$, the topological dual space of $\ma(0)$, and $\varphi$
prescribed in $H_{1/2}(\ntzero)$, \emph{if and only if}
\be
\label{wdp1}
\Phi \in \ma(\varphi)\;\mbox { and } \;  \int_\nst  \nabla  \Phi \cdot \nabla \omega \,d{\bf x} + \int_\nst f \,\omega \,d{\bf x} =0\qquad \forall \omega \in \ma^\infty_c(0)\,,
\ee
with
\be\label{ainfty}
\ma^\infty_c (0) \equiv \left\{\omega\in \mathcal C^\infty_c\left((-\infty, 0]\times \R^n\right)\;:\; \omega = 0 \mbox{ at } \ntzero\right\}\,.
\ee
In comparison with Ref.~\cite{Agmon}, notice that here we do not assume the boundary value and its extension to the interior to be $\mathcal C^\infty$,  in replacement, we elect to prescribe
$\varphi\in H_{1/2}(\ntzero)$; further, we assume  $f\in H_{-1}(\ntzero)$, in replacement of the assumption $f\in L^2(\ntzero)$.

Because  the space $\ma^\infty_c(0)$ is dense in $\ma(0)$ (just as $\mathcal C^\infty_c (\R^n)$ is dense in $H_1(\R^n)$, yielding  $H_1(\R^n)\equiv H_{1;0}(\R^n)$),
when  $\eqref {wdp1}$ is satisfied, it applies also to all $\omega \in \ma(0)$.

Definition \eqref{wdp} implicitly defines the \emph{weak Dirichlet Laplacian on $\ma(\varphi)$} through the identification $\Delta\Phi\simeq \int_\nst  -\nabla  \Phi \cdot \nabla (\cdot) \,d{\bf x}$ (that is, we regard  $\Delta\Phi$, for each $\Phi\in \ma(\varphi)$, as an element of $H_{-1}(\nst)$).

\mn
In analogy, we define a weakly differentiable function
$\Phi$ to be a \emph{weak solution} to the \emph{nonlinear Dirichlet Problem}
\be
\label{wnl}\quad \mathcal{(D^{\,\prime\prime})}\qquad\left\{\begin{array}{ll}
\Lambda  \,\Phi \equiv - \Delta \Phi +\mathcal  Q (\Phi) =0\quad&\mbox{on } \nst  \\
\Phi\in  \ma(\varphi)\,,
\end{array}\right.
\ee
in which $\mathcal Q$ is a given nonlinear function of a single variable, such that $\mathcal Q(\Phi)\in H_{-1}(\nst)$, and  with $\varphi$ prescribed in $H_{1/2}(\ntzero)$,  \emph{if and only if}
\be
\label{wnl1}
\Phi \in \ma(\varphi), \mbox { and } \int_\nst  \left(\nabla  \Phi \cdot \nabla \omega  + \mathcal Q (\Phi) \,\omega \right)\,d{\bf x} =0\; \quad \forall \omega \in \ma^\infty_c(0)
\ee
In the definition above, $\Lambda\equiv - \Delta +\mathcal  Q (\cdot)$  is a nonlinear differential operator with image in  $H_{-1}(\nst)$, namely,
\ba
\label{Lambda}
\Lambda\;:\;\ma(\varphi) &\to H_{-1}(\nst)\,\notag\\
\Phi &\mapsto\Lambda \Phi \equiv - \Delta \Phi + \mathcal Q  (\Phi)\simeq  \int_\nst\bigl(\nabla\Phi\cdot\nabla (\cdot)+\mathcal Q(\Phi)(\cdot)\bigr)\,d{\bf x} \,.
\end{align}
In the context of the polynomial theory under study, we set $\mathcal Q\equiv \mathcal P^\prime$. In fact, the conditions imposed on $\mathcal P$ described in the current section guarantee
 that
\be
\label{estLambdab}
\Vert \Lambda\Phi\Vert_{H_{-1}(\nst)}\equiv\sup_{\omega\in\ma (0)}
\frac{\left| \Lambda\Phi(\omega)\right|}
{\Vert\omega\Vert_{H_1(\nst)}}\leq C\, \max\,\left\{\Vert \Phi\Vert_{H_1(\nst)}, \Vert \Phi\Vert^{k-1}_{H_1(\nst)}\right\}\,,
\ee
for some constant $C$ independent of $\Phi$, thus
$\Lambda\Phi$ is a continuous linear operator on $\ma(0)$.

\sn
In wide generality, given a differentiable function of a real variable $\mathcal P$,  satisfying $\mathcal P^\prime(\Phi)\in H_{-1} (\nst)$ for all $\Phi\in\ma(\varphi)$, setting $\mathcal Q(\Phi) \equiv \mathcal P^\prime(\Phi)$ in \eqref{wnl},
one can show that if the functional
$$\mathcal J[\Phi]\equiv \int_\nst  \left(\nabla\Phi\cdot\nabla\Phi + \mathcal P(\Phi)\right)\,d{\bf x}$$ possesses a  stationary point $\Phi\in\ma(\varphi)$, such $\Phi$ would
also be a weak solution  to the nonlinear Dirichlet problem \eqref{wnl}, and vice versa (by a density argument).

In fact, for $\Phi\in \ma(\varphi)$, $\omega \in\ma(0)$, one has
\ba & D\mathcal J[\Phi](\omega)\equiv\lim_{\lambda\to 0} \frac{1}{\lambda}\left\{\int_\nst  \left(\nabla(\Phi + \lambda\omega) \cdot\nabla (\Phi + \lambda\omega)+ \mathcal P(\Phi+ \lambda\omega)\right)\,d{\bf x} \right.\notag\\
&\left. -\int_\nst  \left(\nabla\Phi\cdot\nabla\Phi + \mathcal P(\Phi)\right)\,d{\bf x} \right\}=
\int_\nst\bigl(  \nabla\Phi\cdot\nabla\omega  + \mathcal P^\prime (\Phi) \,
\omega\bigr)  \,d{\bf x}\,;\notag\\
\end{align}
thus, $\Phi\in\ma(\varphi)$ is a stationary point  of $\mathcal J[\Phi]$ \emph{if and only if} it satisfies $\int_\nst\bigl(  \nabla\Phi\cdot\nabla\omega  + \mathcal P^\prime (\Phi) \,
\omega\bigr)  \,d{\bf x}=0\;\forall \omega\in \ma(0)$, or, by definition, if it satisfies  the \emph{weak Euler-Lagrange equations}  in \eqref{wnl}, with $\mathcal Q$ substituted by $\mathcal P^\prime$.

In the context of the polynomial theory under study,   $\mathcal P(\cdot)$ is a polynomial satisfying conditions which guarantee the existence of a unique (absolute) minimizer of $\mathcal J=\ie$ on $\ma(\varphi)$, with prescribed boundary value $\varphi\in  \mathcal B$, and  such a minimizer is also its unique stationary point, that is,  the (unique) weak solution to the Euler-Lagrange equations contained in \eqref{wnl}, with $\mathcal Q$ substituted by $\mathcal P^\prime$.

\noindent{\bf \emph{Elliptic regularity and global control.}}  Having ascertained that the unique absolute minimizer of $\Ies$ over $\ma(\varphi)$, with prescribed $\varphi\in  \mathcal B$, satisfies weakly
the nonlinear Dirichlet Problem
\be
\label{wnlP}
\left\{\begin{array}{ll}
\Lambda  \,\Phi \equiv - \Delta \Phi +\mathcal  P^\prime (\Phi) =0\quad&\mbox{on } \nst  \\
\Phi\in  \ma(\varphi)\,,
\end{array}\right.
\ee
one wants to obtain interior regularity of $\Phi$, as well as a certain degree of boundary regularity and integrability, and a global estimate over the unbounded domain $\nst$ (\cf Theorem \ref{T-reg1} below),
by means of bootstrapping and elliptic estimates.
All proofs are more elementary in the \emph{subcritical} cases, i.e. for which the degree of $\mathcal P$ is strictly less than the critical exponent, and for polynomials of any degree in  $2$ dimensions (\cf \eqref{expcrit}).
In fact, in subcritical cases, one can exploit the existence theory, freeze the polynomial term and replace the weak nonlinear Dirchlet problem \eqref{wnlP} by  the linear Dirichlet problem  \eqref{wdp} with $f\equiv -\mathcal P^\prime$.
In the \emph{critical} cases instead, because no improvement would be obtained at the first step of the bootstrapping for \eqref{wdp} with $f\equiv -\mathcal P^\prime$, one needs to implement a special initial step which, by exploiting the existence theory, replaces system  \eqref {wnlP} by the homogeneous linear system
\be
\label{lin}\quad\qquad\left\{\begin{array}{ll}
\Lambda_\Phi u\equiv -\Delta u + g (\Phi)\,u =0
&\mbox{on }\; \nst\\
u\in\ma(\varphi)\,,
\end{array}\right.
\ee
with prescribed $\varphi\in\bb$ and $g (\Phi)\equiv \mathcal P^\prime(\Phi)/\Phi\in L^{{d}/{2}}_{loc}(\nst)$.
The solution  $\Phi\in H_1(\nst)$ to \eqref{wnlP}  satisfies \emph{by construction} the linear Dirichlet-type problem \eqref{lin},  the latter having been derived by freezing the coefficient $g(\Phi)$. The estimate $g (\Phi)\in L^{{d}/{2}}_{loc}(\nst)$ comes from the embeddings
$H_1(\nst)\subset L^{2d/(d-2)}(\nst)$,  and $L^p(\Omega)\subset L^q(\Omega)$, for $p\geq q$, on bounded domains $\Omega$.

System \eqref{lin} still presents itself as `borderline', thus requires a non-straightforward technique. Its regularity is accomplished in \cite{Marini:2016} by applying a `regularity lifting method by means of a contracting operator' \footnote{Regularity of \eqref{lin} could also have been accomplished by an alternate method, perhaps unknown or scarcely utilized outside the field of geometric analysis, based on the regularity theory as developed several years earlier by one of the authors in the context of Yang-Mills connections; \cf \cite{Marini:1992} and
the `Acknowledgements' section in the present article. This is one of some instances in which techniques developed in gauge theories anticipate methods later established and formally codified in classical analysis contexts.}. For the method of contracting operator and more details on its adaptation to the context at hand, see \cite{Chen-Li:2010, Marini:2016}.

At various stages of the proof of regularity, additional difficulties arise, due to the non-triviality of the boundary data $\varphi$ prescribed at $\ntzero$, and to the unboundedness of the domain $\nst$. In fact, the regularity lifting method, as described in \cite{Chen-Li:2010}, directly applies only to linear Dirichlet-type problems with prescribed vanishing boundary data on bounded domains $\Omega$. Further work is also required because, in the context of the  Euclidean-signature semi-classical program under study, we are interested in establishing global control of the solution over the unbounded domain $\nst$ (and over the slices $\{t\}\times \R^n$).

Boundary issues are dealt with by reflection of operators and associated function spaces across the boundary, and introduction of cut-off functions, in a suitable way,  thus transforming Dirichlet-type problems on boundary-type neighborhoods like
\be
\label{bdryU}
\mathcal U_R\equiv\{ {\bf x}=(x^0, {\bf x^\prime} )\in (-\infty, 0]\times \R^n\;:\;\Vert {\bf x} - {\bf a}\Vert < R\}\,,
\ee
in which ${\bf a}=(0, a^1, a^2, a^3)$ is a fixed boundary point, and $R$ a positive number,
into Dirichlet-type problems on balls
$$B_{{\bf a};R}\equiv \{ {\bf x}\in\R^{d}\,:\, \Vert {\bf x} - {\bf a}\Vert < R\}\,,$$
with homogeneous boundary data.
The \emph{doubling} technique may involve technicalities.
In our context, because $\nst$ is endowed with a Euclidean metric, the resulting doubled boundary-type neighborhoods are flat, while in more general cases, the latter  would possess a Lipschitz-bounded metric; \cf for example, \cite{Marini:1992, Maitra:inprep}.

To implement the procedure, we first extend the boundary value $\varphi\in\bb$ to the unique solution
 $\hat \varphi$ to
\be
\label{Laplace}\quad\qquad\left\{\begin{array}{ll}
\Delta u  =  0 \quad&\mbox{ on } \;\nst\\
u\in\ma(\varphi)\,.
\end{array}\right.
\ee
Such $\hat\varphi$ is proven to satisfy the global estimate $\hat\varphi\in H_{{3}/{2}}(\nst)$, besides the regularity property $\hat\varphi\in\mathcal C^\infty(\nst)$,  and for boundary data $\varphi$ specified in $\bb\cap C^\infty(\ntzero)$,  to satisfy $\hat\varphi\in\mathcal C^\infty(\inst)$, that is, smoothness up to and including the boundary; for the technical details, we refer the reader to Lemma 5.1 in \cite{Marini:2016}.

Due to technicalities pertaining to unbounded domains\footnote{These arise due to the additional constraints on the Sobolev exponents  required  for Sobolev embeddings to hold on unbounded domains; namely,
on unbounded domains $\Omega_u$ having the cone property, in  dimension $d> hq$, the embeddings $L^q_h (\Omega_u) \subset  L^p(\Omega_u)$  require the additional hypothesis $q\leq p$ (besides the hypotheses ${h}/{d} +{1}/{p} -{1}/{q} \geq 0$, $p,q\geq 1$, $h\geq 0$, required on bounded domains); \cf Theorems 7.57, 7.58 in \cite{Adams}. 
In the proof of Theorem \ref{T-reg1}, we use repeatedly the embeddings
$H_1(\nst)\subset L^p(\nst)$  $\forall p\in [2, 2d/ (d-2)]$ for $d\geq 3$, and $\forall p\geq 2$ for  $d=2$}, to the purpose of establishing a global estimate for $\Phi$, we also consider the extension $\Phi_L$, unique solution to
\be \label{Linear}\left( L\right) \quad\qquad\left\{\begin{array}{ll}
\Lambda_L u\equiv -\Delta u + 2a_2 u =0\quad&\mbox{on } \nst\\
u\in  \ma ({\varphi})\,.
\end{array}\right.
\ee
The same results hold for $\Phi_L$ as obtained for $\hat\varphi$, thus both those functions are smooth in the interior and satisfy the best possible global estimate on $\nst$; \cf Lemma 5.3 in \cite{Marini:2016}.

After performing the various steps of what presents itself as a somewhat laborious and at times technical procedure,
one ultimately obtains the regularity theorem below (Theorem 5.1 in \cite{Marini:2016}).
\begin{theorem}\label{T-reg1}
Let $\Phi$ be a solution to \eqref{wnlP} with prescribed boundary value $\varphi\in\bb$, in dimension
$n+1\equiv d=2, 3$, or $4$.
Let $\hat\varphi\in H_{1}(\nst)$ be the extension of $\varphi$ satisfying Laplace's equation $\Delta \hat\varphi =0$ on $\nst$, $\Phi_L$ be the extension  of $\varphi$ satisfying the linearized problem \eqref{Linear}.
Then  $\hat\varphi, \Phi_L, \Phi\in H_{{3}/{2}}(\nst)\cap \mathcal C^\infty (\nst)$. If, in addition, the boundary value $\varphi\in \mathcal C^\infty(\ntzero)$, then $\Phi\in H_{{3}/{2}}(\nst)\cap\mathcal C^\infty\left((-\infty, 0]\times \R^n\right)$, that is, $\Phi$ is smooth all the way up to and including the boundary.
\end{theorem}
Thus, also $\Phi$, unique solution to the non-linear problem \eqref{wnlP}, and unique minimizer of $\Ies$ over the space $\ma(\varphi)$, satisfies interior regularity and the best possible global estimate attainable on $\nst$.

In order to implement the initial step to interior regularity, with focus on the critical cases, one shows that $\alpha\Phi$, with $\alpha$ a suitable cut-off function centered at a point $\bf a\in\nst$, is a solution to
\be
\label{cutoff1}\quad\qquad\left\{\begin{array}{ll}
{\Lambda}_\Phi  u \equiv -\Delta u + g (\Phi) u  = f_1&\mbox{on } \nst\\
\quad \quad u\in H_{1; 0} (B_{{\bf a};R})\,,
\end{array}\right.
\ee
with $g(\Phi)\in  L^{{d}/{2}}(B_{{\bf a};R})$, $f_1\equiv -2\nabla\Phi\cdot \nabla \alpha -\Phi \Delta \alpha\in L^2(B_{{\bf a};R})$.

Using $L^2(B_{{\bf a};R})\subset L^{{d}/{2}}(B_{{\bf a};R})$  in $d=3$, or $4$  dimensions, the \emph {regularity lifting theorems}, Theorems 3.3.1 and 3.3.2 of Ref.~\cite{Chen-Li:2010} can then be applied directly, with no adaptations needed, yielding $\alpha\Phi$  in $L^p(B_{{\bf a};R})$  for all $p>1$.
Thus, $\Phi\in L^p(B_{{\bf a};R_1})$   for all $p>1$, for $0<R_1<R$.

Then, bootstrapping on systems such as
\be
\label{sys-int}\quad\qquad\left\{\begin{array}{ll}
-\Delta u = f_2\quad&\mbox{on } \nst\\
\quad \quad u\in H_{1; 0} (B_{{\bf a};R_1})\,,
\end{array}\right.
\ee
in which  $f_2\equiv-\alpha_2 \,\mathcal P^\prime(\Phi)+f_1 (\Phi)\in L^2(\nst)$ (using $\mathcal P^\prime(\Phi)\in L^p(B_{{\bf a};R_1})\,\forall p>1$ and $\mathcal L_{\alpha_2} (\Phi)\in L^2(\nst))$, will yield $\Phi\in \mathcal C^\infty(\nst)$, because the point ${\bf a}\in\nst$ was arbitrarily chosen.

The initial step to boundary regularity, still with focus on the critical cases, is not so-straightforward. One first observes that
the function $\Phi-\hat\varphi$ is a weak solution to the Dirichlet problem with homogeneous boundary data,
\be
\label{lb}\quad\qquad\left\{\begin{array}{ll}
\Lambda_\Phi u\equiv -\Delta u + g(\Phi) u=  - g(\Phi)\hat \varphi&\mbox{ on } \;\nst\\
u \in\ma(0)\,,
\end{array}\right.
\ee
with $g(\Phi)\in L^{{d}/{2}}_{loc}(\nst)$ in dimension $3$ and $4$, and $\hat\varphi\in  H_{{3}/{2}}(\nst)\subset L^{p} (\nst)$ for all $2\leq p\leq {2d}/{d-3}$ in dimension $d>3$, and for all $p\geq 2$ in dimension $d=3$.
By H\"older's inequality,  $g(\Phi)\hat \varphi\in L^{{2d}/{(d+1)}}_{loc} (\nst)$ if $d>3$, whereas for the three-dimensional case $g(\Phi)\hat \varphi\in L^p_{loc}(\threest)$ $\forall  p< 3/2$. If the boundary value $\varphi$ is further assumed to be smooth, one has $\hat \varphi\in L^\infty_{loc}(\nst)$, thus  $g(\Phi)\hat \varphi\in  L^{{d}/{2}}_{loc}(\nst)$ in $d\geq 3$ dimensions.

System \eqref{lb} reflected across the boundary yields
\be
\label{doubl1}
{\tilde \Lambda}_\Phi u\equiv -\Delta u + \overline{g (\Phi)}\,u = \check {r} \quad\mbox{on } \R^{d}\,,
\ee
in which $\overline {g(\Phi)}$ is the even extension of $g (\Phi)$ and  $\check r$ is the odd extension of $-g(\Phi)\hat \varphi$.
Because  even and odd extensions both preserve membership in $L^p$,
$\overline {g(\Phi)}\in L^{{d}/{2}}_{loc}(\R^{d})$,  and $\check r\in
L^{{2d}/{(d+1)}}_{loc}(\R^{d})$ if $d>3$,   $\check r\in L^{p}(\R^{3})$ $\forall p<3/2$ if $d=3$.

The function $\check \Phi$, defined as the odd extension of $\Phi-\hat\varphi$, is then a weak solution in $H_1 (\R^{d})$ to Eq.  \eqref{doubl1}.

Let then ${\bf a}=(0, a^1, a^2, a^3)$ be a fixed boundary point, $\mathcal U_R$ a boundary-type neighborhood as defined in \eqref{bdryU},  $B_{{\bf a};R}$ its double, and $\alpha$  a smooth cut-off function, compactly supported in $B_{{\bf a};R}$, with value $1$ on $B_{{\bf a};R_1}$, with $0<R_1<R$.
Then,
\be
\label{cutoff}
{\tilde\Lambda}_\Phi (\alpha\check\Phi) = \alpha {\tilde\Lambda}_\Phi (\check\Phi) + \mathcal L_{\alpha} (\check\Phi) = \alpha \check r + \mathcal L_{\alpha} (\check\Phi)\equiv h_1\,,
\ee
in which $\mathcal L_{\alpha} (\check\Phi)\equiv -2\nabla(\check\Phi) \cdot \nabla \alpha -(\check\Phi) \Delta \alpha$ is a lower order linear differential operator with smooth coefficients determined by the cut-off
$\alpha$, applied to $\check\Phi$. In $d>3$ dimensions then, $\check r\in
L^{{2d}/{(d+1)}}_{loc}(\R^{d})$, and
$\mathcal L_{\alpha_1} (\check\Phi)\in L^2(B_{{\bf a};R})\subset  L^{{2d}/({d+1})}(B_{{\bf a};R})$. Thus,  $h_1\in L^{{2d}/({d+1})} (\R^{d})$ in $d>3$ dimensions (because $\alpha$ is supported in $B_{{\bf a};R}$); in particular,  in $4$ dimensions
$h_1\in L^{{8}/{5}} (\R^{4})$.
A similar analysis yields  $h_1\in L^{p}(\R^{3})$ $\forall p<3/2$ in $d=3$ dimensions. In conclusion,
$\alpha_1\check\Phi$ solves the linear system
\be
\label{lin3}\quad\qquad\left\{\begin{array}{ll}
{\tilde \Lambda}_\Phi u \equiv-\Delta u + \overline{g(\Phi)} u = h_1 &\mbox{ on } \R^{d}\\
 \quad u\in H_{1; 0} (B_{{\bf a};R})\,.
\end{array}\right.
\ee
with  coefficients $\overline{g(\Phi)}\in L^{{d}/{2}}_{loc}(\R^{d})$,  $ h_1\in L^{{2d}/({d+1})} (\R^{d})$ in $d=n+1>3$ dimensions, $h_1\in L^{p}(\R^{3})$ $\forall p<3/2$ in $d=n+1=3$ dimensions.

Unfortunately,  in dimensions $d>3$, the term $\check r$, odd extension across $\ntzero$ of $\Phi-\hat\varphi$, accounts for the lesser regularity of the right hand side in \eqref{lin3}, which now no longer satisfies the hypothesis satisfied by its counterpart $f_1$, namely, $f_1\in L^2(B_{{\bf a};R})\subset L^{d/2}(B_{{\bf a};R})$ in dimensions $3$ and $4$, in System \eqref{cutoff1}, utilized for the proof of interior regularity. The latter hypothesis is explicitly formulated, and required, in Theorem 3.3.2 of Ref.~\cite{Chen-Li:2010}.
The need to work with lesser regularity would often
arise when studying boundary regularity for Dirichlet-type problems such as \eqref{lin} in which non-vanishing boundary data in a border-line case are prescribed, and it seems worthwhile to extend the regularity lifting theorem to include those boundary cases. That is the purpose of the following lemma.
\begin{lemma} \label{L-RL} Regularity Lifting Theorem.
Let $u$ be a solution to the Dirichlet problem with homogeneous boundary data
\be
\label{C-1}
-\Delta u = a({\bf x}) u + b({\bf x})\,,\; u\in H_{1; 0}(\Omega)\,,
\ee
on a  smooth bounded domain $\Omega$  in $d>2$ dimensions,
$a\in L^{\frac{d}{2}}(\Omega)$ and  $b\in L^{\frac{2d}{d+1}} (\Omega)$ if $d>3$, $b\in L^{p} (\Omega)$ $\forall p< 3/2$ if $d=3$.  Then $u\in L^{{2d}/{d-3}}(\Omega)$ in $d>3$ dimensions; $u\in L^p(\Omega)$ $\forall p\in (1,\infty)$ in $3$ dimensions.
\end{lemma}
The corresponding hypotheses in Theorem 3.3.2 of Ref.~\cite{Chen-Li:2010} are $a({\bf x})\in L^{\frac{d}{2}}(\Omega)$, and $b({\bf x})\in L^{\frac{d}{2}}(\Omega)$, $d>2$. Because, in all dimensions $d>3$, $L^{\frac{d}{2}}(\Omega)\subset L^{\frac{2d}{d+1}} (\Omega)$ on bounded domains $\Omega$, our hypotheses on the coefficient $b$ are weaker in all dimensions $d>2$ (and they become comparatively even more so as the dimension increases); in dimension $3$ our hypothesis on $a$ is weaker as well.

The proof of Lemma \ref{L-RL} is based, as that of Theorem 3.3.2 in \cite{Chen-Li:2010}, on decomposing the domain $\Omega$ in the two subdomains
$$\Omega_A\equiv \{{\bf x}\in \Omega \;:\;\vert a({\bf x})\vert\geq A\}\,, \mbox{ and } \Omega_B\equiv (\Omega_A)^c\,,$$
for $A$ a positive fixed constant. Note that the measure of $\Omega_A$ satisfies
\be
\label{asympt}
\lim_{A\to \infty} m(A)\to 0\,.
\ee
Applying $T\equiv (-\Delta)^{-1}$,   Eq.\eqref{C-1} becomes
$$u= T(a\mathcal X_{A} u) + T(a\mathcal X_{B} u) + T(b)\,,$$
in which $\mathcal X_{A}$, $\mathcal X_{B}$ are the characteristic functions of $\Omega_A$, and $\Omega_B$.
As in Ref. \cite{Chen-Li:2010}, one can show that, if $A$ is  a sufficiently large constant,  the operator
$T(a\mathcal X_{A}\cdot)\,:\, L^p(\Omega) \to L^p(\Omega)$ is a contracting operator $\forall p\in (\frac{d}{d-2}, \infty)$, making use of the asymptotic behavior \eqref{asympt}. There are only minor changes, in comparison with Ref.~\cite{Chen-Li:2010}, in the procedure to estimate the term
$T(a \mathcal X_B u)$, yielding
$$\Vert T(a \mathcal X_B u)\Vert_p\leq C \Vert u\Vert_{2;1}<\infty\,,\quad \forall p\in (\frac{d}{d-2}, \infty)\,;\quad 3\leq d\leq 6\,.$$

In $d>3$ dimensions,  our weaker hypothesis $b\in L^{\frac{2d}{d+1}} (\Omega)$  yields the estimate  $T(b)\in L^p$  for $p\in \left(\frac{d}{d-2}, \frac{2d}{d-3}\right]$ which constitutes a first improvement in all dimensions $d>3$, as $L^{2d/ d-3}(\Omega)\subset L^{2d/d+1}(\Omega)$ for $d>3$. Such upper constraint on the Sobolev exponent $p$ is  not present in Ref.~\cite{Chen-Li:2010}. Nevertheless, despite this limitation, a first improvement  is all that is needed in a variety of cases, analogous to ours, in which the non-homogeneous term $b$ depends on the solution $\Phi$ to a non-linear system related to \eqref{lb}, and one can feed that initial improvement obtained for $u$ into the term $b$ and iterate. Thus, Lemma \ref{L-RL} may be instrumental to a somewhat general extent, 
if one wishes to achieve boundary regularity by our doubling technique, that is, by transforming a Dirichlet-type problem with non-homogeneous boundary data on a boundary-type neighborhood,  into an interior neighborhood with a (possibly) Lipschitz-bounded metric and homogeneous boundary data.
(In our specific context,  recall that $b$ is the sum of the odd extension of $-g(\Phi)\hat \varphi$ with $-2\nabla(\check\Phi) \cdot \nabla \alpha -(\check\Phi) \Delta \alpha$, derived from \eqref{lb} and \eqref{cutoff}.)

Note also that, for applications to boundary value problems, one can further weaken the requirements  on the term $b$,  as long as those are still sufficient to guarantee a first improvement in the regularity of $u$; a sensible hypothesis is  $b\in L^{2d/(d-1)}_{ -1}(\Omega)$, on bounded domains $\Omega$; \cf  Ref.~\cite{Marini:unpub}.

Having established Lemma \ref{L-RL}, a repeated application of doubling techniques, and procedures   conceptually similar to those applied so far, yield the results stated in Theorem \ref{T-reg1}; \cf \cite{Marini:2016} for technical details, in particular with regard to the proof of higher order regularity. In the latter reference, we show at an early step that $\tilde\Phi\equiv\Phi - \hat \varphi$ satisfies $\tilde\Phi\in\mathcal C^1( (-\infty, 0]\times \R^2 )$ in $3$ dimensions, while  $\tilde\Phi\in\mathcal C^0( (-\infty, 0]\times \R^3 )$ in $4$ dimensions, for general prescribed boundary data in $\varphi\in\bb$.
If the boundary value satisfies $\varphi\in\mathcal C^\infty(\threetzero)$, by differentiating  the auxiliary equations satisfied by $\tilde\Phi$, obtained in \cite{Marini:2016},  at first with respect to tangential derivatives, then relating those to normal derivatives, we eventually obtain $\tilde\Phi$ and  $\Phi$ in $\mathcal C^\infty((-\infty,0]\times \R^d)$.

\mn To the purpose of establishing global control of $\Phi$ over the unbounded domain $\nst$, having already proved, besides interior regularity,
the global estimates $\hat\varphi, \Phi_L\in H_{3/2}(\nst)$, for $\hat\varphi$ a solution to System \eqref{Laplace}, and $\Phi_L$ a solution to System \eqref{Linear},  one wishes to prove that the conditions imposed on the coefficients of the polynomial $\mathcal P$  are sufficient to yield an analogous global estimate for the solution $\Phi$ to the Dirichlet problem \eqref{wnlP}  in dimension  $2\leq d\equiv n+1\leq 4$. On unbounded domains, a mild complication arises (\cf footnote 3). The coefficients appearing in System \eqref{lb} belong only \emph{locally} to certain Sobolev spaces; that is because their  various summands fail to lie in $L^p(\nst)$ for any shared value $p$. That issue can be easily circumvented in our case. In fact, replacing $-\Delta$ with
$\Lambda_L$,  allows one to rewrite System \eqref{wnlP} as
\be
\label{wnlPL}
\left\{\begin{array}{ll}
\Lambda_L \Phi +\mathcal  R(\Phi) =0\quad&\mbox{on } \nst  \\
\Phi\in  \ma(\varphi)\,,
\end{array}\right.
\ee
in which the polynomial $\mathcal  R$ no longer contains the lower degree term in $\mathcal P^\prime$.
Replacing $\hat\varphi$ by  $\Phi_L$, one  derives that
$\Phi -\Phi_L$ satisfies
\be
\label{Phi-Phil}
\Lambda_L u + g_1(\Phi;\Phi_L) u = r_1(\Phi;\Phi_L)\,,
\ee
in which  $g_1\in L^{d/2}(\nst )$, $r_1\in L^{2d/d+1}(\nst)$. This leads, by doubling, to the study of regularity for systems
 \be
\label{RLFu}
\left\{\begin{array}{ll}
\Lambda_L \Phi +a({\bf x}) =b({\bf x})\quad&\mbox{on } \Omega_u  \\
u\in H_{1;0}(\Omega_u)\,,
\end{array}\right.
\ee
with $a\in L^{d/2}(\Omega_u)$, $b\in L^{2d/d+1}(\Omega_u)$,   $\Omega_u\subset \R^d$ a smooth unbounded domain. Notice that $H_{1;0}(\R^d)\equiv H_1(\R^d)$, and a cutoff function should not be introduced for the case at hand.
To the purpose of establishing global control, the cases $d\equiv n+1=4$, $d=3$, and $d=2$ are treated separately in \cite{Marini:2016}. One ultimately obtains the following
\begin{lemma} \label{L 5.4} (Lemma 5.4 in \cite{Marini:2016}) The unique solutions $\Phi$ of the non-linear Dirichlet problem \eqref{wnlP} satisfies the global control estimate $\Phi\in H_{3/2}(\nst)$.
\end{lemma}
The proof relies on an extension to unbounded domains of the regularity lifting method previously shown. 
Such extension relies on constructing contracting operators
$$T_1(a\mathcal X_A\cdot)\,:\,H_k(\Omega_u)\to H_{k+2}(\Omega_u)\,;$$
here,
$\mathcal X_A$ is the characteristic function on $\Omega_A$, the definition of which, adapted to the unbounded domain case, is
$$\Omega_A\equiv \{{\bf x}\in \Omega \;:\;\vert a({\bf x})\vert\geq A\,,\;\vert {\bf x}\vert>A\}\,,$$
for $A$ a positive fixed constant, and $T_1$ is the solution operator for
\be\label{controlG}
\quad\qquad\left\{\begin{array}{ll}
\Lambda_L u \equiv -\Delta u + 2a_2  u=  f \quad&\mbox{ on } \;\Omega_u\\
u\in H_1(\Omega_u)\,.
\end{array}\right.
\ee
Applying $T_1$ to
$$ \Lambda_L + \bar{g_1} u = \check r_1\,,$$
obtained by taking suitable extensions across $\ntzero$ of the coefficients in Eq. \eqref{Phi-Phil}, one obtains that the odd extension, $\check\Phi_1$,  of $\Phi-\Phi_L$ satisfies
\be
\label{integral}
u =  T_1(- \overline {g_1}\mathcal X_A u) + T_1(- \overline {g_1}\mathcal X_B u) + T_1 (\check r_1)\,.
\ee
One has that  $T_1(- \overline {g_1}\mathcal X_B (\check\Phi_1)\in H_3(\R^d)$, $T_1 (\check r_1)\in L^{2d/d+1}_2(\R^d)\subset H_{3/2}(\R^d)$, and  that
\be
T_1(- \overline {g_1}\mathcal X_A \cdot)\,:\, H_{{3}/{2}}(\R^d) \to H_{{3}/{2}}(\R^d)
\ee
is a contracting operator, yielding the estimate $\check \Phi_1\in H_{{3}/{2}}(\R^d)$, thus $\Phi -\Phi_L\in H_{3/2}(\nst)$, and
$\Phi\in H_{{3}/{2}} (\nst)$,  in dimension  $d\equiv n+1=4$.

\noindent{\bf\emph {Smoothness of the Hamilton-Jacobi functional ${S}$.}}
There are in general two obstructions to smoothness of \emph{Hamilton-Jacobi} functionals $G$ defined, in analogy to the way $S$ has been defined in the present section,
as
\be
\label{Ggen}
G[\varphi] \equiv \inf_{\Phi\in\ma(\varphi)} \mathcal G[\Phi]\,,
\ee
in which $\mathcal G$ is a given action functional, $\varphi$ are specified boundary data in a function space $\bb$ and $\ma(\varphi)$ is the function space in which minimization takes place, all satisfying suitable assumptions, apt to guarantee the well-posedness of \eqref{Ggen} and the other main results obtained so far, such as the existence of a minimizer and ellipticity of the Euler-Lagrange equations it satisfies. A first obstruction is the non-uniqueness of such a minimizer, a second one is the existence of non-trivial \emph {Jacobi fields} along the minimizer.
One such example is the length-squared functional on a Riemannian manifold, for which one end-point is fixed and the other plays the role of the boundary value $\varphi$. In that case the failure of minimizing geodesics between fixed endpoints to be unique, or the existence of nontrivial Jacobi fields along such geodesics, yield singularities. Occurrence of these obstructions signals a non-empty Ôcut locusÕ for the manifold. Such cut loci however are
always lower dimensional subsets of the ambient manifold so that smoothness of the length-squared
functional at least holds almost everywhere. They have moreover the interesting
property of encoding practically the entire nontrivial topology of the ambient manifold and thus
play a central geometric role in its analysis;   with regard to materials on Jacobi fields and cut loci, see for example Chapters 1 and 2 of \cite{Klingenberg:1982}.
An immediate argument explaining a reason for non-uniqueness being an obstruction to global differentiability of a Hamilton-Jacobi functional such as $G$ is that the gradient of a solution to a Hamilton-Jacobi equation should produce the complementary momentum for the trajectory through a given point in configuration space. Thus, if the trajectory is not unique, the existence of the gradient at the chosen point is compromised.
In the context of the Yang-Mills theory, discussed in the next section, the lack of uniqueness for the absolute minimizer for the Euclidean action corresponds to a lack of `everywhere' differentiability of the analogue to the functional ${S}$. In that context, the relevant
Riemannian space will be the ÔmoduliÕ space of spatial (i.e., boundary data) connections modulo gauge transformations with its (weak) Riemannian structure induced from the Yang-Mills kinetic
energy in a well-known way; \cf, for example, \cite{Singer:1981} and \cite{Babelon:1981}.
In that context, wave functionals exhibit \emph{scars} localized along lower
dimensional subsets of the (moduli-space) configuration manifold, to some extent analogous to the scars that arise, in a totally different setting, in the study of quantum chaos (see, for example, \cite{Ozorio:1988}).
In the present section, we rule out those obstructions for the polynomial theories, thus showing smoothness of the functional \eqref{H-J funct} as an application of the
implicit function theorem between Banach spaces.
More precisely, the approach used in \cite{Marini:2016} entails proving Theorem \ref{sd} below,  under the additional strict convexity assumption  on $\mathcal P$,
\be\label{strictpoint-convex}
\mathcal P^{\prime\prime}(z)> 0\,.
\ee
\begin{theorem}\label{sd}
The solution $\Phi_\varphi$ to the nonlinear Dirichlet problem \eqref  {wnlP}
depends smoothly on the boundary data; more precisely,
\begin{equation}\begin{array}{cc}
& \bb \to H_1(\nst)\\
&\varphi\mapsto \Phi_\varphi\\
\end{array}
\end{equation}
is a smooth map.
\end{theorem}

Theorem \ref{sd}, combined with smoothness of the functional $\ie\,:\, H_1(\nst)\to\R$, yields smoothness of
\[
S \,:\, \varphi\in\bb\mapsto\Phi_\varphi\in \ma(\varphi)\mapsto \ie[\Phi_\varphi]\in\R\,.
\]

The proof of Theorem \ref{sd} is essentially an application of the Implicit Function Theorem to the functional
\begin{align} \me\,:\, &\bb\times \ma(0)\to H_{-1}(\nst)\notag\\
& (\varphi, h)\to \Lambda \left( \fiL + h\right) \,,\notag\\
\end{align}
in which $\Lambda\equiv -\Delta +\pp^\prime (\cdot)$, and $\fiL$ is the \emph{unique} solution to the linear boundary value problem \eqref{Linear} previously defined. Such functional is smooth because the map $\varphi\in\bb\to \fiL\in H_1 (\nst)$ is smooth.

For fixed $\varphi_0\in\bb$, we define
\be h_0\equiv \Phi_{{\varphi}_0}
-{\Phi}_ {{\varphi}_0}^L\,,\ee
so to obtain
$h_0\in\ma(0)$ and
$\me ({\varphi}_0, h_0) = \Lambda (\fioL +
h_0) = \Lambda (\fio) =0$.
We linearize $\me(\varphi, h)$ at the point $(\varphi_0, h_0)$ along the direction of the second variable $h$, thus deriving the linear operator
\ba
& D_2\me_0 \,:\, T_{h_0}\ma(0)\simeq\ma(0) \to T_0\,H_{-1}(\nst)\simeq H_{-1}(\nst) \,,\notag\\
&\xi\mapsto \lim_{\lambda\to 0}\frac{\me({\varphi}_0, h_0 +
\lambda\xi) - \me({\varphi}_0, h_0)}{\lambda} = \lim_{\lambda\to
0}\frac{\Lambda(\fio + \lambda \xi)}{\lambda}
= -\Delta \xi+ \mathcal P^{\prime\prime}(\fio)\,
\xi \,.
\end{align}
Here ${D_2\me_0}(\xi)$ is to be interpreted as an element of $H_{-1}(\nst)$ via the identification
\be
\label{def D2xi}
{D_2\me_0}(\xi)\,:\, f\in \ma(0)\mapsto \int_\nst f \,D_2\me_0 (\xi)\,d{\bf x} \equiv \int_\nst \nabla f \cdot \nabla \xi \,d{\bf x}+ \int_\nst \mathcal P^{\prime\prime} (\fio)\,f\,\xi \, d{\bf x}\,,
\ee
in coherence with the definitions of \emph{weak Laplacian} and \emph{weak Dirichlet problem} (linear and non-linear) given earlier.

Under the strict convexity and coerciveness
conditions imposed on $\mathcal P$,  one can show that
the linear operator
$D_2\me_0$ is a bicontinuous vector space isomorphism between
$\ma(0)$ and $H_{-1}(\nst)$.
That is achieved by showing that the standard inner product in $H_1(\nst)$ and the inner product
induced by $\mathcal P$ given by
\be
\label{ipa}
\langle f, g\rangle_{\mathcal P} \equiv \int_\nst \nabla f \cdot \nabla g \,d{\bf x}+ \int_\nst \mathcal P^{\prime\prime} (\fio)\,f\,g \, d{\bf x}
\ee
determine  \emph {equivalent} norms,  thus induce the same natural topology.
\sn
Such equivalence, yields in particular that  $D_2\me_0 (\xi)$ is a bounded linear operator on $\ma(0)$.
Note that, for any $g\in\ma(0)$, one has by definition $D_2\me_0 (\xi)(g) = \langle \xi, g\rangle_{\mathcal P} $, and
$D_2\me_0$ is a one-to-one,  self-adjoint operator.
The implicit function theorem between Banach spaces then states that
there exist
neighborhoods $I\subset \bb$, $J\subset\ma(0)$ of ${\varphi}_0$, $h_0$ respectively, such that $\forall
\varphi\in I$ there exists a unique $h (\varphi)\in J$ for which $\me (\varphi,
h(\varphi))=0$ and that such map, $\varphi\in\bb\to h(\varphi)\in\ma(0)$, is $\mathcal C^\infty$. Because $\Phi_\varphi = \fiL+ h(\varphi)$, this ultimately yields smoothness for the functional $S$.

\noindent{\bf\emph{The Hamilton-Jacobi equations for the absolute minimizer of the Euclidean signature action functional $\ie$.}}
For any given distribution $\Phi\in \ma(\varphi)$, finiteness of the integral $\int_{\R^-\times \R^n} \mathcal P(\Phi) \,dt\,d {\bf x^\prime}$, together with finiteness of the Euclidean action,  guarantee that the \emph{energy functional}
\be
\label{mathcal F}
\mathcal F [\Phi]\equiv
\int_\nst \varepsilon(\Phi) \bigl(t, {\bf x^\prime}\bigr) \, dt \,d {\bf x^\prime}\,,
\ee
having \emph{energy density}
\be
\label{density}
\varepsilon(\Phi) \equiv \frac{1}{2}{ \dot\Phi}^2 - \frac{1}{2}\vert\nabla^\prime\Phi\vert^2 -\mathcal P (\Phi)\,,
\ee
be finite. Here and throughout the text, $\nabla^\prime$ denotes the gradient with respect to the spatial variables only,  and we identify $x^0\equiv t$, $\partial_0  \Phi\equiv \partial \Phi/\partial x^0 \equiv \dot{\Phi}\,$.

Thus, by Fubini's theorem, the \emph {energy}  of a distribution $\Phi\in\ma(\varphi)$,
\be\label{energy}
e [\Phi] (t) \equiv\int_{\R^n}\varepsilon(\Phi)(t,{\bf x^\prime}) \,d{\bf x^\prime}\,,
\ee
 is finite for almost all $t$'s in  $\R^{-}$ and integrable over $\R^-$.

Further, for $\Phi$ a minimizer of $\Ies$ over the affine space $\ma(\varphi)$, a simple calculation, using the Euler-Lagrange equations satisfied by $\Phi$ as well as the interior regularity proved for such equations in this context,  shows that
\be
\label{calc}
\p_0\varepsilon (\Phi) = - \nabla^\prime \cdot (\dot\Phi \nabla^\prime\Phi)\,.
\ee
As a result, by interior regularity, if $\Phi$ is a minimizer,
the Fundamental Theorem of Calculus applied to the smooth functions of time
$$t\in\R^-\mapsto \partial_0\varepsilon (\Phi)(t,  {\bf x^\prime})\in\R\,,\quad \mbox{  with } {\bf x^\prime} \mbox{ fixed}\,,$$
yields
\begin{equation}
\label{smoothenergy}
\varepsilon (\Phi) (t_2,{\bf x^\prime}) - \varepsilon (\Phi) (t_1,{\bf x^\prime})= \int_{t_1}^{t_2} \p_0\varepsilon (\Phi) (t, {\bf x^\prime})\,dt =- \int_{t_1}^{t_2}\nabla^\prime \cdot (\dot\Phi \nabla^\prime\Phi)\,dt\,,\quad t_1< t_2<0\,.
\end{equation}
If the boundary value $\varphi\in\bb$ is in addition prescribed to be smooth, ensuring smoothness of the minimizer $\Phi$ all the way up to and including the boundary, the equalities above hold for all $t_1< t_2\leq 0$.
Integrating \eqref{smoothenergy} over  $\R^n$, one obtains
\begin{equation}
\label{smooth1}
 e[\Phi] ({t_2}) - e[\Phi]({t_1})= - \int_{\R^n} d{\bf x^\prime}\int_{t_1}^{t_2}\nabla^\prime \cdot (\dot\Phi \nabla^\prime\Phi)\,dt\,,\quad \mbox{ for } t_1< t_2<0\,,
\end{equation}
as long as the quantities on the left hand side are finite, that is, outside possibly a subset of
$\R^-$ of zero measure.

A computation, involving approximation of the integral on the right hand side of \eqref{smooth1} via integrals over the regions $\{\Vert {\bf x^\prime}\Vert = R\}\times [t_1, t_2]$ and an application of
Green's theorem, yields
\be
\label{smooth*}
e[\Phi] ({t_2}) - e[\Phi]({t_1}) = -\lim_{R\to\infty}\int_{\{\Vert {\bf x^\prime}\Vert = R\}\times [t_1, t_2]}
\dot\Phi \nabla^\prime\Phi\cdot \frac{{\bf x^\prime}}{\Vert{\bf x^\prime}\Vert}\,
d\sigma\,,\quad  t_1< t_2<0
\,,
\ee
in which $d\sigma$ is the surface element on the cylinder $\{\Vert {\bf x^\prime}\Vert = R\}\times [t_1, t_2]$, and $t_1$, $t_2$ can be taken arbitrarily outside possibly a set of zero measure.
At that point, one shows that
\be\label{int van}
\lim_{R\to\infty}\int_{\{\Vert {\bf x^\prime}\Vert = R\}\times [t_1, t_2]}
\dot\Phi \nabla^\prime\Phi\cdot \frac{{\bf x^\prime}}{\Vert{\bf x^\prime}\Vert}\,
d\sigma=0\,,\quad  t_1< t_2<0\,,
\ee
yielding
\be
\label{en const}
e[\Phi] (t) \equiv \int_{\R^n} \varepsilon(\Phi)(t, {\bf x^\prime})\, d {\bf x^\prime}= C  \quad a.e. \ t\in\R^-
\,,
\ee
in which $C$ is a constant,
whenever $\Phi$ is a minimizer of the action functional $\ie$ over the space $\ma(\varphi)$, with  boundary value $\varphi\in \bb$; \cf \cite{Marini:2016} for details.
Such constant $C$ depends on $\Phi$ only and, ultimately, by uniqueness of the minimizer in the case of the polynomial scalar field theory, is uniquely determined by the prescribed boundary value $\varphi\in\bb$.

Further, the global control established for the action minimizer $\Phi$, namely $\Phi\in H_{{3}/{2}}(\nst)$ (the best possible global estimate for prescribed boundary conditions $\varphi$ in $H_1(\{0\}\times \R^n)$,
ensures
\be
\label{C}
e[\Phi] (t) \equiv \int_{\R^n} \varepsilon(\Phi)(t, {\bf x^\prime})\, d {\bf x^\prime}= C\quad \forall t\leq 0\,,
\ee
thus energy is preserved along the flow.
Equation \eqref{C} and integrability of $e[\Phi] (t)$ (consequence of finiteness of the energy functional $\mathcal F [\Phi]$) further yield that the constant $C$ equals zero.

Having shown already that the solution $\Phi_\varphi$ to \eqref{wnlP}, unique minimizer of $\ie$ with prescribed boundary value $\varphi\in\bb$,  depends smoothly on the latter, one has, in particular,
that the function
\[v(\lambda)\equiv \Phi_{\varphi +\lambda\psi} - \Phi_\varphi\quad \mbox{ with } \varphi\in \bb, \psi\in \bb\cap\mathcal C^\infty(\ntzero) \mbox{ fixed }\,,
\]
is differentiable with respect to $\lambda$. One also has the relation  $v^\prime(0)_{\left |_{\ntzero}\right.} =\psi$.

The variational derivative with respect to
$\varphi\in\bb$, evaluated at  $\psi$, of the functional $S$ defined in \eqref{H-J funct}
 is therefore
\ba\label{var-der}
&\int_{\R^n}\frac{\delta S}{\delta \varphi}\,\psi\,d{\bf x^\prime}\equiv 
\lim_{\lambda\to 0} \frac{S[\varphi +\lambda\psi] - S[\varphi]}{\lambda}=
\lim_{\lambda\to 0} \frac{\ie(\Phi_{\varphi +\lambda\psi}) - \ie(\Phi_\varphi)}{\lambda}=\notag\\
&\notag\\
&\int_\nst\left( \nabla\Phi_\varphi\cdot\nabla v^\prime(0)  + \mathcal P^\prime (\Phi_\varphi) \,
v^\prime(0)  \right)  \,d{\bf x} =\int_\nst\left(-\Delta \Phi_\varphi +\mathcal P^\prime (\Phi_\varphi) \right)\,v^\prime(0)  \,d{\bf x} \,+\notag\\
&\notag\\
&+\int_\ntzero
\psi\,\nabla\Phi_\varphi\cdot (1, {\bf 0}) \,d{\bf x^\prime} + \lim_{R\to\infty} \int_\rinfty v^\prime(0) \,\nabla\Phi_\varphi \cdot
\frac{{\bf x}}{\Vert  {\bf x}\Vert}\,d\sigma \notag\\
&\notag\\
& = \int_\ntzero
\psi\,\nabla\Phi_\varphi\cdot (1, {\bf 0}) \,d{\bf x^\prime} = \int_\ntzero
\psi\,\dot{\Phi}_\varphi \,d{\bf x^\prime} \,,\notag\\
\end{align}
in which
 $d\sigma$ denotes the surface element on the cylinder
$\rinfty$, and  we have used  the Euler-Lagrange equations satisfied by $\Phi_\varphi$ as well as the vanishing of
$$\lim_{R\to\infty} \int_\rinfty v^\prime(0) \,\nabla\Phi_\varphi \cdot
\frac{{\bf x}}{\Vert  {\bf x}\Vert}\,d\sigma\,.$$
Because  $\psi\in \bb\cap\mathcal C^\infty(\ntzero)$ is arbitrarily fixed, \eqref{var-der} entails
\be\label{vardot}
\frac{\delta S}{\delta \varphi}=  \dot \Phi_\varphi(0)
\ee
Combining the vanishing of the energy at time $t=0$,
$$
e[\Phi](0) = \int_\ntzero \left(\frac{1}{2}  (\dot \Phi_\varphi)^2 -\frac{1}{2}\nabla^\prime\varphi({\bf x^\prime})\cdot\nabla^\prime\varphi({\bf x^\prime}) - \mathcal P(\varphi({\bf x^\prime}) ) \right) \,d{\bf x^\prime}=0\,,$$
   with \eqref{vardot}, one  derives  the Hamilton-Jacobi equation
\be\label{HJeq}
\int_\ntzero \left(\frac{1}{2}\frac{\delta S}{\delta\varphi( {\bf x^\prime}) }\,\frac{\delta S}{\delta\varphi( {\bf x^\prime}) } -\frac{1}{2}\nabla^\prime\varphi({\bf x^\prime})\cdot\nabla^\prime\varphi({\bf x^\prime}) - \mathcal P(\varphi({\bf x^\prime}) ) \right) \,d{\bf x^\prime}=0\,.
\ee

\noindent{\bf\emph{Decay of the approximate ground state wave functional.  A `virial' argument.}}
A first straightforward estimate for $\Omega_0$, the tree approximation to the ground state functional (\cf Eq. \eqref{Omega}),
of the type
\be\label{aim}
\vert\Omega_0(\varphi)\vert \leq \mathcal N \exp \left\{
\frac{    -\Vert \varphi\Vert^2_  {H_{\frac{1}{2}}(\R^n)}} {C}
\right\}\,,
\ee
in which $C$ is a constant independent of $\varphi$, can be proven as a consequence of coerciveness of $\Ies$ and
of the Trace Theorem, which guarantees an estimate of   $\Vert \varphi\Vert_{H_{\frac{1}{2}}(\R^n)}$ in terms of $\Vert \Phi_\varphi\Vert_{H_1(\nst)}$.
(This argument  does not apply to the massless $\Phi^4$ theory on $\st$ because $\ie$ is not coercive in that case.)

A heuristic argument for a more refined estimate, yielding a better decay, is based on a  so-called \emph{virial argument},  which takes into account the presence of the higher order polynomial  term in $\ie$. A rigorous argument would require one taking into account the higher order corrections of the ground state functional, $S_{(1)}, S_{(2)} \dots$, which we are disregarding at the moment. Such quantum corrections all vanish in the case of (massive) \emph{free fields}, but do not vanish for the more general
polynomial scalar field theory under study.

Having made the coerciveness assumption $\mathcal P(\Phi) \geq C\Phi^2$
for some positive constant $C$, by defining  $m_0 \equiv \sqrt {2C}$ and considering the corresponding
massive \emph{free field} functional
\[{S}_{(0)}^{free}\,[\varphi]=\mathcal I_{es}^{free} \,[\Phi_\varphi] \equiv
\int_{\R^n} \int_{-\infty}^0\left( 1/2 \,\dot\Phi^2  + 1/2\, \nabla^\prime\Phi_\varphi\cdot\nabla^\prime\Phi_\varphi +1/2\, m_0^2\,\Phi^2_\varphi \right) \,dt\,d{\bf x^\prime}\,,\]
one obtains
\[{S}_{(0)}[\varphi] \geq {S}_{(0)}^{free}\,[\varphi] \qquad\forall \varphi\in\bb\,.\]
(For an explicit calculation of  ${S}_{(0)}^{free}\,[\varphi]$, see  \cite{Maitra:2007}.)
Thus,
$\exp \{- {S}_{(0)}[\varphi]/\hbar\}$ decays at least as rapidly as some specific Gaussian. Here, we are still working under the assumption that the ground state quantum corrections to ${S}_{(0)}[\varphi]$  may be legitimately disregarded for the purpose of estimating the decay of the true ground state functional.

In \cite{Marini:2016} we make a conjecture on the behavior of the ${S}_{(0)}$ functional under a constant rescaling $\varphi\to \varphi_\lambda=  e^\lambda \varphi$, in the limit of large $\varphi$. To that purpose, we consider the ratio
\be\label{ratio}
\mathcal R =\left.\frac{  d{S}_{(0)}[\varphi_\lambda]/d\lambda  }
                           {{S}_{(0)}[\varphi_\lambda]}\right| _{\lambda=0}
                        = \frac{\int_{\R^n}  \varphi(\mathbf x^\prime)\,\delta{S}_{(0)}[\varphi]/\delta  \varphi(\mathbf x^\prime)\,d\mathbf x^\prime}
                           {{S}_{(0)}[\varphi]}
                           \ee
and observe that, in the case of the massive \emph{free field} functionals ${S}_{(0)}[\varphi ]^{free}$ described above, this ratio would simply be given by $\mathcal R = 2$ for all  $\varphi$.

	  Assuming that
	  $\lim_{\Vert\varphi\Vert_{H_{1/2}(\R^n)}\to \infty}\mathcal R(\varphi)$,
	   exists, we evaluate
	   \be
	   \label{phi_t}
	   \lim_{t\to t^*} \mathcal R(\varphi_t)\,,
	   \ee
	   in which $\varphi_t$ is a curve satisfying $\Vert\varphi_t\Vert_{H_{1/2}(\R^n)}\to\infty$, as $t$ approaches its limiting value $t^*$. Using the fact that both numerator and denominator of \eqref{ratio} tend to infinity as $\Vert\varphi\Vert_{H_{1/2}(\R^n)}\to\infty$, we appeal to L'Hospital's rule and differentiate with respect to $t$ numerator and denominator in \eqref{phi_t}.
	 To that purpose, we further use the technique of estimating such limit along `solution curves' of the `gradient semi-flow' of ${S}_{(0)}[\varphi ]$. We compute formal time derivatives along the flow by applying the functional differential operator	
	\be\label{ell}
\mathcal L =\int_{\R^n} \left(\frac{ \delta{S}_{(0)}[\varphi]}{\delta  \varphi(\mathbf y^\prime)}\right)
\frac{\delta}{\delta  \varphi(\mathbf y^\prime)}\,d\mathbf y^\prime \,.
                           \ee
The Hamilton-Jacobi equation
	 \be\label{hj}
\frac{1}{2}\int_{\R^n} \left(\frac{ \delta{S}_{(0)}[\varphi]}{\delta  \varphi(\mathbf z^\prime)}\right)
\left(\frac{\delta {S}_{(0)}[\varphi]}{\delta \varphi(\mathbf z^\prime)}\right)\,d \mathbf z^\prime
  =      \int_{\R^n} \left(\frac{1}{2} \nabla^\prime   \varphi(\mathbf z^\prime)\cdot \nabla^\prime\varphi(\mathbf z^\prime)   +   \mathcal P( \varphi(\mathbf z^\prime))\right)\,d   \mathbf z^\prime \,,         \ee
then simplifies the formula for  this ratio of `time' derivatives to
\be\label{timeder}
\mathcal T = 1 +
\frac
{\int_{\R^n} \left(\nabla^\prime   \varphi(\mathbf y^\prime)\cdot \nabla^\prime\varphi(\mathbf y^\prime) + 2 a_2 \varphi^2(\mathbf y^\prime)+\dots + ka_k\varphi^k(\mathbf y^\prime)\right)\, d\mathbf y^\prime}
{\int_{\R^n} \left( \nabla^\prime   \varphi(\mathbf z^\prime)\cdot \nabla^\prime\varphi(\mathbf z^\prime) + 2 a_2 \varphi^2(\mathbf z^\prime)+\dots + 2a_k\varphi^k(\mathbf z^\prime)\right)\, d\mathbf z^\prime}
\ee
Luckily the formula above is independent of ${S}_{(0)}(\varphi )$ and only depends upon the given `potential energy' from the Hamilton-Jacobi equation. This simplification is the main reason for proposing to compute the flow along the (Hamilton-Jacobi) solution curves instead of along the `rescaling curves'  $\varphi_\lambda=  e^\lambda \varphi$. Note also that in the absence of the higher order terms eq. \eqref{timeder} immediately reproduces the free field result  $\mathcal T\to 2$  without the need for taking $\varphi$  `large'.
In the general case, Sobolev estimates show that each polynomial term except the top order ones will tend asymptotically to $0$ as $\Vert\varphi (\cdot)\Vert_{H_1(\R^n)}\to\infty$,  leaving the result $\mathcal T\to 1+k/2$. This is the `intuitively expected' result since, from the form of the Hamilton-Jacobi equation satisfied by ${S}_{(0)} [\varphi ]$, we seem to need ${S}_{(0)} [\varphi]$ scaling like $\varphi^{1+k/2}$ for large  $\varphi$   in order to match the behavior of the given potential energy for large arguments.

A difficulty in making the above argument rigorous arises through the fact that extension of the formal `integral curves' of the `gradient semi-flow' of the functional ${S}_{(0)}[\varphi]$
to positive $t$'s (for data specified at $t = 0$), does not necessarily make sense. Using our regularity results there is a clear mathematical sense to such `curves' for $t < 0$ but, to extend them in the opposite temporal direction seems problematic in general, especially when the boundary data chosen is as `rough' as possible. However, such rough data cannot arise at an interior point of such a hypothetically extendible curve. In fact, the global control and interior regularity, established in \cite{Marini:2016} and outlined in the present section for a solution $\Phi_\varphi$ to the nonlinear Dirichlet problem \eqref{wnlP}, ensure $\Phi_\varphi\in H_{3/2}(\nst)\cap \mathcal C^\infty(\nst)$. Thus, $\varphi_t\in H_{1}(\R^n)\cap \mathcal C^\infty(\R^n)\subset\bb$,  $\forall t<0$, and one could regard the smoothed interior data at some $t < 0$ as new `initial data' for a curve that is in fact extendible (at least back to the original $t = 0$ starting point) and
make presumably precise sense of the argument for a dense subset of the full space of initial data.

\noindent{\bf\emph{Final remarks.}} As described in the previous chapter, our fundamental solution $S_{(0)}(\mathbf{x})$, to the (inverted-potential-vanishing-energy) Hamilton-Jacobi equation for a coupled system of nonlinear oscillators has a natural geometric interpretation; \cf the discussion preceding Eqs. (\ref{eq430},\ref{eq431}).  Further,
the first quantum `loop correction', \(\mathcal{S}_{(1)}(\mathbf{x})\) also has a natural geometric interpretation in terms of `Sternberg coordinates' for the gradient (semi-) flow generated by $S_{(0)}(\mathbf{x})$, because the latter linearize the Hamilton-Jacobi flow equations
\eqref{eq432} to the form \eqref{eq433}), and generate a Jacobian determinant for the Hilbert-space integration measure that cancels out the contribution of the first quantum `loop correction' to inner product calculations (\cf eqs. (\ref{eq436}, \ref{eq437}, \ref{eq438})).
For purely \textit{harmonic} oscillators the original (Cartesian) coordinates are already of Sternberg type, \(\mathcal{S}_1(\mathbf{x})\) accordingly vanishes, and Hilbert space inner product integrals reduce to Cartesian form. For \textit{free fields}, on the other hand, such formal, stand-alone `Lebesgue measures', although, of course, mathematically undefined, when combined with the universally appearing convergence factors, \(N_\hbar\; e^{-\mathcal{S}_{(0)}[\varphi]/\hbar}\), arising in all of the associated wave functionals, can be interpreted as providing rigorously defined gaussian measures for Fock space.

To compute such higher order `loop' corrections for bosonic field theories, one will first need to regularize the formal functional Laplacian that arises in the Schr\"{o}dinger operator \eqref{23} and that will reoccur in each of the transport equations which result from substituting ansatz \eqref{25}  into the time independent Schr\"{o}dinger equation \eqref{26} and expanding formally in powers of \(\hbar\). Solving these transport equations for the `loop corrections', \(\left\lbrace\mathcal{S}_{(1)} [\varphi], \mathcal{S}_{(2)} [\varphi ], \cdots\right\rbrace\), to the ground state wave functional simply amounts to \textit{evaluating} sequentially computable, smooth functionals on the Euclidean action minimizers, \(\Phi_\varphi\), for arbitrary chosen boundary data, \(\varphi\).

Solving the transport equations for excited states is somewhat more involved since these equations entail a lower order term in the unknown but the technology for handling this, is well understood \cite{Moncrief:2012, Dimassi:1999, Helfer:1988}. If, for example, a Sternberg diffeomorphism could be shown to exist, then the leading order, excited state transport equation could be solved in closed form (as already established for nonlinear oscillators).

 Otherwise, one could simply fall back on the machinery developed in Refs.~\cite{Moncrief:2012, Dimassi:1999, Helfer:1988}, which does not assume the existence of Sternberg coordinates, and solve this and the corresponding higher order excited state equations in a less direct fashion. In either case it is intriguing to note that the excited states for \textit{interacting} field theories would be naturally labeled by sequences of (integral) `particle excitation numbers' in much the same way that the Fock-space excited states of a free field are characterized.

One often hears that the fundamental particle interpretation of \textit{interacting} quantized fields hinges upon their approximation by corresponding \textit{free} fields. This is unsatisfactory at best since, of course, an elementary particle cannot `turn off' its self-interactions in order to conveniently behave, even asymptotically, like a Fock-space, free field quantum. As we have already emphasized one of the natural features of this (Euclidean signature-semi-classical) program is that it maintains the dynamical nonlinearities of an interacting quantum system intact at every level of the analysis rather than attempting to reinstate nonlinear effects gradually through a perturbative expansion.

\subsection{An application to Yang-Mills fields}
\label{subsec:ym}

As in the foregoing models, construction of a Euclidean-signature-semi-classical wave functional for Yang-Mills theory proceeds by showing the existence of a solution to the Euclidean signature Dirichlet problem for initial data prescribed on $\tzero$.  The existence proof takes the form of a localizing and counting argument, based on the collective work of Uhlenbeck \cite{Uhlenbeck:1982a, Uhlenbeck:1982b}, Sedlacek \cite{Sedlacek:1982}, and Marini \cite{Marini:1992}.  In particular, two results of Uhlenbeck are crucial:  first, that an $L^{n/2}$ bound on the curvature of a connection yields a representative in Hodge gauge, the $L^p_1$ norm of which is bounded by the $L^p$ norm of its curvature \cite{Uhlenbeck:1982a}, and second, the removability of singularities for Yang-Mills connections in dimension 4 through application of an appropriate gauge transformation \cite{Uhlenbeck:1982b}.  The Sobolev bound on the connection in Hodge gauge allows the use of the direct method in the calculus of variations, invoking weak relative compactness of bounded sets in Sobolev space.  Meanwhile, the Hodge gauge condition itself enables the Yang-Mills equation to be reformulated as an elliptic partial differential equation, making available the use of powerful regularity results.  The removability of singularities advances two purposes:  first the removal of potentially singular points which occur in the solution because of the localization and counting procedure, and second the establishment of bounds on the decay of Yang-Mills connections at infinity, in a suitable global gauge choice.

In \cite{Marini:1992}, smooth solutions to the Dirichlet and Neumann problems for Euclidean signature Yang-Mills theory are shown to exist on a compact manifold with smooth boundary.  Because our base manifold in the present program is $\st$ (Euclidean spacetime), we extend this work to accommodate a noncompact base manifold with smooth boundary.  The counting argument in \cite{Sedlacek:1982} depends on compactness of the base manifold; thus we give a new proof of the relevant result.  Explicitly allowing for a noncompact base manifold obviates the need to restrict gauge transformations to approach a consistent value at spatial infinity, as would be necessary if one proceeded by compactifying spacetime and invoking results in \cite{Marini:1992}.  If gauge transformations were required to approach a coherent limit at spatial infinity, the space of physical connections on spacetime would divide into disjoint topological sectors, and a distinction could be introduced between `large' and `small' gauge transformations as in \cite{Jackiw:1980} according to homotopy class.  Under the approach in \cite{Jackiw:1980}, this distinction gives rise to the `vacuum angle' of quantum Yang-Mills theory, whereby physical states are only invariant up to a phase under large (i.e. non-null-homotopic) gauge transformations.  However as noted in \cite{Jackiw:1980} and \cite{Khoze:1994}, a vacuum angle can alternatively be introduced at the level of the  Lagrangian, leaving this avenue open to our approach if a vacuum angle is physically indicated.

We present our analysis for the case of Yang-Mills theory in 4 spacetime dimensions, but we note here that our results allow analogous construction of a candidate leading-order ground state wave functional in lower dimensions.  Because the quantization of Yang-Mills theory in 3 spacetime dimensions has been well addressed by the methods of Karabali, Kim, and Nair (see e.g. \cite{Karabali:1998}), in future work we aim to make contact with this program as a potentially illuminating comparison.

Similar to the preceding sections, we show the existence of a solution to the Euclidean signature Yang-Mills Dirichlet problem using the direct method in the calculus of variations.  Hence our first step is to prove lower semicontinuity of the (Euclidean signature) Yang-Mills action functional with respect to a suitably defined Sobolev topology of connections on $\st$.  Next we show that for a minimizing sequence of connections on $\st$ having initial data specified within a (local) Sobolev space of connections on the boundary $\tzero$, we can construct an open cover of $\st$ (possibly missing a finite collection of points) and a subsequence of the minimizing sequence for which we can locally transform to Hodge gauge.  Thanks to relative weak compactness of bounded sets in Sobolev space and lower semicontinuity of the Yang-Mills functional, a further subsequence of the minimizing sequence converges to a minimizer of the Yang-Mills functional.  Demonstrating that this minimizer is a smooth solution to the Yang-Mills Dirichlet problem is achieved by local arguments which apply unchanged from results in \cite{Marini:1992}.  Removability of singularities allows the solution to be extended to the points of $\st$ possibly missing from our open cover.  The removability of singularities is also differently applied to achieve decay results for Yang-Mills connections (see Proposition \ref{Decay} below, and the result it extends, Corollary 4.2 in \cite{Uhlenbeck:1982b}).

Having established the existence of a solution to the Yang-Mills Dirichlet problem, we turn to the properties of the leading-order semiclassical state constructed therefrom.  Using the Banach space version of Rademacher's theorem, we show that the natural logarithm of the leading-order semiclassical wave functional (the functional we denote as $S[\cdot]$) is G\^ateaux differentiable outside a Gaussian null set on the Sobolev space of connections on $\tzero$.  We discuss the possible application of the Banach space implicit function theorem to establish Fr\'echet differentiability of $S$ to all orders.

Although our leading-order ground state wave functional will be constructed on $\st$, we carry out the minimization of the Euclidean signature Yang-Mills action more generally over a smooth $n$-dimensional Riemannian manifold $(M,\mathbf{g})$, not necessarily compact, with smooth boundary $\partial M$.  The Yang-Mills structure group is taken to be a compact semisimple real Lie group $G$, with Lie algebra $\g$.  We denote by $P$ a principal $G$-bundle over $M$.  Following \cite{Fr-Uhl:1991}, associated bundles $\eta$ having fiber $V$ are constructed by specifying a representation $\rho: G \to \Aut V$ and forming the twisted product\footnote{Note that, as in \cite{Bredon:1972}, for a topological space $X$ admitting a continuous right action by a topological group $G$ and a topological space $Y$ admitting a continuous left action by $G$, the \emph{twisted product} $X \times_G Y$ is defined as the Cartesian product $X \times Y$ modulo the equivalence relation $(x.g ,y) \sim (x, g.y)$. } $\eta = P \times_\rho V$.  In particular, the automorphism bundle of $\eta$ is $\Aut \eta \equiv P \times_\ad G$, where $\ad:G \to \Aut G$ denotes the action of $G$ on itself by conjugation.  Gauge transformations are given by sections $\Gamma (\Aut \eta)$.  The adjoint bundle $\Ad \eta \equiv P \times_\Ad \g$, where $\Ad$ denotes the adjoint representation of $G$ on $\g$, is used to construct $\g$-valued differential $k$-forms on $M$ as sections of $\Ad \eta \otimes \Lambda^k M$.  A connection on an associated bundle $\eta$ is a first-order differential operator $D : \Gamma(\eta) \to \Gamma (\eta \otimes T^*M)$, which can be expressed in terms of a base connection $D_0$ as $D_0 + g \stA$ for $\stA \in \Gamma ( \Ad \eta \otimes T^*M )$.  The coefficient $g$ represents the Yang-Mills coupling constant.  Connections are gauge transformed through conjugation by elements of $\Gamma(\Aut \eta)$:
\begin{equation}
\label{gt}
\begin{split}
\sigma^*(D) &= \sigma^{-1} D  \sigma = D_0 + \sigma^{-1}D_0
\sigma + g\, \sigma^{-1} \stA \sigma \\
&= D_0 + g\, \left( \frac{1}{g} \sigma^{-1}D_0
\sigma + \sigma^{-1} \stA \sigma \right) .
\end{split}
\end{equation}
The curvature of a connection $\stF(D) \in
\Gamma (\Ad\,\eta\otimes \Lambda^2 M)$ is defined by
\begin{equation*}
\stF \left( v,w \right) = \frac{1}{g}\, \left( \left[ D_v , D_w \right] - D_{\left[ v,w \right]} \right),
\end{equation*}
and transforms by conjugation under gauge transformations:
\begin{equation*}
\sigma^*(\stF) = \sigma^{-1} \stF \sigma.
\end{equation*}
Locally, we can write
\begin{equation*}
\stF = d \stA+\frac{g}{2}\,\left[ \stA,\stA\right] \,.
\end{equation*}

Central to defining the Euclidean signature Yang-Mills action and obtaining its minimizer for given boundary data on $M$ is the construction of Sobolev spaces for $\g$-valued differential forms.  An $L^2$ inner product on sections of $\Ad \eta \otimes \Lambda^kM$ is given by
\begin{align}
\label{FormProd}
\langle \phi ,\theta \rangle  \equiv
\int_M \tr \left( \phi \wedge \ast \theta \right) \, ,
\end{align}
where for $\theta = T \otimes \mu \in \Gamma (\Ad \eta \otimes \Lambda^kM )$, $\ast \theta \equiv T^\dagger \otimes \ast \mu$, with $T^\dagger$ the adjoint in $\g$ and $\ast \mu$ the Hodge dual with respect to the metric $\textbf{g}$.  A $G$-invariant positive-definite inner product results, since in the adjoint representation of $\g$, $\tr XY^\dagger = - \tr XY$, the negative of the Killing form, which thanks to compactness and semisimplicity of $\g$ is negative definite.  The $L^{p}$ norm of a $\g$-valued differential $k$-form $\phi$ can then be defined by
\begin{align}
\label{Lpnorm} \left\| \phi \right\| _{p}=\left( \int_{M}\left|
\phi \right| ^{p}\right) ^{\frac{1}{p}} = \left( \int_{M}\left[ \tr \left( \phi \wedge \ast \phi \right) \right]
^{p/2}\right) ^{\frac{1}{p}} \, .
\end{align}
Accordingly, the Euclidean signature Yang-Mills action is defined as
\begin{align}
\label{yme}
\Ies [ \stA ] &= \frac{1}{2}\left\| \stF \right\| _{2}^{2} =
\frac{1}{2} \int_{M} \tr \left( \stF \wedge \ast \stF \right) \notag \\
&= \frac{1}{4} \int_{M} \textbf{g}^{\mu \rho }\textbf{g}^{\nu \sigma}\bigl(\stF_{\mu \nu}, \stF_{\rho \sigma}\bigr) \sqrt{\det \textbf{g}} \ d{\bf x}_M =
\frac{1}{4} \int_{M} \stF^I_{\mu \nu} \stF_I^{\mu \nu} \sqrt{\det \textbf{g}} \ d{\bf x}_M\,,
\end{align}
where $\bigl(\stF_{\mu \nu}, \stF_{\rho \sigma}\bigr)\equiv \tr \stF_{\mu \nu} {\stF_{\rho \sigma}}^\dagger$, and the index $I$ runs over a basis of the Lie algebra $\g$ normalized with respect to $\left( \cdot , \cdot \right)$.  Summation with respect to repeated up and down indices is implicit (Einstein summation convention).  Gauge invariance of the action functional follows from $G$-invariance of the form inner product.  The Yang-Mills Dirichlet problem results from varying this action with fixed initial data to derive Euler-Lagrange equations:
\begin{equation*}
\label{dirichlet} (\mathcal{D})\quad\left\{\begin{array}{ll}
D^{\ast} \stF = 0 \quad & \mbox{ on } M \\
i^{\ast} \stA \sim \A \quad & \mbox{ on } \partial M\, .
\end{array}\right. \, ,
\end{equation*}
where $\sim$ denotes a gauge transformation on $\partial M$ which can be extended to the interior.

Before elaborating our results, we note that the definition of Sobolev spaces of sections of a vector bundle over a noncompact base manifold must be undertaken relative to a choice of base connection.  In contrast to the situation over a compact base manifold, this choice affects not only the values of Sobolev norms, but also whether or not a given connection belongs to the Sobolev space.  Relatedly, working over a noncompact manifold introduces a distinction between local and global Sobolev spaces of sections.  Membership of connections in local Sobolev spaces is independent of the choice of connection, since the condition is that the section have finite Sobolev norm when restricted to any compact subset of the manifold.  For details on defining Sobolev spaces of sections of vector bundles, in particular over noncompact manifolds, see \cite{Wehrheim:2004}.  

In proving the existence of a minimizer for the Euclidean signature Yang-Mills functional $\Ies [ \cdot]$ given arbitrary boundary data $A$ on $\tzero$, we work over the local Sobolev space of connections
\be
\connLloc (\A) \equiv \left\{ D = d + g \stA \, : \, \stA \in L^2_{1;loc}(\zerost , \Ad \eta \otimes T^*(\zerost)), \;  i^* \stA \sim \A \right\} \, .
\ee
Here $i^*$ is the pullback of the inclusion map $i: \tzero \xhookrightarrow{} \zerost$, so that the condition $i^* \stA \sim \A $ requires the tangential component of $\stA$ restricted to the boundary to be gauge equivalent to $\A$ on the restriction of the bundle to the boundary, via a gauge transformation which extends with suitable regularity to the interior.  The boundary value $\A$ is from the space $\bb$ consisting of $L^2_{1;loc} \left(\tzero , \Ad \eta \otimes T^*( \tzero ) \right)$ connections on the boundary extending to $\st$ with finite Euclidean action.  Gauge transformations must be one degree more regular than connections owing to the form of the transformation \eqref{gt}; hence they belong to $L^2_{2;loc}$.

Having defined the relevant Sobolev space over which to minimize, we prove that the Yang-Mills functional is lower semicontinuous with respect to its weak topology:
\begin{theorem}
\label{semicont}
The Euclidean signature Yang-Mills functional on a 4-dimensional Riemannian manifold $M$ (with or without boundary) is lower semicontinuous with respect to the weak topology on $L^2_{1;loc}\left( M, \Ad \eta \otimes T^*M \right)$; i.e. any sequence of connections $\{\stA_i\}$ such that $\stA_i\rightharpoonup \stA_\infty$ in $L^2_{1;loc}\left( M, \Ad \eta \otimes T^*M \right)$ satisfies the inequality
\be
\left.\Ies \right|_U\,\left[ \stA_\infty \right] \equiv  \frac{1}{2} \Vert \stF_{\infty} \Vert^2_{L^2(U)}
\leq \lim \inf_{i\rightarrow \infty }\left. \Ies \right|_U\,\left[ \stA_{i} \right]
\equiv \lim \inf_{i\rightarrow \infty } \frac{1}{2} \Vert \stF_i \Vert^2_{L^2(U)}\,,
\ee
for all bounded open sets $U \subset M$.
\end{theorem}

Although the setting is different, our proof is similar in method to Sedlacek's proof of Lemma 3.6 in \cite{Sedlacek:1982}.  Since norms are weakly lower semicontinuous, we show that the sequence $\left\{ \stF_i  \right\}$ has a subsequence $\left\{ \stF_{i_k}  \right\}$ weakly convergent in $L^2(U,\Ad \eta \otimes \Lambda^2U)$ such that $\liminf \| \stF_{i_k} \|_{L^2(U)} = \liminf \| \stF_i \|_{L^2(U)}$.  Weak convergence of the first term in $\stF_i = dA_i + \frac{g}{2} \left[ \stA_i , \stA_i \right]$ is immediate from our hypotheses, while for the second, weak convergence of a suitable subsequence follows by using the Principle of Uniform Boundedness together with continuity of the embedding $L^2_1 \hookrightarrow L^4$ and the multiplication $L^4 \times L^4 \to L^2$ to show that $\left\{ \| \left[ \stA_i , \stA_i \right] \|_{L^2(U)} \right\}$ is bounded.

\noindent\textit{\textbf{Existence of a minimizer for $\Ies$ for initial data $A \in \bb$.}}  The construction of the functional $S[A] \equiv \Ies \left[ \stA_A \right]$ hinges upon the existence of a minimizer $\stA_A$ of $\Ies$ for initial data $\A$.  Although we could simply define $S[A]$ as $\inf_{\stA \in \connLloc(\A)} \Ies[\stA]$, failure of the minimizer to exist would be an obstruction to the continuity of $S[A]$, while nonuniqueness of the minimizer would be an obstruction to differentiability.  

As in \cite{Marini:1992}, we proceed by proving the existence of a cover of $M$ (possibly missing finitely many points) on which the choice of Hodge gauge ($d^\ast \stA=0$ on the interior, $d^\ast_\tau \stA_\tau = 0$ on the boundary, where $\tau$ denotes tangential components) effectively renders the Yang-Mills equation elliptic.  Indeed, the Hodge gauge allows us to interpret the highest-order term in the Yang-Mills equation
\begin{equation*}
d^{\ast} d \stA + \frac{g}{2} d^{\ast} \left[ \stA, \stA \right] + g \left( \ast \left[ \stA, \ast \stF \right] \right) = 0
\end{equation*}
as the Laplace-de Rham operator $\Delta = d^{\ast} d + d d^{\ast} $, where $ d^{\ast} = \ast d \ast $.  Owing to a theorem proved by Uhlenbeck on interior neighborhoods (\cite{Uhlenbeck:1982b} Thm 2.1) and by Marini (\cite{Marini:1992} Thms 3.2 and 3.3) on neighborhoods at the boundary, such a gauge choice is locally possible provided the $L^{n/2}$ norm of the curvature is sufficiently small:
\begin{theorem}
\label{localgauge}
Let the local neighborhood $U$ be a ball in the interior of $M$ or a half-ball centered at a point on the boundary of $M$.  Let $D = d + g \stA$ be a connection such that $\stA \in L^p_1(U)$ (and for a boundary neighborhood, additionally $i^\ast \stA \equiv \stA_\tau \vert_{U} \in L^p_1(\partial U \cap \partial M)$), where $ n/2 \leq p < n$.  Then there exists $K \equiv K(n)/g > 0$ and $c \equiv c(n)$
such that every connection $D = d + g \stA$ satisfying $\| \stF \|_{L^{n/2}(U)}<K$ (and for a boundary neighborhood, additionally $\| \stF_\tau \|_{L^{n/2}(\partial U \cap \partial M)}<K$)
is gauge equivalent to a connection $d + g \hat{\stA},$ $\hat{\stA} \in L^p_1(U, \Ad \eta \otimes T^* U),$ satisfying
\begin{enumerate}[(i)]

\item $d^\ast \hat{\stA}=0$ on $U$

\item For interior neighborhoods, $\hat{\stA}_{\nu }=0 \ on \ \partial U$, where $\nu$ denotes the normal component to $\partial U$

\item For neighborhoods at the boundary, $d_{\tau } ^\ast \hat{\stA}_{\tau }=0\ \ on\ \partial U \cap \partial M$, and $\hat{\stA}_{\nu }=0 \ on \ \partial U \setminus \partial M$

\item $\| \hat{\stA} \| _{1,n/2}<c(n) \| \hat{\stF} \| _{n/2}$

\item $\| \hat{\stA} \| _{1,p}<c(n) \| \hat{\stF} \|_{p}$

\end{enumerate}
Moreover, the gauge transformation $s$ used to obtain $\hat{\stA} = \frac{1}{g} s^{-1} d s+s^{-1}\stA s$ can be taken in $L^{n/2}_2(U)$
($s$ will in fact always have one more degree of regularity than $\stA$; see Lemma 1.2 in Ref. \cite{Uhlenbeck:1982b}).
\end{theorem}

As observed in \cite{Marini:1992}, conformal invariance of the condition $\| \stF \|_{n/2} < K$ allows us to achieve the simultaneous condition $\| \stF_\tau \|_{L^{n/2}(\partial U \cap \partial M)} < K$ for a half-ball $U$ at the boundary by means of a dilation $\tilde{\mathbf{x}} = r \mathbf{x}$, since $\| \stF \|_{L^{n/2}(\partial U \cap \partial M)}$ acquires a factor of $r$.  For the case $n=4$ as in the physical problem, the requisite $L^{n/2}(U)$ bound on $\stF$ is the same as a bound on the Euclidean signature Yang-Mills action restricted to $U$.  Thus to apply Theorem \ref{localgauge}, we seek a cover of $M$ by balls on the interior and half-balls centered at boundary points such that the Yang-Mills action restricted to each ball $U$ in the cover meets the condition  $\| \stF \|_{L^{n/2}(U)}<K$.  We prove a general result for a sequence of positive $L^1$-bounded functions on $\R^n$ (or indeed an arbitrary Riemannian manifold, with or without boundary) assuring the existence of a cover (missing at most a finite collection of points) for which the $L^1$ norm of the functions restricted to balls in the cover is less than an arbitrary fixed bound $\varepsilon$.  For the case of Yang-Mills theory, the possibility that our cover may miss finitely many points will be addressed by removability of singularities, as discussed below.

For simplicity, we state and prove our `good cover' theorem on $\R^n$, but the modifications for a general Riemannian manifold are obvious.

\begin{theorem} \label{non-compact}
Consider a sequence of positive functions $\{f_i\}_{i\in\N}$ on $\Rn$
such that
\begin{equation}
 \label{bound}
 F_i \equiv \int_{\Rn} f_i \, d{\bf x}\leq C\,, \quad\; \forall i\in \N\,,
 \end{equation}
for some constant $C$.

\noindent Then for any arbitrarily fixed $\epsilon>0$,  there exists a subsequence of $\{f_i\}$ and an open cover $\mathscr{G}$ made up of balls $B \subset \R^n$, covering $\R^n$ except at most a finite number of points $P_1, P_2, \dots , P_N$ and satisfying
\be \int_B f_i \, d{\bf x}\leq\epsilon\,,\quad\;\forall i\geq i_B \,\ee
where the index $i_B$ depends on the ball  $B\in\mathscr{G}.$
\end{theorem}

For the proof, we construct a countable cover $\mathscr{C}$ of $\R^n$ as the union of countable covers of $\R^n$ by balls of radius $\frac{1}{j}$ for each $j \in \mathbb{N}$.  Fixing an enumeration of the balls in the cover, we extract a subsequence $\{ f_i^{(1)}\}$ of $\{ f_i \}$ such that for the first ball $B_1$, $\left\{ \int_{B_1} f_i^{(1)} \, d \mathbf{x} \right\}_{i=1}^\infty$ converges.  From this subsequence, we extract a further subsequence $\{ f_i^{(2)}\}$ so that on the second ball $B_2$, $\left\{ \int_{B_1} f_i^{(1)} \, d \mathbf{x} \right\}_{i=1}^\infty$ converges, and so on.  We observe that for the diagonal subsequence $\{ f_i^{(i)}\}$, the sequence $\left\{ \int_{B_k} f_i^{(i)} \, d \mathbf{x} \right\}_{i=1}^\infty$ converges on each ball $B_k$ in $\mathscr{C}$.  If the limit of this sequence for some $B_k$ is greater than or equal to the bound $\varepsilon$, we discard the ball $B_k$ from $\mathscr{C}$.  Finally, we show that the resulting collection $\mathscr{G}$ of balls, for each of which the sequence $\left\{ \int_{B_k} f_i^{(i)} \, d \mathbf{x} \right\}_{i=1}^\infty$ converges to a value less than $\varepsilon$, covers all but finitely many points of $M$.  To do so, we use the uniform bound $C$ on the $L^1$ norms of the $f_i^{(i)}$ to deduce that the number of pairwise disjoint balls of radius $\frac{1}{j}$ which have been discarded for each $j \in \N$ cannot exceed $[C/\varepsilon]$.  For each $j \in \N$ we then select a maximal set of pairwise disjoint discarded balls of radius $\frac{1}{j}$.  By expanding their radii to $\frac{3}{j}$, we ensure that their union $S^j$ contains all the discarded balls of radius $\frac{1}{j}$.  This fact together with the bound on the number of discarded balls for each $j \in \N$ allows us to argue by contradiction that the set of points $\cap_{j=1}^\infty S^j$ missed by the cover is limited by the same bound $[ C / \varepsilon ]$ as the number of discarded balls of each radius $\frac{1}{j}$.  (For full details, refer to \cite{Maitra:inprep}.)

A crucial difference between the above proof and that of the analogous result in \cite{Sedlacek:1982} (namely Proposition 3.3) is that in our version the underlying manifold $M$ need not be compact.  In \cite{Sedlacek:1982}, compactness is used in two ways:  first to guarantee that any cover of $M$ by balls of radius $\frac{1}{j}$ can be assumed finite, and second to argue that the centers of balls $B$ on which $\int_{B} f_i^{(i)} \, d \mathbf{x} $ exceeds $\varepsilon$ must be fixed, using the fact that a sequence of centers of such balls must have a convergent subsequence.  Because we first pass to a subsequence of $\{ f_i \}$ on which balls can be consistently labeled as good or bad, we avoid the need to argue after diagonalizing that bad balls stabilize.

A direct application of Theorem \ref{non-compact} to sequences of connections with uniformly $L^2$-bounded curvature yields
\begin{theorem}
\label{goodcover} Let $\left\{ \stA(j)\right\}_{j \in \mathcal{J}} $ be a sequence of connections in $G$-bundles $P_{j}$ over a smooth Riemannian n-dimensional manifold $M$ with boundary, with uniformly $L^2$-bounded curvature $\left(
\int_{M}\left| \stF(j)\right| ^{2} \ d {\bf x} \right)^{1/2} < B \ \forall \ j$.  For any $\varepsilon > 0$, there exists a countable collection
$\left\{ U_{\alpha }\right\} $ of balls in the interior and half-balls at the boundary, a collection of indices $J_{\alpha }$, a subsequence $\left\{ \stA(j) \right\} _{j \in \mathcal{J}^{\prime }}\subset \left\{ \stA(j)\right\}
_{j \in \mathcal{J}}$, and at most a finite number of points $\left\{
{\bf x}_{1}, \dots ,{\bf x}_{k}\right\} \in M$ such that
\begin{align*}
\bigcup U_{\alpha } \supset M \setminus \left\{ {\bf x}_{1}, \dots ,{\bf x}_{k} \right\} \\
\| \stF (j) \|_{L^2 \left( U_{\alpha} \right)} < \varepsilon \ \ \forall j \in \mathcal{J}^{\prime }, \ \ j>J_{\alpha }.
\end{align*}
\end{theorem}

Since an action-minimizing sequence of connections with given boundary data necessarily satisfies a uniform $L^2$-bound on curvature, Theorem \ref{goodcover} ensures the existence of a subsequence and a countable cover of $M$ (possibly missing finitely many points) for which the connections in the subsequence locally satisfy the $L^2$ bounds on curvature prescribed by Theorem \ref{localgauge} to allow transformation to Hodge gauge.  Using the bound on Sobolev norms of connections in local Hodge gauge as given by condition \textit{(iv)} of Theorem \ref{localgauge}, we can diagonalize over the countable cover to obtain a further subsequence of connections convergent to a weak limit in $L^2_1$ on each neighborhood of the cover.  To justify this localization, we must establish that the collection of limiting connections on neighborhoods of the cover can be patched into a global connection which satisfies the Yang-Mills equation and preserves the correct boundary data.  This result appears as Theorem 3.4 in \cite{Marini:1992} and applies unchanged to the present problem.

Lower semicontinuity of the Yang-Mills functional (Theorem \ref{semicont}) now implies that the limiting connection $\stA_\infty \equiv \{ \stA_\alpha \}$ indeed minimizes the Euclidean signature action, which in turn allows a proof by contradiction to establish that $\stA_\infty$ is a weak solution of the Yang-Mills Dirichlet problem with the prescribed initial data (\cite{Marini:1992}, Proposition 3.5).  These results culminate in Theorem 3.6 of \cite{Marini:1992}, reproduced here for completeness (with appropriate changes for the notation and function spaces of the current problem defined over a noncompact manifold):
\begin{theorem}
\label{YMexistence}
Let $\{ \stA_i \}$ be a sequence of $L^2_1$ connections on fiber bundles $P_i \to M$ with prescribed smooth tangential boundary components $(\stA_i)_\tau \vert_{\partial M} \equiv \A$, which also minimizes the Euclidean signature Yang-Mills action, i.e. $\Ies \left[ \stA_i \right] \to \inf_{\stA \in \connLloc(\A)} \Ies \left[ \stA \right]$.  Then there exists a collection of neighborhoods $\{ U_\alpha \}$ covering $M$ except at most a finite number of points $\{ \mathbf{x}_1 , \dots , \mathbf{x}_k \}$ and trivializations $\sigma_\alpha (i)$, such that a subsequence can be found that admits a weak limit in $L^2_1$, on each $U_\alpha$, called $\stA_\alpha$.  The collection $\{ \stA_\alpha \}$ makes a connection $\stA_\infty$ on a bundle over $M \setminus \{ \mathbf{x}_1 , \dots , \mathbf{x}_k \}$, with transition functions in $L^2_2$.  This connection solves the Yang-Mills Dirichlet problem with boundary data $\hat A$, gauge equivalent to $A$ via a smooth gauge transformation.  The connection $\stA_\infty$ satisfies conditions (i) - (iii) of Theorem \ref{localgauge}.
\end{theorem}

Regularity of $\stA_\infty$ on a bundle over $M \setminus \{ \mathbf{x}_1 \dots , \mathbf{x}_k \}$ follows due to ellipticity of the Yang-Mills equation in Hodge gauge.  The limiting connection can be extended to the points $\{ \mathbf{x}_1 , \dots , \mathbf{x}_k \}$ by means of the removable singularity theorems due to Uhlenbeck for interior points (Theorem 4.1 in Ref. \cite{Uhlenbeck:1982b}), and to Marini for boundary points (Theorem 4.6 in Ref. \cite{Marini:1992}), so that the connection $\stA_\infty$ extends to a smooth connection.

\noindent\textit{\textbf{{Decay estimates on Yang-Mills connections.}}} Just as existence of a regular solution to the Yang-Mills Dirichlet problem depends on transforming locally to Hodge gauge, decay properties of Yang-Mills connections in a suitably chosen global gauge depend on the removability of singularities.  We prove a decay result for Yang-Mills connections which concludes that any smooth boundary data for a Yang-Mills connection has a gauge representative in $L^2_1(\tzero)$.  The first part of the proposition is a version of Uhlenbeck's Corollary 4.2 in \cite{Uhlenbeck:1982a} for the base manifold $M=\st$ while the second part applies the same principle to bound the growth of the connection 1-form $\stA$;

\begin{proposition}
\label{Decay}
Let $D = d + g \stA$ be a Yang-Mills connection in a bundle $P$ over an exterior
region $\mathcal{V}=\left\{ {\bf y} \in \st : \left| {\bf y} \right| \geq N \right\} $ with finite Euclidean action and smooth boundary data $\A = i^* \stA$ on $\tzero$. 
Then
\begin{enumerate}[(a)]
\item $\left| \stF ({\bf y}) \right| \leq C \left| {\bf y} \right| ^{-4}$ for some constant $C$ (not uniform);
\item There exists a gauge in which $D = d + g \tilde{\stA}$ satisfies $\left| \tilde{\stA} ({\bf y}) \right| \leq K \left| {\bf y} \right| ^{-2}$;
\item $\A$ is gauge equivalent to a connection in $L^2_1 (\tzero)$.
\end{enumerate}
\end{proposition}
Our proof of part \textit{(a)} proceeds by pulling back a conformal mapping $f:U_\ast = \{ \mathbf{x} \in \st \ : \ 0 < \lvert \mathbf{x} \rvert \le 1 \} \to \mathcal{V}$ defined by ${\bf x} \mapsto {\bf y} \equiv N \mathbf{x}/\lvert \mathbf{x} \rvert^2$ to act on the connection $D$, so that $f^\ast D$ is a Yang-Mills connection on $U_\ast$ and the removable singularities result for half-balls on the boundary (Theorem 4.6 in \cite{Marini:1992}) guarantees a gauge transformation $\sigma$ under which $f^\ast D$ which extends smoothly to the origin.  The transformation law for 2-forms then allows us to bound $\lvert \stF (\mathbf{y}) \rvert$ as follows:
\begin{align*}
\left| \stF \left( {\bf y}\right) \right| &=\left| f^{\ast } \stF ({\bf x}) \right| \left|
d f \left( {\bf x} \right) \right| ^{-2} \\
& \leq \max_{{\bf x} \in U} \left| f^{\ast } \stF ({\bf x}) \right| \cdot \left( N / \left|
{\bf x} \right| ^{2}\right) ^{-2} \\
& = C^{\prime }N^{2}\left| {\bf y}\right| ^{-4}
\end{align*}

For part \textit{(b)}, we construct the appropriate gauge transformation $s$ as the composition $\sigma \circ f^{-1}$ of the maps $\sigma$ and $f^{-1}$ from part \textit{(a)}.  An estimate analogous to that in part \textit{(a)} using the transformation law for 1-forms yields the desired bound on the decay of $\tilde \stA \equiv \stA^s$.

To complete the proof of part \textit{(c)}, we simply compute derivatives of the transformation law to show componentwise that $\lvert \partial \stA^s / \partial y^\alpha \rvert \le K' \lvert \mathbf{y} \rvert^{-2}$ for $\mathbf{y} \in \mathcal{V}$.  We then define a simple extension of $s$ to all of $\st$ by extrapolation into $\st \setminus \mathcal V$ of its values on $\partial \mathcal V$, so that $\stA^s$ is now a global gauge representative for $\stA$ with decay of the connection and its componentwise partial derivatives bounded by a constant multiple of $\lvert \mathbf{y} \rvert^{-2}$.  Since $i^\ast \stA$ obeys the same decay condition and since the volume element on $\tzero$ has weight $\lvert \mathbf{y} \rvert^2$, we have $i^* \stA^s \in L^2_1 \left( \tzero \right)$.

Note that as a corollary, any smooth connection $A$ on $\tzero$ which serves as boundary data for a Yang-Mills connection approaches pure gauge at spatial infinity.  This follows simply because the decay condition implies that the curvature of $A$ approaches zero at spatial infinity, and hence $A$ must approach pure gauge (although it need not approach the same value in all spatial directions).

\noindent\textit{\textbf{{Almost-everywhere G\^ateaux differentiability of $S$.}}}  To show that the functional $S[\cdot]$ is G\^ateaux differentiable almost everywhere, we employ the Banach space version of Rademacher's theorem derived in \cite{Phelps:1978}, Theorem 6.  This requires us to view $S\left[ \cdot \right]$ as a map from an open subset $U$ of a separable real Banach space $X$ to a real Banach space $Y$ satisfying the Radon-Nikodym property (i.e. every function of bounded variation from $\left[ 0,1 \right]$ to $Y$ must be differentiable almost everywhere), and to show that as such it is locally Lipschitz.  Thus our first task is to show that with suitable interpretation of its domain and range, $S \left[ \cdot \right]$ is locally Lipschitz.  The Banach space version of Rademacher's theorem then states that such a locally Lipschitz map $T$ from $U \subset X$ to $Y$ is G\^ateaux differentiable outside a Gaussian null subset of $U$.  A Gaussian null set in a separable Banach space is a Borel subset which has measure zero under every nondegenerate Gaussian measure (for details of defining Gaussian measures on Banach spaces, see \cite{Phelps:1978}).

To achieve Banach space structure on the domain $\bb$ of $S\left[ \cdot \right]$, we view the local Sobolev space $L^{2}_{1;loc} (\tzero , \Ad \eta \otimes T^*(\tzero))$ as the union of all affine Sobolev spaces $\underaccent{\tilde}{A} + L^2_1(\tzero , \Ad \eta \otimes T^*(\tzero) )$ defined in terms of a reference connection $\underaccent{\tilde}{A} \in L^2_{1;loc}(\tzero , \Ad \eta \otimes T^*(\tzero) )$ determining the covariant derivative $D_{\underaccent{\tilde}{A}}$.  The global Sobolev space $L^2_1(\tzero , \Ad \eta \otimes T^*(\tzero) )$ consists of sections $\omega \in \Gamma \left( \tzero , \Ad \eta \otimes T^*(\tzero) \right)$ having finite Sobolev norm $\| \omega \|_{2,1} \equiv \left( \| \omega \|_2^2 + \| D_{\underaccent{\tilde}{A}} \omega \|_2^2 \right)^{1/2}$ (see \cite{Wehrheim:2004}).

Given a fixed reference connection $\underaccent{\tilde}{A}$ on $\{0\} \times \R^3$, we interpret $S \left[ \cdot \right]$ as a functional defined on the Banach space $L^2_1(\tzero , \Ad \eta \otimes T^*(\tzero) )$, namely $\omega \mapsto S \left[ \underaccent{\tilde}{A} + \omega \right]$.  We prove that thus considered, $S \left[ \cdot \right]$ is locally Lipschitz:
\begin{theorem}
Let $\underaccent{\tilde}{A} \in L^{2}_{1;loc} (\tzero , \Ad \eta \otimes T^*(\tzero))$ be a reference connection such that $\underaccent{\tilde}{F} \equiv F_{\underaccent{\tilde}{A}}$ satisfies $\int_{\tzero} \underaccent{\tilde}{F} ^2 \, d \mathbf{x'} < \infty$, and such that there exists an extension $\underaccent{\tilde}{\mathscr{A}} \in L^{2}_{1;loc} (\st , \Ad \eta \otimes T^*(\st))$ of $\underaccent{\tilde}{A}$ into $\st$ with finite Euclidean signature Yang-Mills action.  Then for all sections $\omega \in L^2_1(\tzero , \Ad \eta \otimes T^*(\tzero) )$, the functional $\omega \mapsto S \left[ \underaccent{\tilde}{A} + \omega \right]$ is well defined and locally Lipschitz with respect to the norm $\| \cdot \|_{2,1}$ on $L^2_1(\tzero , \Ad \eta \otimes T^*(\tzero) )$.
\label{locallipschitz}
\end{theorem}

We first verify that $S\left[ \underaccent{\tilde}{A} + \omega \right]$ is well defined, namely that $\underaccent{\tilde}{A} + \omega$ has an extension $\stA$ into $\st$ with finite Euclidean signature Yang-Mills action.  To construct $\stA$, we represent the extension $\underaccent{\tilde}{\mathscr{A}}$ of $\underaccent{\tilde}{A}$ into $\st$ in a choice of gauge such that $i^* \underaccent{\tilde}{\mathscr{A}} = \underaccent{\tilde}{A}$ (as opposed to merely $i^* \underaccent{\tilde}{\mathscr{A}} \sim \underaccent{\tilde}{A}$).  We then define $\stA$ piecewise over the time component of $\st$ as a linear interpolation between $\underaccent{\tilde}{A} + \omega$ and $\underaccent{\tilde}{A}$, followed by $\underaccent{\tilde}\stA$:
\begin{equation*}
\mathscr{A} = \begin{cases} \underaccent{\tilde}{A} + (1+t)\omega, & 0 \ge t \ge -1 \\ \underaccent{\tilde}{\mathscr{A}
} \vert_{t+1} , & -1 > t  > -\infty  \end{cases} \,
\end{equation*}
(define the $dt$ component of $\mathscr{A}$ to be 0 for $0 > t \ge -1$).  Straightforward computation of the curvature of $\stA$ followed by Sobolev estimates confirms that $S\left[ \underaccent{\tilde}{A} + \omega \right]$ is well defined.

To show that $\omega \mapsto S \left[ \underaccent{\tilde}{A} + \omega \right]$ is locally Lipschitz on $L^{2}_1 (\tzero , \Ad \eta \otimes T^*(\tzero))$, we must show that for every $\omega_0 \in L^{2}_1 (\tzero , \Ad \eta \otimes T^*(\tzero))$, there exists $\delta > 0$ such that for all sections $\omega_1, \omega_2$ satisfying $\| \omega_j - \omega_0 \|_{2,1} < \delta$, $j=1,2$,
\begin{align}
\left| S [ \omega_1] - S [ \omega_2 ] \right| \le M \| \omega_1 - \omega_2 \|_{2,1}
\end{align}
for some constant $M$ allowed to depend on $\omega_0$.  Observe that for sections $\omega_1 , \ \omega_2$ satisfying $\| \omega_j - \omega_0 \|_{2,1} < \frac{1}{2}$, $j=1,2$, we have $\| \omega_1 - \omega_2 \|_{2,1}  < 1$ and $\| \omega_j \|_{2,1}  < \| \omega_0 \|_{2,1} + \frac{1}{2}$.

Similar to the argument that $S$ is well defined, we define an extension $\stA_1^\prime$ for $\underaccent{\tilde}{A}+\omega_1$ by linearly interpolating to $\underaccent{\tilde}{A}+\omega_2$ and appending its Yang-Mills minimizer $\stA_2$ (such a minimizer exists since $\underaccent{\tilde}{A}+\omega_2$ can be extended into $\st$ with finite Euclidean signature Yang-Mills action).  In the definition below, we use $t_{12} \equiv \|\omega_1 - \omega_2 \|_2<1$:
\begin{align*}
\mathscr{A}'_1 \equiv \begin{cases} \underaccent{\tilde}{A} + ( 1 + \frac{t}{t_{12}} )  \omega_1  -  \frac{t}{t_{12}} \omega_2 , & 0 \ge t \ge - t_{12}   \\
\mathscr{A}_2 \vert_{ t + t_{12} } , & -t_{12}>t>-\infty \, , \end{cases}
\end{align*}
as before defining the $dt$ component of $\mathscr{A}'_1$ to be 0 for $0 \ge t \ge -t_{12}$.  We define an analogous extension $\stA_2^\prime$ for $\underaccent{\tilde}{A}+\omega_2$.

Using the fact that $S[\underaccent{\tilde}{A} +\omega_1] \le \Ies [\mathscr{A}'_1] = \int_{-t_{12}}^0 \int_{\R^3} ( \mathscr{F}' )_1^2 d \mathbf{x}' dt + S[ \underaccent{\tilde}{A} + \omega_2]$ and analogously with the indices 1 and 2 reversed (in fact, $\int_{-t_{12}}^0 \int_{\R^3} \left( \mathscr{F}'_1 \right)^2 \, d \mathbf{x}' dt = \int_{-t_{12}}^0 \int_{\R^3} \left( \mathscr{F}'_2 \right)^2 \, d \mathbf{x}' dt$)), we have the estimate
\begin{equation}
\lvert S [ \underaccent{\tilde}{A} + \omega_1 ] - S [ \underaccent{\tilde}{A} + \omega_2 ]  \rvert \le \int_{-t_{12}}^0 \int_{\R^3} \left( \mathscr{F}'_1 \right)^2 \, d \mathbf{x}' dt = \int_{-t_{12}}^0 \int_{\R^3} \left( \mathscr{F}'_2 \right)^2 \, d \mathbf{x}' dt
\label{diffbound}
\end{equation}
As before, a computation of $\mathscr{F}'_j$ allows us to conclude using Sobolev estimates that $\int_{-t_{12}}^0 \int_{\R^3} \left( \mathscr{F}'_j \right)^2 \, d \mathbf{x}' dt \le Ct_{12} = C \| \omega_1 - \omega_2 \|_2$ (by choice of $t_{12}$), where the constant $C$ depends on $\|\omega_0\|_{2,1}$.  This concludes local Lipschitz continuity of $\omega \mapsto S \left[  \underaccent{\tilde}{A} + \omega \right]$, and together with Theorem 6 in \cite{Phelps:1978} yields
\begin{theorem} \label{Gateaux} Let $\underaccent{\tilde}{A}$ be a reference connection in $L^{2}_{1;loc} (\tzero , \Ad \eta \otimes T^*(\tzero))$ such that $\underaccent{\tilde}{F} \equiv F_{\underaccent{\tilde}{A}}$ satisfies $\int_{\tzero} \underaccent{\tilde}{F} ^2 \, d \mathbf{x'} < \infty$ and such that there exists an extension $\underaccent{\tilde}{\mathscr{A}} \in L^{2}_{1;loc} (\st , \Ad \eta \otimes T^*(\st))$ of $\underaccent{\tilde}{A}$ into $\st$ with finite Euclidean signature Yang-Mills action. The functional $\omega \mapsto S \left[ \underaccent{\tilde}{A} + \omega \right]$ on $L^{2}_1 (\tzero , \Ad \eta \otimes T^*(\tzero))$ is G\^ateaux differentiable outside of a Gaussian null set.
\end{theorem}
For further details, please see \cite{Maitra:inprep}.

\noindent\textit{\textbf{{Potential application of the implicit function theorem to smoothness of $S$.}}}  As in the case of polynomial field theories discussed in the preceding section, one wishes to use the Banach space version of the implicit function theorem to conclude Fr\'echet differentiability of $S\left[ \cdot \right]$ to all orders.  However as discussed in Section \ref{subsec:phi} regarding smoothness of the Hamilton-Jacobi functional, the gradient of $S[\cdot]$ (where it exists) should be given by the complementary momentum to the initial data for the minimizing trajectory on which $\Ies$ is evaluated.  Thus non-uniqueness of the absolute minimizer for a given boundary data $\A$ would be an obstruction to G\^ateaux differentiability of $S[\cdot]$ at $\A$.  Results on nonunique minimizers for the Euclidean signature Yang-Mills action on the unit ball $B^4 = \left\{ x \in \R^4 \ : \ \lvert x \rvert \le 1 \right\}$ are derived in \cite{Isobe:1997}; in particular, these authors observe that, having identified $B^4$ with the southern hemisphere $S^-$ of $S^4$, if one prescribes as boundary data a connection $\A$ which is the pull-back to the equator of an anti-self dual connection $\mathscr{I}_-$ on $S^4$, then one can at the same time regard $\A$ as boundary data of $\mathscr{I}_-$  restricted to $S^-\simeq B^4$, and as boundary data of the connection obtained by restricting $\mathscr{I}_-$ to the northern hemisphere and reflecting across the equator.  While the former is anti-self dual, the latter is self dual, as reflection reverses orientation. 

The fact that on $\st$ the functional $S\left[ \cdot \right]$ is G\^ateaux differentiable outside of a Gaussian null set (Theorem \ref{Gateaux}) suggests the prospect that an implicit function theorem argument may be applicable almost everywhere (in a suitable sense) on the set of boundary data.  Accordingly, we define a map which evaluates the Euclidean signature Yang-Mills operator on varying extensions of given boundary data, and we compute its linearization with respect to the variation of the extension.  In order to define the map, denote by $\bb^\ast$ the set of boundary data in (local) Hodge gauge, $\bb^\ast \equiv \left\{ A \in \bb \ : \ d^\ast A = 0 \right\}$.  Note that by the existence of a Hodge-gauge representative for connections with sufficient $L^{n/2}$ bounds on curvature \cite{Uhlenbeck:1982a}, $\bb^\ast$ includes a representative for each set of boundary data in $\bb$.  Let $\stA^L$ be the solution to the linearized Yang-Mills Dirichlet problem for boundary data $A \in \bb^\ast$:
\begin{equation} \label{LinDirichlet}
\left\{ \begin{array}{l} d^\ast d \stA^L = 0 \\
d^\ast \stA^L = 0 \\
i^\ast \stA^L = \A \\
d_\tau^\ast \left( \stA^L \right)_\tau =0 \, ,
\end{array} \right.
\end{equation}
where the last equality is satisfied because $A \in \bb^\ast$.  Arguments from Hodge theory imply that the solution to \eqref{LinDirichlet} is unique and that the map $\A \mapsto \stA^L$, $A \in \bb^\ast$ is smooth. 
Thus we define the smooth map
\begin{align*}
\mathcal{E} : \bb^\ast \times \connLloc(0) &\to \connLloc(0)^\ast \\
(\A , h ) & \mapsto D^\ast_{\stA^L +h} \stF_{\stA^L +h}  \, ,
\end{align*}
where $\connLloc(0)^\ast$ is the dual of $\connLloc(0)$ and $D^\ast_{\stA^L +h} \stF_{\stA^L +h}$ is interpreted in a weak sense as $\langle \stF_{\stA^L + h} , \ D_{\stA^L + h} \ \cdot \ \rangle$.

Suppose that $\left( \A_0 , h_0 \right) \in \bb^\ast \times \connLloc(0)$ satisfies $\mathcal{E} \left( \A_0 , h_0 \right) = 0$, indicating that $\stA^L + h_0$ weakly satisfies the Euclidean signature Yang-Mills equation.  Then we compute the linearization of $\mathcal{E}$ at $\left( \A_0 , h_0 \right)$ with respect to the second variable $h$ as follows:
\begin{align*}
D_h \mathcal{E}\left( \A_0 , h_0 \right) : T_{h_0} \connLloc(0) &\to T_0 \left( \connLloc(0)^\ast \right) \\
\xi &\mapsto \lim_{t \to 0} \frac{\mathcal{E} (\A_0 , h_0 + t \xi) - \mathcal{E} (\A_0 , h_0) }{t} \\
&= \lim _{t \to 0} \frac{D^*_{\stA_0^L + h_0 + t \xi} \stF_{\stA_0^L + h_0 + t \xi}}{t} \\
&= D^\ast_{\stA_0} D_{\stA_0} \xi + \ast \left[ \xi , \ast \stF_0 \right] \, ,
\end{align*}
where in the last line we denote $\stA_0^L+h_0$ by $\stA_0$ and $\stF_{\stA_0^L + h_0}$ by $\stF_0$.

In order to apply the Banach space implicit function theorem and conclude (almost everywhere) smoothness of $S\left[ \cdot \right]$, we must show that except for $\A_0$ belonging to a set of measure zero in $\bb^\ast$ (with respect to a suitable measure), the kernel of the map $D_h \mathcal{E} \left( \A_0 , h_0 \right)$ is trivial (modulo gauge transformations).  To this end, we assume that the absolute minimizer of $\Ies[\cdot]$ in $\connLloc(\A_0)$ is unique up to gauge equivalence, removing the obstruction to differentiability presented by the existence of inequivalent absolute minimizers.  The set of boundary values for which inequivalent absolute minimizers exist should be of measure zero, thanks to almost-everywhere G\^ateaux differentiability of $S[\cdot]$.  The form of the kernel derived above bears clear similarity to terms occurring in the full covariant derivative Laplacian, and thus in ongoing work we seek to leverage results from elliptic regularity theory for covariant differential operators to conclude almost-everywhere triviality of the kernel.

\noindent\textit{\textbf{Future work.}}  To complete the application of our program to Yang-Mills theory we must consider, after resolving the issue of almost-everywhere Fr\'echet differentiability of the Hamilton-Jacobi functional $S[\cdot]$, the construction of higher-order quantum corrections thereto, as modeled in the quantum mechanical case discussed in Section \ref{sec:microlocal}.  For field theories such as Yang-Mills as well as the polynomial field theories considered in Section \ref{subsec:phi}, this will require regularization of the relevant functional Laplacian operator.  Indeed, the higher-order quantum corrections $S_{(k)}[\cdot]$ must satisfy transport equations analogous to the quantum mechanical versions in Section \ref{sec:microlocal} and \cite{Moncrief:2012}, each involving a Laplacian of the next lower order correction $S_{(k-1)}$.

As mentioned in Section \ref{subsec:b-e} below, such a process of regularization is expected to have the favorable side effect of introducing a length scale into quantum Yang-Mills theory in 4 dimensions.   Without a length scale, 4-dimensional Yang-Mills theory does not admit the construction of a mass gap from the constants at hand (namely Planck's constant, the speed of light and the Yang-Mills coupling constant), seeming to defeat at the outset any hope of fulfilling this anticipated hallmark of a quantized Yang-Mills theory.  In the following section we discuss in detail our proposal for the application of bounds involving the `Bakry-Emery Ricci tensor' to this important issue.



\subsection{An intended application to the mass gap problem}
\label{subsec:b-e}

	A fundamental question in quantum gauge theory is whether the Schr\"{o}dinger operator for certain non-abelian Yang-Mills fields admits a spectral gap. Such a gap, if it exists, could represent the energy difference between the actual vacuum state and that of the lowest energy, excited `glueball' states and confirm the expectation that massless gluons (the fundamental quanta of Yang-Mills dynamics) cannot propagate freely as photons do but must instead exhibit a form of `color confinement'. It seems to be well understood that this question lies beyond the scope of conventional perturbation theory and will require a more global analytical treatment for its ultimate resolution.

	Many years ago I. M. Singer proposed an elegant, geometrical approach to this fundamental problem based on the fact that the classical, reduced configuration space for (Lorentzian signature) Yang-Mills dynamics --- namely the `orbit space' of spatial connections modulo gauge transformations --- has a naturally induced, \textit{curved} Riemannian metric with everywhere non-negative sectional curvature \cite{Singer:1981}. The classical Hamiltonian for the reduced dynamics --- a real-valued functional defined on the cotangent bundle of this orbit space --- consists of a `kinetic' term induced from the spatial integral of the square of the vectorial electric component of the full, spacetime Yang-Mills curvature tensor and a `potential' term induced from the spatial integral of the square of its complementary, vectorial magnetic component. The non-vanishing curvature of the Riemannian metric defined by the kinetic term arises from the implementation of the Gauss-law constraint during the process of reduction to the quotient, orbit space and was independently computed by several investigators \cite{Singer:1981,Babelon:1981,Vergeles:1983}. The classical reduced dynamics is thus that for a system point (namely a gauge equivalence class of spatial connections) moving on a positively curved, infinite dimensional manifold under the influence of a (non-negative) potential energy.

	Upon canonical quantization the Schr\"{o}dinger operator for this (pure Yang-Mills) dynamical system will thus include a kinetic term that, formally at least, encompasses the (negative\footnote{We here adopt the usual physicists' sign convention for the definition of a Laplacian.}) Laplace-Beltrami operator for an infinite-dimensional, curved Riemannian manifold --- namely the orbit space alluded to above. Whereas the (covariant) Hessian of sufficiently smooth (wave) functionals can be rigorously defined in such infinite-dimensional contexts, its associated trace need not make sense without some suitable regularization since the Hessian will not, in general, be trace class. Singer, in particular proposed an elegant zeta-function regularization scheme to define the needed Laplacian \cite{Singer:1981}.

	A classical result in Riemannian geometry due to A. Lichnerowicz \cite{Lichnerowicz:1958} shows that the Laplace operator for a complete, connected (finite-dimensional) Riemannian manifold necessarily exhibits a spectral gap provided that the Ricci tensor of this manifold is bounded, positively, away from zero\footnote{It follows from the Bonnet-Myers theorem that such a manifold is necessarily compact \cite{Myers:1941}.}. Such a result however cannot be expected to extend in any straightforward way at least, to the infinite-dimensional manifolds arising in quantum Yang-Mills theory. First of all, as Singer pointed out, their Ricci tensors, which would result from taking traces of corresponding (rigorously computable) curvature tensors, are not in general well-defined --- the curvature tensors in question not being trace-class --- and would require a suitable regularization for their meaningful formulation. Again Singer proposed a zeta function regularization scheme as an elegant means of accomplishing this. Some such regularization, however, is actually a desirable feature of the quantization procedure, at least in 4 spacetime dimensions, since it allows the introduction of a length scale into the quantum formalism. In the absence of such a scale no hypothetical spectral energy gap could even be expressed in terms of the naturally occurring parameters of the theory (Planck's constant, the speed of light and the Yang-Mills coupling constant).

	Another difficulty with attempting to extend the Lichnerowicz argument to the infinite-dimensional setting of interest here is that, thanks to the Bonnet-Myers theorem, one knows that a complete, finite dimensional Riemannian manifolds with positive Ricci curvature bounded away from zero is necessarily compact \cite{Myers:1941}. For a connected such manifold the lowest eigenvalue of its associated (negative) Laplacian always vanishes and corresponds to a globally constant eigenfunction. That such an eigenfunction is nevertheless always normalizable follows from the manifold's compactness. The spectral gap referred to in Lichnerowicz's theorem is thus simply the lowest non-vanishing eigenvalue of the manifold's (negative) Laplacian which, in view of compactness, necessarily has a discrete spectrum.

	Generalizations of Lichnerowicz's theorem have been established under less stringent conditions on the Ricci tensors provided that the manifolds under study have finite diameters \cite{Ling:2006,Yang:1999}. L. Andersson has proven that Riemannian Hilbert manifolds have finite diameters whenever their full sectional curvatures are positively bounded away from zero \cite{Andersson:1986} but this result does not apply to the orbit space sectional curvatures of interest here since these latter admit (infinite dimensional) families of 2-planes on which they actually vanish. In any case the diameters of these Yang-Mills orbit spaces are known to be infinite \cite{orbit-misc01}.

	The true, normalizable ground state wave functional must necessarily reflect the presence of the potential energy term in the Schr\"{o}dinger operator. In a recent paper \cite{Moncrief:inprep} the authors showed how to modify the original Lichnerowicz argument (in a finite dimensional setting) to allow for the occurrence of such a potential energy term and show that a corresponding gap estimate follows therefrom provided that a suitably defined `Bakry-Emery Ricci tensor' is bounded positively away from zero. This Bakry-Emery Ricci tensor differs from the actual Ricci tensor by a term in the (covariant) Hessian of the logarithm of the true ground state wave function. Its positivity could hold on a flat or even negatively curved space and thus its applicability is not limited to manifolds of finite diameter.

	Furthermore the natural integration measure arising in this (generalized Lichnerowicz) analysis includes the squared modulus of the ground state wave function itself so that the total space, even if it has infinite diameter, now has finite measure simply by virtue of the normalizability of the vacuum state. This should prove to be especially significant for any potential extensions to infinite dimensional problems wherein formal Lebesgue measures no longer make sense but for which normalizable vacuum state wave functional are nevertheless expected to exist.

	Our Euclidean signature semi-classical program, when applied to Yang-Mills fields, has the significant advantage over conventional, Rayleigh- Schr\"{o}dinger perturbation theory of keeping the non-linearities and \textit{non-abelian} gauge invariances fully intact at every level of the analysis. Our expectation is that it should yield an asymptotic expansion for the needed, fully gauge invariant, logarithm of the ground state wave functional that is far superior to any attainable by conventional perturbation methods. The latter, by requiring an expansion in the Yang-Mills coupling constant, disturb both the nonlinear structure and the closely associated (non-abelian) gauge invariance of the Yang-Mills dynamical system at the outset and attempt to reinstate those vital features only gradually, order-by-order in the expansion.

	Though our main focus in Ref. \cite{Moncrief:inprep} was on the Yang-Mills system we proved therein that (non-vanishing) orbit space curvature also arises naturally through the (minimal) coupling of a Maxwell field to a charged scalar field. In this case curvature arises only for the scalar factor of the (product) orbit space and not for the Maxwell factor which remains flat. We were thus led to conjecture that orbit space curvature could even serve as an independent source of mass for matter fields themselves provided that they are (minimally) coupled to (abelian or non-abelian) gauge fields. In the \textit{standard model} of elementary particle physics such masses arise exclusively (in view of the necessity of maintaining gauge invariance) through the interaction of matter fields with the Higgs field. It is intriguing to speculate whether orbit space curvature effects could supply an alternative source of such masses.

	Variation of the action functional for Yang-Mills fields on Minkowski space with respect to the (Lorentz frame) time component of the spacetime connection field yields the so-called Gauss-law constraint equation which, at each fixed time t, may be viewed as an elliptic equation on \(\mathbb{R}^3\) for this time component --- a Lie-algebra valued function. If, with suitable boundary conditions imposed, one solves this constraint and substitutes the solution back into the action, the resulting reduced kinetic term (a quadratic form in the `velocity' of the spatial connection) is found to be degenerate along gauge orbit directions but smooth, gauge invariant and positive definite in the transversal directions \cite{Singer:1981,Babelon:1981,Vergeles:1983}. It thus follows that this kinetic term defines a smooth, Riemannian metric on the natural `orbit space' of spatial connections modulo gauge transformations. This orbit space is (at least almost everywhere) itself a smooth, infinite dimensional manifold and provides the geometrically natural (reduced) configuration space for (classical) Yang-Mills dynamics.

	A corresponding smooth \textit{potential energy} function is induced on this orbit space by the integral over \(\mathbb{R}^3\) (at fixed \textit{t}) of the square of the curvature of the spatial connection field --- the `magnetic' component of the curvature of the full spacetime connection field. A Legendre transformation leads in turn to the Hamiltonian functional for the classical dynamics which takes the `standard' form of a sum of (curved space) kinetic and potential energies defined on the cotangent bundle of the aforementioned orbit space.

	The sectional curvature of this reduced configuration space was independently computed in \cite{Singer:1981,Babelon:1981,Vergeles:1983} and shown to be everywhere non-negative but almost everywhere non-vanishing whenever the gauge group is non-abelian. Though Singer discussed the need for a suitable regularization scheme to make sense of the formally (positively) divergent Ricci tensor of the orbit space metric, the actual form of such a regulated Ricci tensor seems still to be unknown. It would be most interesting if a suitably defined Ricci tensor could be shown to be bounded, positively away from zero on this orbit space, especially inasmuch as we think it quite unlikely that our proposed Bakry-Emery `enhancement' of this tensor would nullify its (hypothetical) positivity properties but perhaps, more likely, complement them\footnote{This would be true for example if the relevant logarithm were (almost everywhere) convex.}. Furthermore, as we discussed near the end of Section II of Ref. \cite{Moncrief:inprep}, it is quite plausible that strict positivity of the Bakry-Emery Ricci tensor, though seemingly sufficient for the implication of a spectral gap, is not absolutely necessary for this conclusion to hold.

\subsection{Euclidean-Signature Asymptotic Methods and the Wheeler-DeWitt Equation}
\label{subsec:w-dw}
Globally hyperbolic spacetimes, \(\lbrace{}^{(4)}V,{}^{(4)}g\rbrace\), are definable over manifolds with the product structure, \({}^{(4)}V \approx M \times \mathbb{R}\). We shall focus here on the `cosmological' case for which the spatial factor \textit{M} is a compact, connected, orientable 3-manifold without boundary. The Lorentzian metric, \({}^{(4)}g\), of such a spacetime is expressible, relative to a time function \(x^0 = t\), in the 3+1-dimensional form
\begin{equation}\label{eq2155}
\begin{split}
{}^{(4)}g &= {}^{(4)}g_{\mu\nu}\; dx^\mu \otimes dx^\nu\\
 &= -N^2 dt \otimes dt + \gamma_{ij} (dx^i + Y^idt) \otimes (dx^j + Y^jdt)
\end{split}
\end{equation}
wherein, for each fixed \textit{t}, the Riemannian metric
\begin{equation}\label{eq2156}
\gamma = \gamma_{ij} dx^i \otimes dx^j
\end{equation}
is the first fundamental form induced by \({}^{(4)}g\) on the corresponding \(t = \mathrm{constant}\), spacelike hypersurface. The unit, future pointing, timelike normal field to the chosen slicing (defined by the level surfaces of \textit{t}) is expressible in terms of the (strictly positive) `lapse' function \textit{N} and `shift vector' field \(Y^i \frac{\partial}{\partial x^i}\) as
\begin{equation}\label{eq2157}
{}^{(4)}n = {}^{(4)}n^\alpha \frac{\partial}{\partial x^\alpha} = \frac{1}{N} \frac{\partial}{\partial t} - \frac{Y^i}{N} \frac{\partial}{\partial x^i}
\end{equation}
or, in covariant form, as
\begin{equation}\label{eq2158}
{}^{(4)}n = {}^{(4)}n_\alpha dx^\alpha = -N\; dt.
\end{equation}
The canonical spacetime volume element of \({}^{(4)}g,\; \mu_{{}^{(4)}g} := \sqrt{-\det{{}^{(4)}g}}\), takes the 3+1-dimensional form
\begin{equation}\label{eq2159}
\mu_{{}^{(4)}g} = N\mu_\gamma
\end{equation}
where \(\mu_\gamma := \sqrt{\det{\gamma}}\) is the volume element of \(\gamma\).

In view of the compactness of \textit{M} the Hilbert and ADM action functionals, evaluated on domains of the product form, \(\Omega = M \times I\), with \(I = \lbrack t_0,t_1\rbrack \subset \mathbb{R}\), simplify somewhat to
\begin{equation}\label{eq2160}
\begin{split}
I_{\mathrm{Hilbert}} &:= \frac{c^3}{16\pi G}\; \int_\Omega \sqrt{-\det{{}^{(4)}g}}\; {}^{(4)}R({}^{(4)}g)\; d^4x\\
 &\hphantom{:}= \frac{c^3}{16\pi G}\; \int_\Omega \left\lbrace N\mu_\gamma \left( K^{ij} K_{ij} - (tr_\gamma K)^2\right) + N\mu_\gamma {}^{(3)}R(\gamma)\right\rbrace d^4x\\
 &\hphantom{:=} + \frac{c^3}{16\pi G}\; \int_M \left(-2\mu_\gamma tr_\gamma K\right) d^3x\Big|_{t_0}^{t_1}\\
 &:= I_{\mathrm{ADM}} + \frac{c^3}{16\pi G}\; \int_M \left(-2\mu_\gamma tr_\gamma K\right) d^3x\Big|_{t_0}^{t_1}
\end{split}
\end{equation}
wherein \({}^{(4)}R ({}^{(4)}g)\) and \({}^{(3)}R(\gamma)\) are the scalar curvatures of \({}^{(4)}g\) and \(\gamma\) and where
\begin{equation}\label{eq2161}
K_{ij} := \frac{1}{2N} \left(-\gamma_{ij,t} + Y_{i|j} + Y_{j|i}\right)
\end{equation}
and
\begin{equation}\label{eq2162}
tr_\gamma K := \gamma^{ij}K_{ij}
\end{equation}
designate the second fundamental form and mean curvature induced by \({}^{(4)}g\) on the constant \textit{t} slices. In these formulas spatial coordinate indices, \(i,j,k, \ldots,\) are raised and lowered with \(\gamma\) and the vertical bar, `\(|\)', signifies covariant differentiation with respect to this metric so that, for example, \(Y_{i|j} = \nabla_j (\gamma) \gamma_{i\ell} Y^\ell\). When the variations of \({}^{(4)}g\) are appropriately restricted, the boundary term distinguishing \(I_{\mathrm{Hilbert}}\) from \(I_{\mathrm{ADM}}\) makes no contribution to the field equations and so can be discarded.

Writing
\begin{equation}\label{eq2163}
I_{\mathrm{ADM}} := \int_\Omega\; \mathcal{L}_{\mathrm{ADM}} d^4x,
\end{equation}
with Lagrangian density
\begin{equation}\label{eq2164}
\mathcal{L}_{\mathrm{ADM}} := \frac{c^3}{16\pi G}\; \left\lbrace N\mu_\gamma \left( K^{ij} K_{ij} - (tr_\gamma K)^2\right) + N\mu_\gamma {}^{(3)}R(\gamma)\right\rbrace ,
\end{equation}
one defines the \textit{momentum} conjugate to \(\gamma\) via the Legendre transformation
\begin{equation}\label{eq2165}
p^{ij} := \frac{\partial\mathcal{L}_{\mathrm{ADM}}}{\partial\gamma_{ij,t}} = \frac{c^3}{16\pi G}\; \mu_\gamma \left(-K^{ij} + \gamma^{ij} tr_\gamma K\right)
\end{equation}
so that \(p = p^{ij} \frac{\partial}{\partial x^i} \otimes \frac{\partial}{\partial x^j}\) is a symmetric tensor density induced on each \(t = \mathrm{constant}\) slice.

In terms of the variables \(\lbrace\gamma_{ij}, p^{ij}, N, Y^{i}\rbrace\) the ADM action takes the Hamiltonian form
\begin{equation}\label{eq2166}
I_{\mathrm{ADM}} = \int_\Omega \left\lbrace p^{ij}\gamma_{ij,t} - N \mathcal{H}_\perp (\gamma,p) - Y^i\mathcal{J}_i (\gamma,p)\right\rbrace d^4x
\end{equation}
where
\begin{equation}\label{eq2167}
\mathcal{H}_\perp (\gamma,p) := \left(\frac{16\pi G}{c^3}\right) \frac{\left(p^{ij}p_{ij} - \frac{1}{2}(p_m^m)^2\right)}{\mu_\gamma} - \left(\frac{c^3}{16\pi G}\right) \mu_\gamma\; {}^{(3)}R (\gamma)
\end{equation}
and
\begin{equation}\label{eq2168}
\mathcal{J}_i(\gamma,p) := -2\; p_{i\hphantom{j}|j}^{\hphantom{i}j}.
\end{equation}
Variation of \(I_{\mathrm{ADM}}\) with respect to \textit{N} and \(Y^i\) leads to the Einstein (`Hamiltonian' and `momentum') \textit{constraint equations}
\begin{equation}\label{eq2169}
\mathcal{H}_\perp (\gamma,p) = 0,\quad \mathcal{J}_i(\gamma,p) = 0,
\end{equation}
whereas variation with respect to the \textit{canonical variables}, \(\lbrace\gamma_{ij},p^{ij}\rbrace\), gives rise to the complementary Einstein \textit{evolution equations} in Hamiltonian form,
\begin{equation}\label{eq2170}
\gamma_{ij,t} = \frac{\delta H_{\mathrm{ADM}}}{\delta p^{ij}},\quad p^{ij}_{\hphantom{ij},t} = -\frac{\delta H_{\mathrm{ADM}}}{\delta\gamma_{ij}}
\end{equation}
where \(H_{\mathrm{ADM}}\) is the `super' Hamiltonian defined by
\begin{equation}\label{eq2171}
H_{\mathrm{ADM}} := \int_M \left( N\mathcal{H}_\perp (\gamma,p) + Y^i\mathcal{J}_i(\gamma,p)\right) d^3x.
\end{equation}
The first of equations (\ref{eq2170}) regenerates (\ref{eq2161}) when the latter is reexpressed in terms of \textit{p} via (\ref{eq2165}). Note that, as a linear form in the constraints, the super Hamiltonian vanishes when evaluated on any solution to the field equations. There are neither constraints nor evolution equations for the lapse and shift fields which are only determined upon making, either explicitly or implicitly, a choice of spacetime coordinate \textit{gauge}. Bianchi identities function to ensure that the constraints are preserved by the evolution equations and thus need only be imposed `initially' on an arbitrary Cauchy hypersurface. Well-posedness theorems for the corresponding Cauchy problem exist for a variety of spacetime gauge conditions \cite{Choquet-Bruhat:2009,Andersson:2003}.

A formal `canonical' quantization of this system begins with the substitutions
\begin{equation}\label{eq2172}
p^{ij} \longrightarrow \frac{\hbar}{i} \frac{\delta}{\delta\gamma_{ij}},
\end{equation}
together with a choice of operator ordering, to define quantum analogues \(\hat{\mathcal{H}}_\perp (\gamma,\frac{\hbar}{i} \frac{\delta}{\delta\gamma})\) and \(\hat{\mathcal{J}}_i(\gamma,\frac{\hbar}{i}\frac{\delta}{\delta\gamma})\) of the Hamiltonian and momentum constraints. These are then to be imposed, \`{a} la Dirac, as restrictions upon the allowed quantum states, regarded as functionals, \(\Psi [\gamma]\), of the spatial metric, by setting
\begin{equation}\label{eq2173}
\hat{\mathcal{H}}_\perp \left(\gamma,\frac{\hbar}{i}\frac{\delta}{\delta\gamma}\right) \Psi [\gamma] = 0,
\end{equation}
and
\begin{equation}\label{eq2174}
\hat{\mathcal{J}}_i \left(\gamma,\frac{\hbar}{i}\frac{\delta}{\delta\gamma}\right) \Psi [\gamma] = 0.
\end{equation}
The choice of ordering in the definition of the quantum constraints \(\lbrace\hat{\mathcal{H}}_\perp,\hat{\mathcal{J}}_i\rbrace\) is highly restricted by the demand that the \textit{commutators} of these operators should `close' in a natural way without generating `anomalous' new constraints upon the quantum states.

While a complete solution to this \textit{ordering problem} does not currently seem to be known it has long been realized that the operator, \(\hat{\mathcal{J}}_i (\gamma,\frac{\hbar}{i}\frac{\delta}{\delta\gamma})\), can be consistently defined so that the quantum constraint equation (\ref{eq2174}), has the natural geometric interpretation of demanding that the wave functional, \(\Psi [\gamma]\), be invariant with respect to the action (by pullback of metrics on \textit{M}) of \(\mathcal{D}\mathit{iff}^0(M)\), the connected component of the identity of the group, \(\mathcal{D}\mathit{iff}^+(M)\), of orientation preserving diffeomorphisms of \textit{M}, on the space, \(\mathcal{M}(M)\), of Riemannian metrics on \textit{M}. In other words the quantized momentum constraint (\ref{eq2174}) implies, precisely, that
\begin{equation}\label{eq2176}
\Psi [\varphi^*\gamma] = \Psi [\gamma]
\end{equation}
\(\forall\; \varphi \in \mathcal{D}\mathit{iff}^0(M)\) and \(\forall\; \gamma \in \mathcal{M}(M)\). In terminology due to Wheeler wave functionals can thus be regarded as passing naturally to the quotient `superspace' of Riemannian \textit{3-geometries} \cite{Fischer:1970,Giulini:2009,misc:08} on \textit{M},
\begin{equation}\label{eq2177}
\boldsymbol{\mathbb{S}}(M) := \frac{\mathcal{M}(M)}{\mathcal{D}\mathit{iff}^0(M)}.
\end{equation}

Insofar as a consistent factor ordering for the Hamiltonian constraint operator, \(\hat{\mathcal{H}}_\perp (\gamma,\frac{\hbar}{i}\frac{\delta}{\delta\gamma})\), also exists, one will be motivated to propose the (Euclidean-signature, semi-classical) ansatz
\begin{equation}\label{eq2178}
\overset{(0)}{\Psi}_{\!\hbar} [\gamma] = e^{-S_\hbar [\gamma]/\hbar}
\end{equation}
for a `ground state' wave functional \(\overset{(0)}{\Psi}_{\!\hbar} [\gamma]\). In parallel with our earlier examples, the functional \(S_\hbar [\gamma]\) is assumed to admit a formal expansion in powers of \(\hbar\) so that one has
\begin{equation}\label{eq2179}
S_\hbar [\gamma] = S_{(0)} [\gamma] + \hbar S_1 [\gamma] + \frac{\hbar^2}{2!} S_{(2)} [\gamma] + \cdots + \frac{\hbar^k}{k!} S_{(k)} [\gamma] + \cdots .
\end{equation}
Imposing the momentum constraint (\ref{eq2174}) to all orders in \(\hbar\) leads to the conclusion that each of the functionals, \(\lbrace S_{(k)} [\gamma]; \; k = 0, 1, 2, \ldots\rbrace\), should be invariant with respect to the aforementioned action of \(\mathcal{D}\mathit{iff}^0(M)\) on \(\mathcal{M}(M)\), ie, that
\begin{equation}\label{eq2180}
S_{(k)} [\varphi^*\gamma] = S_{(k)} [\gamma],\; k = 0, 1, 2, \ldots
\end{equation}
\(\forall\; \varphi \in \mathcal{D}\mathit{iff}^0(M)\) and \(\forall\; \gamma \in \mathcal{M}(M)\).

Independently of the precise form finally chosen for \(\hat{\mathcal{H}}_\perp (\gamma,\frac{\hbar}{i}\frac{\delta}{\delta\gamma})\), the leading order approximation to the \textit{Wheeler-DeWitt equation},
\begin{equation}\label{eq2181}
\hat{\mathcal{H}}_\perp \left(\gamma,\frac{\hbar}{i}\frac{\delta}{\delta\gamma}\right) e^{-S_{(0)} [\gamma]/\hbar - S_{(1)}[\gamma] - \cdots} = 0,
\end{equation}
for the ground state wave functional will, inevitably reduce to the Euclidean-signature Hamilton-Jacobi equation
\begin{equation}\label{eq2182}
\left(\frac{16\pi G}{c^3}\right)^2 \frac{\left(\gamma_{ik}\gamma_{j\ell} - \frac{1}{2} \gamma_{ij} \gamma_{k\ell}\right)}{\mu_\gamma} \frac{\delta S_{(0)}}{\delta\gamma_{ij}} \frac{\delta S_{(0)}}{\delta\gamma_{k\ell}} + \mu_\gamma {}^{(3)}R(\gamma) = 0.
\end{equation}
This equation coincides with that obtained from making the canonical substitution,
\begin{equation}\label{eq2183}
p^{ij} \longrightarrow \frac{\delta S_{(0)} [\gamma]}{\delta\gamma_{ij}},
\end{equation}
in the Euclidean-signature version of the Hamiltonian constraint,
\begin{equation}\label{eq2184}
\mathcal{H}_{\perp\mathrm{Eucl}} := -\left(\frac{16\pi G}{c^3}\right)\; \frac{\left(p^{ij} p_{ij} - \frac{1}{2} (p_m^m)^2\right)}{\mu_\gamma} - \left(\frac{c^3}{16\pi G}\right)\; \mu_\gamma\; {}^{(3)}R(\gamma) = 0,
\end{equation}
that, in turn, results from repeating the derivation sketched above for \(I_{\mathrm{ADM}}\) but now for the Riemannian metric form
\begin{equation}\label{eq2185}
{}^{(4)}g\Big|_{\mathrm{Eucl}} = {}^{(4)}g_{\mu\nu}\Big|_{\mathrm{Eucl}} dx^\mu \otimes dx^\nu = N\Big|_{\mathrm{Eucl}}^2 dt \otimes dt + \gamma_{ij} (dx^i + Y^i dt) \otimes (dx^j + Y^j dt)
\end{equation}
in place of (\ref{eq2155}). The resulting functional \(I_{\mathrm{ADM}\; \mathrm{Eucl}}\) differs from \(I_{\mathrm{ADM}}\) only in the replacements \(\mathcal{H}_\perp (\gamma,p) \longrightarrow \mathcal{H}_{\perp\mathrm{Eucl}} (\gamma,p)\) and \(N \longrightarrow N\Big|_{\mathrm{Eucl}}\).

The essential question that now comes to light is thus the following:
\medskip

\begin{quote}
\textit{Is there a well-defined mathematical method for establishing the existence of a \(\mathcal{D}\mathit{iff}^0(M)\)-invariant, fundamental solution to the Euclidean-signature functional differential Hamilton-Jacobi equation (\ref{eq2182})?}
\end{quote}
\medskip

\noindent In view of the field theoretic examples discussed above one's first thought might be to seek to minimize an appropriate Euclidean-signature action functional subject to suitable boundary and asymptotic conditions. But, as is well-known from the Euclidean-signature path integral program \cite{Gibbons:1993}, the natural functional to use for this purpose is \textit{unbounded from below} within any given conformal class --- one can make the functional arbitrarily large and negative by deforming any metric \({}^{(4)}g\Big|_{\mathrm{Eucl}}\) with a suitable conformal factor \cite{Gibbons:1979,Gibbons:1993}.

But the real point of the constructions above was \textit{not} to minimize action functionals but rather to generate certain `fundamental sets' of solutions to the associated Euler-Lagrange equations upon which the relevant action functionals could then be evaluated. But the Einstein equations, in vacuum or even allowing for the coupling to conformally invariant matter sources, encompass, as a special case, the \textit{vanishing} of the 4-dimensional scalar curvature, \({}^{(4)}R({}^{(4)}g\Big|_{\mathrm{Eucl}})\). Thus there is no essential loss in generality, and indeed a partial simplification of the task at hand to be gained, by first restricting the relevant, Euclidean-signature action functional to the `manifold' of Riemannian metrics satisfying (in the vacuum case) \({}^{(4)}R({}^{(4)}g\Big|_{\mathrm{Eucl}}) = 0\) and then seeking to carry out a \textit{constrained minimization} of this functional.

Setting \({}^{(4)}R({}^{(4)}g\Big|_{\mathrm{Eucl}}) = 0\) freezes out the conformal degree of freedom that caused such consternation for the Euclidean path integral program \cite{Gibbons:1979,Gibbons:1993}, wherein one felt obligated to integrate over \textit{all possible} Riemannian metrics having the prescribed boundary behavior, but is perfectly natural in the present context and opens the door to appealing to the \textit{positive action theorem} which asserts that the relevant functional is indeed positive when evaluated on arbitrary, asymptotically Euclidean metrics that satisfy \({}^{(4)}R({}^{(4)}g\Big|_{\mathrm{Eucl}}) \geq 0\) \cite{Schoen:1979,Schoen:1979b,Zhang:1999,Dahl:1997}.

Another complication of the Euclidean path integral program was the apparent necessity to invert, by some still obscure means, something in the nature of a `Wick rotation' that had presumably been exploited to justify integrating over Riemannian, as opposed to Lorentzian-signature, metrics. Without this last step the formal `propagator' being constructed would presumably be that for the Euclidean-signature variant of the Wheeler-DeWitt equation and not the actual Lorentzian-signature version that one wishes to solve. In ordinary quantum mechanics the corresponding, well-understood step is needed to convert the Feynman-Kac propagator, derivable by rigorous path-integral methods, back to one for the actual Schr\"{o}dinger equation.

But in the present setting no such hypothetical `Wick rotation' would ever have been performed in the first place so there is none to invert. Our focus throughout is on constructing asymptotic solutions to the original, Lorentz-signature Wheeler-DeWitt equation and not to its Euclidean-signature counterpart. That a Euclidean-signature Einstein-Hamilton-Jacobi equation emerges in this approach has the very distinct advantage of leading one to specific problems in Riemannian geometry that may well be resolvable by established mathematical methods. By contrast, path integral methods, even for the significantly more accessible gauge theories discussed above, would seem to require innovative new advances in measure theory for their rigorous implementation. Even the simpler scalar field theories, when formulated in the most interesting case of four spacetime dimensions, seem still to defy realization by path integral means. It is conceivable, as was suggested in the concluding section of \cite{Moncrief:2012}, that focusing predominantly on path integral methods to provide a `royal road' to quantization  may, inadvertently, render some problems more difficult to solve rather than actually facilitating their resolution.

The well-known `instanton' solutions to the Euclidean-signature Yang-Mills equations present a certain complication for the semi-classical program that we are advocating in that they allow one to establish the existence of \textit{non-unique minimizers} for the Yang-Mills action functional for certain special choices of boundary data \cite{Maitra:inprep}. This in turn can obstruct the global smoothness of the corresponding solution to the Euclidean-signature Hamilton-Jacobi equation. While it is conceivable that the resulting, apparent need to repair the associated `scars' in the semi-classical wave functionals may have non-perturbative implications for the Yang-Mills energy spectrum --- of potential relevance to the `mass-gap' problem --- no such corrections to the spectrum are expected or desired for the gravitational case. Thus it is reassuring to note that analogous `gravitational instanton' solutions to the Euclidean-signature Einstein equations have been proven \textit{not to exist} \cite{Gibbons:1979}.

We conclude by noting that other interesting, generally covariant systems of field equations exist to which our (`Euclidean-signature semi-classical') quantization methods could also be applied. Classical relativistic `membranes', for example, can be viewed as the evolutions of certain embedded submanifolds in an ambient spacetime --- their field equations determined by variation of the volume functional of the timelike `worldsheets' being thereby swept out. The corresponding Hamiltonian configuration space for such a system is comprised of the set of spacelike embeddings of a fixed \(n - 1\) dimensional manifold \textit{M} into the ambient \(n + k\) dimensional spacetime, each embedding representing a possible spacelike slice through some \textit{n}-dimensional membrane worldsheet. Upon canonical quantization wave functionals are constrained (by the associated, quantized momentum constraint equation) to be invariant with respect to the induced action of \(\mathcal{D}\mathit{iff}^0(M)\) on this configuration space of embeddings. The corresponding quantized Hamiltonian constraint, imposed \`{a} la Dirac, provides the natural analogue of the Wheeler-DeWitt equation for this problem.

A solution to the operator ordering problem for these quantized constraints, when the ambient spacetime is Minkowskian, was proposed by one of us in \cite{Moncrief:2006}. For the compact, codimension one case (i.e., when \textit{M} is compact and \(k = 1\)) it is not difficult to show that the relevant \textit{Euclidean-signature} Hamilton-Jacobi equation has a fundamental solution given by the volume functional of the maximal, spacelike hypersurface that uniquely spans, \`{a} la Plateau, the arbitrarily chosen embedding \cite{Moncrief:unpub}. It would be especially interesting to see whether higher-order quantum corrections and excited state wave functionals can be computed for this system in a way that realizes a quantum analogue of general covariance.

\section*{Acknowledgements}

Marini is grateful to Karen Uhlenbeck for the refined techniques and methodologies in the field of elliptic partial differential equations and geometric analysis learned from her as a graduate student and throughout the years. Her inspired teachings, by providing a profound geometrical understanding of fundamental problems in functional analysis and partial differential equations, have promoted this author's deep appreciation for gauge theories, a field at the intersection of analysis, geometry, topology, and mathematical physics. Moncrief is grateful to the Swedish Royal Institute of Technology (KTH) for hospitality and support during a visit in March 2015 and especially to Lars Andersson for pointing out the relevance of mathematical literature on Bakry-Emery Ricci tensors to the research discussed herein. The authors are also grateful to the Albert Einstein Institute in Golm, Germany, the Institut des Hautes {\'E}tudes Scientifiques in Bures-sur-Yvette, France, the Erwin Schr{\"o}dinger Institute and the University of Vienna in Vienna, Austria for the hospitality and support extended to several of us during the course of this research.

The reader will not have failed to note the very strong sense in which our work builds on the fundamental contributions of Karen Uhlenbeck.

\bibliographystyle{unsrt} 
\bibliography{CAG_refs}   

\begin{thebibliography}{10}

\bibitem{Moncrief:2012}
V.~Moncrief, A.~Marini, and R.~Maitra.
\newblock Modified semi-classical methods for nonlinear quantum oscillations
  problems.
\newblock {\em J. Math. Phys.}, 53:103516--103567, 2012.

\bibitem{Dimassi:1999}
M.~Dimassi and J.~{Sj\"{o}strand}.
\newblock {\em Spectral Asymptotics in the Semi-Classical Limit}.
\newblock Cambridge University Press, 1999.
\newblock see especially Chapters 1--3 and related references cited therein.

\bibitem{Helfer:1988}
B.~Helfer.
\newblock {\em Semi-Classical Analysis for the {Schr\"{o}dinger} Operator and
  Applications}.
\newblock Springer-Verlag, Berlin Heidelberg, 1988.
\newblock For more advanced topics, see \cite{Helfer:1984}.

\bibitem{Helfer:1984}
B.~Helfer and J.~{Sj\"{o}strand}.
\newblock Multiple wells in the semi-classical limit {I}.
\newblock {\em Communications in Partial Differential Equations}, 9:337--408,
  1984.

\bibitem{Brack:2008}
M.~Brack and R.~K. Bhaduri.
\newblock {\em Semiclassical Physics (Frontiers in Physics)}.
\newblock Westview Press, 2008.

\bibitem{Ozorio:1988}
A.~M.~Ozorio de~Almeida.
\newblock {\em Hamitonian Systems: Chaos and Quantization}.
\newblock Cambridge Press, 1988.
\newblock MR0985103.

\bibitem{Stone:2005}
A.~D. Stone.
\newblock {Einstein's} unknown insight and the problem of quantizing chaos.
\newblock {\em Physics Today}, 58:37--43, 2005.

\bibitem{misc-orbit02}
These results were derived by {S}hilong ({P}eter) {T}ang during the course of
  completing a senior physics project at {Y}ale {U}niversity (unpublished,
  {May} 2016).

\bibitem{Maslov:1981}
V.~P. Maslov and M.~V. Fedoriuk.
\newblock {\em Semi-Classical Approximation in Quantum Mechanics}.
\newblock Reidel, 1981.

\bibitem{Abraham:1978}
R.~Abraham and J.~Marsden.
\newblock {\em Foundations of Mechanics}.
\newblock Benjamin/Cummings, 1979.
\newblock see Section 7.2.

\bibitem{Marini:2016}
A.~Marini, R.~Maitra, and V.~Moncrief.
\newblock Euclidean signature semi-classical methods for bosonic field
  theories: interacting scalar fields.
\newblock {\em Annals of Mathematical Sciences and Applications}, 1:3--55,
  2016.

\bibitem{Maitra:inprep}
R.~Maitra, A.~Marini, and V.~Moncrief.
\newblock Euclidean signature semi-classical methods for bosonic field
  theories: {Yang-Mills} fields (title tentative).
\newblock In preparation.

\bibitem{Moncrief:inprep2}
V.~Moncrief, A.~Marini, and R.~Maitra.
\newblock Microlocal semi-classical methods for higher-dimensional
  supersymmetric quantum mechanics (title tentative).
\newblock In preparation.

\bibitem{Moncrief:2015}
V.~Moncrief.
\newblock Euclidean-signature semi-classical methods for quantum cosmology.
\newblock In Lydia Bieri and Shing-Tung Yau, editors, {\em Surveys in
  {Differential} {Geometry} 2015: {One} hundred years of general relativity}.
  International Press, Boston, 2015.

\bibitem{Bae:2015}
J.~H. Bae.
\newblock Mixmaster revisited: wormhole solutions to the {Bianchi IX
  Wheeler-DeWitt} equation using the {Euclidean}-signature semi-classical
  method.
\newblock {\em Class. Quantum Grav.}, 32:075006, 2015.

\bibitem{GlimmJaffeI}
J.~Glimm and A.~Jaffe.
\newblock A $\lambda \phi^4$ quantum field theory without cutoffs. i.
\newblock {\em Physical Review}, 176(5):1945--1951, 1968.

\bibitem{GlimmJaffeII}
J.~Glimm and A.~Jaffe.
\newblock The $\lambda \phi^4_2$ quantum field theory without cutoffs: {II.
  The} field operators and the approximate vacuum.
\newblock {\em The Annals of Mathematics}, 91(2):362+, 1970.

\bibitem{GlimmJaffeIII}
J.~Glimm and A.~Jaffe.
\newblock The $\lambda \phi^4_2$ quantum field theory without cutoffs: {III.
  The} physical vacuum.
\newblock {\em Acta Mathematica}, 125(1):203--267, 1970.

\bibitem{GlimmJaffeIV}
J.~Glimm and A.~Jaffe.
\newblock The $\lambda \phi^4_2$ quantum field theory without cutoffs: {IV.
  Perturbations} of the {Hamiltonian}.
\newblock {\em J. Math. Phys.}, 13(10):1568--1584, 1972.

\bibitem{Feldman:1976}
J.S. Feldman and K.~Osterwalder.
\newblock The {Wightman} axioms and the mass gap for weakly coupled $\phi^4_3$
  quantum field theories.
\newblock {\em Annals of Physics}, 97(1):80--135, 1976.

\bibitem{Podolsky:2010}
D.~I. Podolsky.
\newblock On triviality of $\lambda\phi^4$ quantum field theory in four
  dimensions.
\newblock 2010.
\newblock hep-th/1003.3670.

\bibitem{Suslov:2008}
I.~M. Suslov.
\newblock Is $\phi^4$ theory trivial?
\newblock 2008.
\newblock hep-ph/0806.0789.

\bibitem{Blanchard1992}
P.~Blanchard and E.~Br\"{u}ning.
\newblock {\em Variational Methods in Mathematical Physics}.
\newblock Springer-Verlag, 1992.

\bibitem{Agmon}
S.~Agmon.
\newblock {\em Lectures on Elliptic Boundary Value Problems}, volume~2 of {\em
  Van Nostrand Mathematical Studies}.
\newblock D. Van Nostrand Company, Inc., Princeton, NJ, 2nd ed., rev. 3rd
  printing. edition, 1965.

\bibitem{Gilbarg-Trudinger}
D.~Gilbarg and N.~Trudinger.
\newblock {\em Elliptic Partial Differential Equations of Second Order}.
\newblock Springer Berlin ; New York, 2nd ed., rev. 3rd printing edition, 1998.

\bibitem{Marini:1992}
A.~Marini.
\newblock {Dirichlet} and {Neumann} boundary value problems for {Yang-Mills}
  connections.
\newblock {\em Commun. Pure Appl. Math.}, 45:1015--1050, 1992.

\bibitem{Chen-Li:2010}
W.~Chen and C.~Li.
\newblock {\em Methods on Nonlinear Elliptic Equations}, volume~4 of {\em AIMS
  Series on Differential Equations and Dynamical Systems}.
\newblock American Institute of Mathematical Sciences, 2010.

\bibitem{Adams}
R.~Adams.
\newblock {\em Sobolev Spaces}, volume~65 of {\em Pure and Applied Mathematics
  Series}.
\newblock Academic Press, New York, NY, 2nd ed., rev. 3rd printing edition,
  1975.

\bibitem{Marini:unpub}
A.~Marini.
\newblock A note on elliptic regularity lifting by the method of contracting
  operators. {Unpublished} notes.
\newblock 2013.

\bibitem{Klingenberg:1982}
W.~Klingenberg.
\newblock {\em Riemannian Geometry}.
\newblock Walter de Gruyter, 1982.

\bibitem{Singer:1981}
I.~M. Singer.
\newblock The geometry of the orbit space for non-abelian gauge theories.
\newblock {\em Phys. Scr.}, 24:817--820, 1981.

\bibitem{Babelon:1981}
O.~Babelon and C.M. Viallet.
\newblock The {Riemannian} geometry of the configuration space of gauge
  theories.
\newblock {\em Commun. Math. Phys.}, 81:515--525, 1981.

\bibitem{Maitra:2007}
R.~Maitra.
\newblock {\em Mathematically Rigorous Quantum Field Theories with a Nonlinear
  Normal Ordering of the {Hamiltonian} Operator}.
\newblock PhD thesis, Department of Mathematics, Yale University, 2007.
\newblock MR2711186.

\bibitem{Uhlenbeck:1982a}
K.~K. Uhlenbeck.
\newblock Connections with {L\(^p\)} bounds on curvature.
\newblock {\em Commun. Math. Phys.}, 83:31--42, 1982.

\bibitem{Uhlenbeck:1982b}
K.~K. Uhlenbeck.
\newblock Removable singularities in {Yang-Mills} fields.
\newblock {\em Commun. Math. Phys.}, 83:11--29, 1982.

\bibitem{Sedlacek:1982}
S.~Sedlacek.
\newblock A direct method for minimizing the {Yang-Mills} functional over
  4-manifolds.
\newblock {\em Commun. Math. Phys.}, 86:515--527, 1982.

\bibitem{Jackiw:1980}
R.~Jackiw.
\newblock Introduction to the {Yang-Mills} quantum theory.
\newblock {\em Reviews of Modern Physics}, 52:661--673, 1980.

\bibitem{Khoze:1994}
V.~Khoze.
\newblock The vacuum theta-angle is zero in non-abelian gauge theories.
\newblock {\em Physics Letters B}, 328:387--391, 1994.

\bibitem{Karabali:1998}
D.~Karabali, C.~Kim, and V.P. Nair.
\newblock Planar {Yang-Mills} theory: Hamiltonian, regulators and mass gap.
\newblock {\em Nuclear Physics B}, 524:661--694, 1998.

\bibitem{Fr-Uhl:1991}
D.~Freed and K.~K. Uhlenbeck.
\newblock {\em Instantons and Four-manifolds}.
\newblock Springer-Verlag, 1991.

\bibitem{Bredon:1972}
G.~Bredon.
\newblock {\em Introduction to Compact Transformation Groups}.
\newblock Academic Press, New York, 1972.

\bibitem{Wehrheim:2004}
K.~Wehrheim.
\newblock {\em {Uhlenbeck} compactness}.
\newblock European Mathematical Society Series of Lectures in Mathematics,
  Z\"urich, 2004.

\bibitem{Phelps:1978}
R.~Phelps.
\newblock Gaussian null sets and differentiability of {Lipschitz} map on
  {Banach} spaces.
\newblock {\em Pacific Journal of Mathematics}, 77:523--531, 1978.

\bibitem{Isobe:1997}
T.~Isobe and A.~Marini.
\newblock On topologically distinct solutions of the {Dirichlet} problem for
  {Yang-Mills} connections.
\newblock {\em Calc. Var. Partial Differential Equations}, 5:345--358, 1997.

\bibitem{Vergeles:1983}
S.~N. Vergeles.
\newblock Some properties of orbit space in {Yang-Mills} theory.
\newblock {\em Lett. Math. Phys.}, 7:399--406, 1983.

\bibitem{Lichnerowicz:1958}
A.~Lichnerowicz.
\newblock {\em {G\'eom\'etrie} des groupes de transformations}.
\newblock Dunod, Paris, 1958.
\newblock MR0124009 (23:A1329).

\bibitem{Myers:1941}
S.~B. Myers.
\newblock {Riemannian} manifolds with positive mean curvature.
\newblock {\em Duke Math. J.}, 8:401--404, 1941.

\bibitem{Ling:2006}
Jun Ling.
\newblock The first eigenvalue of a closed manifold with positive {Ricci}
  curvature.
\newblock {\em Proc. Am. Math. Soc.}, 134:3071--3079, 2006.

\bibitem{Yang:1999}
D.~Yang.
\newblock Lower bound estimates of the first eigenvalue for compact manifolds
  with positive {Ricci} curvature.
\newblock {\em Pac. J. Math.}, 190:383--398, 1999.

\bibitem{Andersson:1986}
L.~Andersson.
\newblock The {Bonnet-Myers} theorem is true for {Riemannian} {Hilbert}
  manifolds.
\newblock {\em Math. Scand.}, 58:236--238, 1986.

\bibitem{orbit-misc01}
Note in particular that geodesic curves of infinite proper length of purely
  `polarized' connections (i.e., ones for which the {Lie} brackets all vanish)
  exist along which the magnetic energy monotonically increases without bound.

\bibitem{Moncrief:inprep}
V.~Moncrief, A.~Marini, and R.~Maitra.
\newblock Orbit space curvature as a source of mass in quantum gauge theory.
\newblock arXiv:1809.06318 [hep-th], 2018.

\bibitem{Choquet-Bruhat:2009}
Y.~Choquet-Bruhat.
\newblock {\em General Relativity and the {Einstein} Equation}.
\newblock Oxford University Press, 2009.

\bibitem{Andersson:2003}
L.~Andersson and V.~Moncrief.
\newblock Elliptic-hyperbolic systems and the {Einstein} equations.
\newblock {\em Ann. Henri {Poincar\'{e}}}, 4:1--34, 2003.

\bibitem{Fischer:1970}
A.~E. Fischer.
\newblock The theory of superspace.
\newblock In M.~Carmeli, S.~Fickler, and L.~Witten, editors, {\em Relativity,
  Proceedings of the Relativity Conference in the Midwest}, New York, 1970.
  Plenum Press.

\bibitem{Giulini:2009}
D.~Giulini.
\newblock The superspace of geometrodynamics.
\newblock {\em Gen. Relativ. Gravit.}, 41:785--815, 2009.

\bibitem{misc:08}
c.f., chapter 43 of \cite{Misner:1973}.

\bibitem{Gibbons:1993}
G.~W. Gibbons and S.~W. Hawking.
\newblock {\em Euclidean Quantum Gravity}.
\newblock World Scientific, Singapore, 1993.

\bibitem{Gibbons:1979}
G.~W. Gibbons and C.~N. Pope.
\newblock The positive action conjecture and asymptotically {Euclidean} metrics
  in quantum gravity.
\newblock {\em Commun. Math. Phys.}, 66:267--290, 1979.

\bibitem{Schoen:1979}
R.~M. Schoen and S.-T. Yau.
\newblock Proof of the positive action conjecture in quantum relativity.
\newblock {\em Phys. Rev. Lett.}, 42:547--548, 1979.

\bibitem{Schoen:1979b}
R.~M. Schoen and S.-T. Yau.
\newblock Complete manifolds with non-negative scalar curvature and the
  positive action conjecture in general relativity.
\newblock {\em Proc. Natl. Acad. Sci. USA}, 76:1024--1025, 1979.

\bibitem{Zhang:1999}
X.~Zhang.
\newblock Positive mass conjecture for five-dimensional {Lorentzian} manifolds.
\newblock {\em J. Math. Phys.}, 40:3540--3552, 1999.

\bibitem{Dahl:1997}
M.~Dahl.
\newblock The positive mass theorem for {ALE} manifolds.
\newblock {\em Banach Center Publications}, 41.1:133--142, 1997.

\bibitem{Moncrief:2006}
V.~Moncrief.
\newblock Can one {ADM} quantize relativistic bosonic strings and membranes?
\newblock {\em Gen. Rel. Grav.}, 38:561--575, 2006.

\bibitem{Moncrief:unpub}
V.~Moncrief.
\newblock Unpublished.
\newblock This extends a result contained in \cite{Moncrief:2006} to the case
  of curved, maximal hypersurfaces spanning an artibrary, spacelike embedding.

\bibitem{Misner:1973}
C.~W. Misner, K.~S. Thorne, and J.~A. Wheeler.
\newblock {\em Gravitation}.
\newblock W. H. Freeman and Company, San Francisco, 1973.
\newblock Chapter 21.

\end{thebibliography}

\end{document}